\documentclass[12pt,a4paper]{article}
\usepackage[margin=2cm]{geometry}
\usepackage[margin=5pt,font=small,labelfont=bf]{caption}
\usepackage{graphicx,amsmath,amssymb,fge,natbib,setspace,xspace,color}
\usepackage[displaymath, mathlines]{lineno}
\definecolor{darkblue}{rgb}{0.0,0.0,0.6}
\usepackage[colorlinks=true,breaklinks=true,linkcolor=darkblue,urlcolor=darkblue,anchorcolor=darkblue,citecolor=darkblue]{hyperref}
\usepackage[dvipsnames]{xcolor}
\usepackage{multirow}
\usepackage{authblk}

\newcommand{\T}{\rule{0pt}{3.2ex}} 
\newcommand{\B}{\rule[-1.2ex]{0pt}{0pt}} 

\newcommand{\pdiff}[2]{\dfrac{\partial{#1}}{\partial{#2}}}

\newcommand{\matsdiff}[2]{\dfrac{D_s{#1}}{D{#2}}}
\newcommand{\matldiff}[2]{\dfrac{D_{\ell}{#1}}{D{#2}}}
\newcommand{\matbdiff}[2]{\dfrac{\bar{D}{#1}}{D{#2}}}

\newcommand{\Div}{\Grad\!\cdot}
\newcommand{\Grad}{\mbox{\boldmath $\nabla$}}

\newcommand{\infd}{\textrm{d}}

\newcommand{\vs}{\mathbf{v}_s}
\newcommand{\vl}{\mathbf{v}_{\ell}}
\newcommand{\vbar}{\bar{\mathbf{v}}}

\newcommand{\phis}{(1-\phi)}
\newcommand{\fs}{(1-f)}
\newcommand{\q}{\mathbf{q}}

\newcommand{\taubar}{\bar{\boldsymbol{\tau}}}

\newcommand{\Pbar}{\bar{P}}
\newcommand{\rhos}{\rho_s}
\newcommand{\rhol}{\rho_{\ell}}
\newcommand{\rhobar}{\bar{\rho}}
\newcommand{\Deltarho}{\Delta \rho}
\newcommand{\khat}{\hat{\mathbf{k}}}
\newcommand{\gvec}{\mathbf{g}}

\newcommand{\cbar}{\bar{c}^i}

\newcommand{\I}{\mathbf{I}}
\newcommand{\csi}{c_s^i}
\newcommand{\cli}{c_{\ell}^i}
\newcommand{\cseq}{c_s^{i,eq}}
\newcommand{\cleq}{c_{\ell}^{i,eq}}
\newcommand{\feq}{f^{eq}}

\newcommand{\Pet}{\mathrm{Pe}_T}
\newcommand{\Pec}{\mathrm{Pe}_c}
\newcommand{\Diss}{\mathrm{Di}}
\newcommand{\St}{\mathrm{St}}
\newcommand{\Da}{\mathrm{Da}}
\newcommand{\Tp}{\mathcal{T}}
\newcommand{\Tmi}{T_m^i}
\newcommand{\Ki}{K^i}

\newcommand{\G}{\Gamma}
\newcommand{\Gi}{\Gamma^i}
\newcommand{\Di}{\Delta^i}
\newcommand{\RG}{\mathcal{R}_{\Gamma}}
\newcommand{\RD}{\mathcal{R}_{\Delta}}
\newcommand{\cgi}{c_{\Gamma}^i}

\newcommand{\Ts}{T_\mathrm{sol}}
\newcommand{\Tl}{T_\mathrm{liq}}
\newcommand{\degC}{$^\circ$C}

\newcommand{\rev}[1]{{\color{Black}#1}}

\title{The role of volatiles in reactive melt transport in the asthenosphere}
\author[1*]{Tobias Keller}
\author[1]{Richard F.~Katz}
\affil[1]{Dept.~Earth Sciences, University of Oxford, UK}
\affil[*]{tobias.keller@earth.ox.ac.uk}

\begin{document}

\maketitle

\vspace{2cm}
\begin{center}
\textbf{Keywords:} Channelling instability; magma/mantle dynamics; \\partial melting; reactive flow; volatiles.
\end{center}

\onehalfspace

\begin{abstract}
Experimental studies of mantle petrology find that small concentrations of water and carbon dioxide have a large effect on the solidus temperature and distribution of melting in the upper mantle. However, it has remained unclear what effect small fractions of deep, volatile-rich melts have on melt transport and reactive melting in the shallow asthenosphere. \rev{Here we present theory and computations indicating that low-degree, reactive, volatile-rich melts cause channelisation of magmatic flow at depths approximately corresponding to the anhydrous solidus temperature. These results are obtained with a novel method to simulate the thermochemical evolution of the upper mantle in the presence of volatiles. The method uses a thermodynamically consistent framework for reactive, disequilibrium, multi-component melting. It is coupled with a system of equations representing conservation of mass, momentum, and energy for a partially molten grain aggregate. Application of this method in two-phase, three-component upwelling-column models demonstrates that it reproduces leading-order features of hydrated and carbonated peridotite melting; in particular, it captures the production of low-degree, volatile-rich melt at depths far below the volatile-free solidus. The models predict that segregation of volatile-rich, deep melts promotes a reactive channelling instability that creates fast and chemically isolated pathways of melt extraction. Reactive channelling occurs where volatile-rich melts flux the base of the silicate melting region, enhancing dissolution of fusible components from the ambient mantle. We find this effect to be similarly expressed for models of both hydrated and carbonated mantle melting. These findings indicate that despite their small concentrations, water and carbon dioxide have an important control on the extent and style of magma genesis, as well as on the dynamics of melt transport.}
\end{abstract}

\pagebreak \tableofcontents \pagebreak


\section{INTRODUCTION \label{sect:Introduction}}

\subsection{Magmatism in the mantle with volatiles}
Magmatic processes play a central role in the dynamic, thermal and compositional evolution of the Earth surface and interior. One of the challenges in understanding these magmatic systems is the complexity of thermodynamic phase relations of multi-component silicate materials. Apart from its major element composition, the mantle contains a range of volatile elements that are readily dissolved into magma, where it is present. The mantle contains of the order of 100 parts per million by weight (wt~ppm) of H$_2$O and CO$_2$, the two volatile compounds that are the focus of this study.  Experimental studies have found that even small amounts of water and carbon dioxide significantly lower the solidus temperature \citep{gaetani98, hirschmann99, hirschmann09, dasgupta06, dasgupta07a, dasgupta13}. Thus, peridotite melting sets in at significantly higher pressures than in a volatile-free system, stabilising small fractions of volatile rich melts at depth \citep[e.g.,][]{asimow03}. However, it has remained unclear what effect such low-degree mantle melts beneath the volatile-free melting region have on the dynamics of magma transport through the upper mantle. The ability to systematically study the coupling between the dynamics of melt transport through a deforming host rock and the thermo-chemical evolution of magmatic systems with volatiles is thus key to our understanding of terrestrial magmatism.

Here, we present novel computational simulations of the coupled thermo-chemical and dynamic evolution of a partially molten, multi-component silicate mantle with volatiles. These calculations show that water and carbon dioxide exert an important control on the dynamics of magma transport in the asthenosphere. Volatile flux melting at the base of a decompression melting region induces a reaction-transport feedback, leading to localisation of volatile-enriched melt flux into magmatic channels. In these channels, dissolution of fertile material and magma transport are enhanced by up to an order of magnitude.

\subsection{Magmatic channelling beneath mid-ocean ridges}
Most of the Earth's magmatism is concentrated in the upper mantle along plate boundaries such as subduction zones and mid-ocean ridges (MOR). The latter geodynamic environment represents the most widespread form of magmatism in the upper mantle, where mid-ocean ridge-type basalt (MORB) is produced by decompression of the upwelling mantle beneath the ridge axis \citep[e.g.,][]{langmuir92}. The extraction of these magmas to the ridge axis leads to the formation of oceanic crust and accounts for a significant portion of the heat flow out of the Earth interior \citep{pollack93}. 

Observations of U-series disequilibria in young basaltic lavas suggest that melt transport time scales in the upper mantle are of the order of thousands of years \citep{rubin88, richardson94, jull02, stracke06}. \rev{These data} have been interpreted as an indication that magma must, at least partly, be transported in high-flux, chemically isolated channels, rather than by distributed percolation \citep{elliott03}. \rev{Geologic evidence from ophiolites suggest that dunite observed as channels or tabular bodies, rather than harzburgite, has hosted the majority of melt extraction.} The dunite bodies appear to be chemically equilibrated with MORB and bear evidence of reactive dissolution of minerals \citep{kelemen95a, kelemen97}. These observations have long been interpreted as being a consequence of flow localisation in the partially molten asthenosphere beneath MORs, caused by the reactive infiltration instability \citep{chadam86, chadam88, ortoleva87a, ortoleva87b}. This instability is thought to be driven by the increased solubility of orthopyroxene with decreasing pressure, leading to dissolution of orthopyroxene and precipitation of olivine as melt flows upwards. The hypothesis has been tested experimentally \citep{daines94, morgan03, pec15} and theoretically \citep{aharonov95, kelemen95b, spiegelman01, liang10, hewitt10} and confirmed, at least in simplified systems. As growth of this flow instability is a function of melt flux, locally increased melt flux emanating from fertile inclusions (i.e.~remnants of subducted oceanic crust) has been proposed as a trigger for localisation of reactive flow in the mantle \citep{liang10, weatherley12, katz12}.

Apart from the reactive infiltration instability, there are a number of rheological mechanisms that may lead to channelised melt tranport in the mantle. Nonlinear flow due to the stress- and melt-weakening rock viscosity gives rise to melt localisation in shear bands \citep{stevenson89, richardson98, katz06, butler09}. Furthermore, melt overpressure caused by the buoyancy contrast of melt with respect to the host rock may overcome the low tensile strength of rock ($\sim$10~MPa), leading to distributed inter-granular material failure. Such decompaction failure leads to localised melt transport in vertical tubes or angled bands \citep{connolly07, keller13}. Although these rheological instabilities could interact with reactive channelling in the mantle, they will not be considered further in this work.

\subsection{Reactive magma transport models}
The basic mathematical analysis of the reactive transport feedback process was laid out by \cite{chadam86}. Studies that followed have applied simple reactive flow models to magma transport in a mantle column with increasing solubility of a fertile component with decreasing pressure \citep{aharonov95, spiegelman01, liang10, hesse11}. These models apply a constant solubility gradient, linear kinetic reaction rates, and an imposed, incoming melt flux. More importantly, they exclude effects of compositional depletion and latent heat consumption of melting reactions. More recently, these additional effects have been incorporated \citep{hewitt10, weatherley12}, with results suggesting that the factors neglected in earlier models may inhibit reactive channel formation. Nevertheless, by choosing specific binary phase relations, and by imposing sufficient melt flux either as a boundary condition or through large enough fertile inclusions, reactive channeling was reproduced in these more realistic studies as well. Here, we extend the investigation of reactive melt transport by including the effect of volatiles on melt production and magma transport.

To this end, we introduce a coupled fluid dynamic and thermo-chemical model for reactive two-phase, multi-component flow in magmatic systems. The physics of magmatic systems in the mantle is generally described as two-phase flow of melt through the permeable pore space of a deforming and compacting host rock \citep[e.g.,][]{turcotte78, mckenzie84, fowler85, ribe85b, spiegelman93a, bercovici01a, simpson10}. Our model is composed of the standard equations for conservation of mass and momentum in magmatic two-phase flow \citep{mckenzie84}, along with equations for conservation of energy, component and phase mass for two-phase, multi-component reactive flow \citep{rudge11}. To provide closure conditions for the reactive coupling, we introduce the Reactive Disequilibrium Multi-Component method (R\_DMC method), based on theory developed by \cite{rudge11} (\rev{RBS11}). \rev{RBS11} present an approach based on ideal solutions and nonlinear reaction kinetics; we retain the form of their theory while simplifying some of the details to describe the approximate behaviour of effective components of composition. The following adaptations distinguish our approach from that of \rev{RBS11} and other published models: 
\begin{enumerate}
\item Instead of using well-defined chemical components with empirically known thermodynamic properties \citep{ghiorso95}, we define effective components, representing the leading order, collective behaviour of subsets of a complex, multi-component petrological system (i.e.~\rev{grouping together components by their similar melting points and equilibrium partitioning behaviour}).

\item Whereas \rev{RBS11} write their method specifically for solid solutions, \rev{we use the same general form of expressions to approximate phase relations in terms of component-wise solid--melt partition coefficients. By doing so, we neglect more complicated phase relations including eutectics and peritectics, which are challenging to couple with dynamic models. The advantage of this approximation is to have a continuous and uniquely defined solidus and liquidus temperature across the whole range of bulk compositions in the systems.}

\item Unlike \rev{RBS11}, who express composition as molar fractions $x^i_{s,m}$, we instead use component mass fractions $c^i_{s,m}$ to characterise chemical concentrations. \rev{Here and after, superscripts $i$ denote components in an $n$-component compositional space.} The molar mass $M^i$ of proposed effective components is generally not known. As a consequence, the specific gas constant $R^i = R M^i$ ($R$ is the universal gas constant) arising in some thermodynamic relations in \rev{RBS11} is unknown. It will thus be replaced by a quantity $r^i$ of the same units [J~K$^{-1}$~kg$^{-1}$]. These coefficients $r^i$ are used to adjust the temperature dependence of \rev{partition coefficients} and thus the shape of solidus and liquidus curves in constructed phase diagrams.

\item While \rev{RBS11} discuss thermodynamic forces for reaction rates in terms of chemical potential differences, we use simplified relations that depend on the component-mass difference from equilibrium in the liquid phase. 
\end{enumerate}
These adaptations simplify the treatment of mantle petrology enough to compute it robustly and efficiently in two-phase flow models. However, they retain sufficient complexity and flexibility to capture compositional variability and its consequences.

Direct observational evidence on variability of mantle composition and volatile content is sparse. However, the  distribution of fertile material and volatiles in Earth's mantle is thought to be heterogeneous on a variety of length scales, which follows as a consequence of the recycling of hydrated oceanic crust and lithosphere through subduction zones and convective mixing in the deeper mantle \citep[e.g.,][]{gurnis86, tackley00}. In this study, we take an approach similar to \cite{shorttle14}, treating mantle composition as a mixture of effective components representing a refractory residue component, a fertile basaltic component, and some volatile-enriched silicate components. The R\_DMC method is first presented generally for $n$ components and is later calibrated to represent either hydrated or carbonated peridotite melting in a ternary compositional space. With the calibrated method we perform computational simulations of an upwelling mantle column with volatiles. We investigate the general behaviour of decompression melting in a hydrated or carbonated mantle in a one-dimensional column model. We then consider two-dimensional simulations of a mantle column with compositional heterogeneities in order to investigate reactive melting and reaction transport feedbacks. Results indicate that low degree, high volatile melt formed at depths below the volatile-free melting region cause focusing of melt flow into volatile-flux-driven reactive channels. These findings provide a new perspective on the role of volatiles in mantle magmatism.

The paper is organised as follows. First, the \rev{Method} section introduces the R\_DMC thermodynamic model and discusses some model choices and characteristic properties. This section also states the model equations for conservation of mass, momentum, energy and composition for the two-phase, multi-component, reactive flow problem; it describes the melt model calibration and numerical implementation used to solve the model equations in the context of MORB petrogenesis with volatiles. In the \rev{Results} section we present results of the one- and two-dimensional simulations of reactive flow in an upwelling mantle column. In the \rev{Discussion} section we discuss these results in the context of volatile-flux-driven reactive channelling, and further consider the implications of the results for the petrology and dynamics of magmatic systems.


\section{METHOD \label{sect:Method}}

\subsection{Reactive disequilibrium multi-component method \label{sect:R_DMC_method}}
Reactions in magmatic systems are driven by chemical disequilibria, i.e., differences in chemical potential between phases.  These differences arise as an assemblage of minerals experiences changes in temperature, pressure or bulk composition due to dynamic processes such as tectonic uplift or subsidence, contact or regional metamorphism, or metasomatism due to fluid or melt percolation. Our disequilibrium-driven reactive model, the R\_DMC method, is formulated in terms of the thermodynamic state variables of temperature $T$, pressure $P$ and bulk composition $\cbar$. The latter is defined in a general, $n$-component compositional space, where mass fractions of all components $\cbar$ must sum to unity. Here and throughout the manuscript, an overbar represents a phase mass-averaged quantity, such as the bulk composition $\cbar = f \cli + \fs \csi$, with $f$ the local mass-fraction of melt. The only exception to this notation is the density of the two-phase aggregate, which is phase-averaged by volume as $\rhobar = \phi \rho_{\ell} + \phis \rhos$, with \rev{$\phi=(\rhol/\rhobar) f$} the local volume-fraction of melt.

The R\_DMC method consists of applying the following four steps: (\textit{i}) determine \rev{partition coefficients} $\Ki$ for each thermodynamic component $c^i$ at given $P,T$-conditions; (\textit{ii}) for a given bulk composition $\cbar$, determine the unique melt fraction $\feq$ for which all components are in equilibrium according to $K^i$; (\textit{iii}) determine the equilibrium phase compositions $\cseq$, $\cleq$ at equilibrium melt fraction $\feq$; (\textit{iv}) use linear kinetic constitutive laws to determine equilibration reaction rates driven by compositional disequilibria. In the following, we will describe each of these four steps in turn.

\subsubsection{Thermodynamic equilibrium}
The simplified thermodynamic equilibrium in the R\_DMC method is defined in terms of the equilibrium phase compositions $\cseq$, $\cleq$ together with the melt fraction at equilibrium $\feq$. This equilibrium should be uniquely defined for any given state of the system as characterised by pressure $P$, temperature $T$, and bulk composition $\cbar$,
\begin{linenomath*}\begin{align}
	\label{eq:Equilibrium}
	P, T, \cbar \rightarrow \feq, \cseq, \cleq. \hspace{165pt} \left( 2n+1 \right)
\end{align}\end{linenomath*}
To determine the $(2n+1)$ degrees of freedom of the equilibrium state, we require an equal number of independent constraints. The first set of $n$ constraints are given by the component \rev{partition coefficients, which we define as the ratio of solid to liquid concentration of a component at equilibrium}. The component \rev{partition coefficients} are functions of $T$ and $P$. The functional form of these dependencies is identical to that used in \rev{RBS11}, except that the curvature parameter $r^i$ is substituted for $R^i$ as discussed above,
\begin{linenomath*}\begin{align}
	\label{eq:CompDistrCoeff}
	\Ki &= \dfrac{\cseq}{\cleq} = \exp \left[ \dfrac{L^i}{r^i} \left( \dfrac{1}{T} - \dfrac{1}{\Tmi(P)}\right) \right] \: .
\end{align}\end{linenomath*}
Even though the latent heat of pure-component fusion $L^i$ is not necessarily well defined for every choice of components, \rev{it may usually be constrained by calorimetric experiments to within a reasonable range \citep[e.g.,][]{kojitani97}}. The $P$-dependence of the component melting points $\Tmi(P)$ is given by quadratic polynomials
\begin{linenomath*}\begin{align}
	\label{eq:CompMeltingPoint}
	\Tmi(P) &= T^i_{m,0} + A^i P + B^i P^2 \: ,
\end{align}\end{linenomath*}
with $T^i_{m,0}$ the component melting points at zero pressure, and $A^i$ and $B^i$ the coefficients of the polynomial.

A set of $n$ compositional constraints comes from the lever rules, which are expressions relating the bulk composition to the phase compositions by the phase proportions,
\begin{linenomath*}\begin{align}
	\label{eq:LeverRules}
	\cbar &= (1-\feq) \cseq + \feq \cleq  \: .
\end{align}\end{linenomath*}
The last constraint required to close the thermodynamic equilibrium system is given by the unity sum of all compositions $\sum c^i = 1$. Combining this criterion for the solid and liquid phase, we know that for any partially molten equilibrium state, the following holds,
\begin{linenomath*}\begin{align}
	\label{eq:UnitySum}
	\sum\limits_{i=1}^{n} \cseq - \sum\limits_{i=1}^{n} \cleq &= 0  \: .
\end{align}\end{linenomath*}
Having thus established the required $(2n+1)$ independent constraints, the procedure used to determine the equilibrium state (steps (\textit{i})--(\textit{iii}) of the R\_DMC method) is as follows.
\begin{itemize}
	\item[(\textit{i})] Evaluate equilibrium \rev{partition coefficients} $\Ki$ at given $P,T$.
	\item[(\textit{ii})] Use unity sum constraint \eqref{eq:UnitySum} in terms of $\cbar, \Ki, \feq$ to state that, at equilibrium,
	\begin{linenomath*}\begin{align}
		\label{eq:EquilibriumMelt}
		\sum\limits_{i=1}^{n} \dfrac{\cbar}{\feq / \Ki + (1-\feq)} - \sum\limits_{i=1}^{n} \dfrac{\cbar}{\feq + (1-\feq) \Ki} = 0  \: .
	\end{align}\end{linenomath*}
	Solve \eqref{eq:EquilibriumMelt} numerically (e.g.,~by Newton's method) to find $\feq$.
	\item[(\textit{iii})] Obtain $\cseq, \cleq$ at given $\cbar, \Ki, \feq$ by lever rule as
	\begin{linenomath*}\begin{align}
		\label{eq:EquilibriumComp}
		\cseq &= \dfrac{\cbar}{\feq / \Ki + (1-\feq)}, \\\nonumber
		\cleq &= \dfrac{\cbar}{\feq + (1-\feq) \Ki}.
	\end{align}\end{linenomath*}
\end{itemize}

Although not strictly required by the R\_DMC method, it is useful to identify the solidus and liquidus temperatures at a given $P$, $\cbar$ (for analysis or visualisation). However, no explicit statement for $\Ts$ or $\Tl$ may be given, as temperature only features implicitly in eqn.~\eqref{eq:CompDistrCoeff} for $\Ki$. To obtain $\Ts$, we note that on the solidus, $\cseq = \cbar$, $\feq=0$, and thus $\cleq = \cbar/\Ki$. Stating the unity sum constraint for the liquid phase on the solidus, we write
\begin{linenomath*}\begin{align}
	\label{eq:TsolConstraint}
	\sum\limits_{i=1}^{n} \dfrac{\cbar}{\Ki} = \sum\limits_{i=1}^{n} \dfrac{\cbar}{\exp \left[ \dfrac{L^i}{r^i} \left( \dfrac{1}{\Ts} - \dfrac{1}{\Tmi}\right) \right]} = 1 \: .
\end{align}\end{linenomath*}
This expression is solved numerically for $\Ts$. To find $\Tl$ we apply the same logic, knowing that on the liquidus $\cleq = \cbar$, $\feq=1$ and thus $\cseq = \cbar\Ki$. Therefore we state the unity sum constraint for the solid phase on the liquidus as
\begin{linenomath*}\begin{align}
	\label{eq:TliqConstraint}
	\sum\limits_{i=1}^{n} \cbar\Ki = \sum\limits_{i=1}^{n} \cbar \exp \left[ \dfrac{L^i}{r^i} \left( \dfrac{1}{\Tl} - \dfrac{1}{\Tmi}\right) \right] = 1 \: ,
\end{align}\end{linenomath*}
which is solved numerically for $\Tl$. Matlab routines performing all of the steps above are available for download \rev{from an online repository \citep{rdmc-repo}}.

\subsubsection{Equilibration reaction rates}
Without mass transfer by reaction, any magmatic system will become disequilibrated due to dynamically evolving $P,T$-conditions and changes in $\cbar$ due to segregation of melt with respect to its solid residue. In nature, chemical disequilibrium causes reactions to take place to re-equilibrate the system.

In the R\_DMC method, simplified linear kinetic reaction rates are constructed to simulate this fundamental property of magmatic systems. We refer to this type of model approach as ``reactive disequilibrium'' melting to distinguish it from previous approaches \citep{katz08, hewitt10, weatherley12}, where the computed equilibrium state for melt fraction and phase compositions is instantaneously applied to a dynamic model. Following a reactive disequilibrium approach allows testing of scenarios in which reactive equilibration is more or less efficient relative to melt transport, or where composition of reacted mass depends on various petrological model assumptions.

The following reaction rates need to be defined in a two-phase, multi-component context. $\G$ is the net mass transfer rate from solid to liquid phase (and vice versa with negative sign), also referred to as net melting rate. The set of $n$ rates $\Gi$ are the mass transfer rates of each component from solid to liquid phase. The component melting rates are required to sum to the net melting rate as $\sum \Gi = \G$.

The \rev{component} mass transfer between solid and liquid phase required to equilibrate a disequilibrated system is $\rhobar ( \feq \cleq - f \cli )$. Choosing to apply this mass transfer linearly over a characteristic time scale of reaction $\tau_{\G}$, we write
\begin{linenomath*}\begin{align}
	\label{eq:CompReactDeriv}
	\Gi = \rhobar \dfrac{\feq \cleq - f \cli}{\tau_{\G}}.
\end{align}\end{linenomath*}
By taking $\rhobar \approx \rho_0$ (constant reference density) and substituting the linear kinetic rate factor $\RG = \rho_0 / \tau_{\G}$, we write the closure conditions for the component reaction rates as
\begin{linenomath*}\begin{align}
	\label{eq:BatchReaction}
	\Gi_b &= \RG \left( \feq \cleq - f \cli \right).
\end{align}\end{linenomath*}
The net melting rate $\G$ is the sum of $\Gi_b$. These reaction rates drive the system toward chemical equilibrium. This corresponds to the process known as batch melting, where any new increment of melt is of a composition that allows any present volume of melt and rock to be fully equilibrated. Therefore, we refer to this type of reaction rate $\Gi_b$ as batch reactions.

However, petrological and geochemical observations suggest that mantle melting is better understood in terms of a fractional-assimilative model, where increments of newly formed melt are of a composition that is in local thermodynamic equilibrium with its solid residue only. As these fractional melts are transported away from their residues, they mix with other melts and undergo reactive assimilation with other solids they contact. In order to capture this type of behaviour, we adapt the approach taken by \rev{RBS11}.

We write the general constitutive laws for fractional-assimilative component reaction rates,
\begin{linenomath*}\begin{align}
	\label{eq:CompReaction}
	\Gi &= \cgi \G + \Di \: .
\end{align}\end{linenomath*}
Here, reaction rates are decomposed into two parts. The first term captures phase change reactions (net mass transfer between the phases, $\sum \cgi \G = \G$), whereas the second term captures component exchange reactions (no net mass transfer, $\sum \Di = 0$). The reactive component mass fractions $\cgi$ are thus subject to a unity sum constraint ($\sum \cgi = 1$). It does not follow, however, that $\cgi$ need to \rev{satisfy the constraints $0 \leq \cgi \leq 1$}. If, for example, the first component of $\cgi$ in a two-component system is larger than one, then the second component will be smaller than zero in order for both to sum to unity. In this case, reaction rates of the form given in \eqref{eq:CompReaction} capture what is known as an incongruent melting reaction, a relevant feature in mantle melting processes.

Using the same pattern as above in \eqref{eq:BatchReaction}, we propose a constitutive law for the net melting rate of the form
\begin{linenomath*}\begin{align}
	\label{eq:NetMelting}
	\G &= \RG \left( \feq - f \right) \: .
\end{align}\end{linenomath*}
Depending on the choice of $\cgi$, this net melting rate will not lead to complete reactive equilibration. The choice of component exchange rates $\Di$ should therefore be such that in the limit of very fast reactions, the fractional-assimilative reactive model \eqref{eq:CompReaction} will result in full equilibration, and therefore should be equivalent to the batch reactive model \eqref{eq:BatchReaction}. Following from this, we assert that
\begin{linenomath*}\begin{align}
	\label{eq:ExchangeReactDeriv}
	\Di = \Gi_b - \cgi \G \: .
\end{align}\end{linenomath*}
By substituting \eqref{eq:BatchReaction} and \eqref{eq:NetMelting} into \eqref{eq:ExchangeReactDeriv} and grouping terms, we find a suitable constitutive law for $\Di$,
\begin{linenomath*}\begin{align}
	\label{eq:ExchangeReactFinal}
	\Di = \RD \left( \feq (\cleq - \cgi) - f (\cli - \cgi) \right) \: .
\end{align}\end{linenomath*}
Note that here we introduced a separate rate factor for exchange reactions, $\RD = \rho_0 / \tau_{\Delta}$. Thus we allow exchange reactions to occur with a different reactive time scale  from phase change reactions, reflecting the possibility that different physics may be involved. In theory, a different rate factor could be applied for each chemical component. \rev{Natural multi-component systems are known to exhibit a large range of component diffusivities both in the solid and melt, which lead to significantly different time scales of equilibration for various components. Even though this effect is expected to be particularly important for hydrated systems, where hydrogen is more diffusive than most other species, we will leave this issue to be considered in future work.}

What remains to be determined for a complete description of the reactive model is the reactive composition $\cgi$; that is, the concentration of components in the reactively transferred mass. We follow \rev{RBS11} and choose $\cgi$ to be the fractional melt composition
\begin{linenomath*}\begin{align}
	\label{eq:ReactComposition}
	\cgi &= \left\lbrace \dfrac{\csi}{\Ki} \right\rbrace \: ,
\end{align}\end{linenomath*}
where $\lbrace \cdot \rbrace$ indicates normalisation to unity sum of components. More sophisticated choices could be applied here, but for the context of near-equilibrium decompression melting to which we will apply this method below, this simple choice is sufficient.

Given the reaction rates above, any melt, once it is formed, is allowed to react with any volume of rock it comes into contact with. The amount of assimilation depends on the relative magnitude of $\RG$ and $\RD$ and the rate of melt extraction, as determined by the fluid dynamics of the magmatic system.

In summary, we present a reactive model that, given the appropriate choice of parameters, will recover either batch or fractional-assimilative melt evolution. Even though the parameter choices may be difficult to constrain from experiments, this model allows for systematic variation of reactive behaviour of a magmatic system in order to quantitatively assess parameter sensitivities in the system.

\subsection{Two-phase, multi-component flow model  \label{sect:2phase_multicomp_model}}

A physical model description of a two-phase, multi-component flow must be based on conservation equations for mass, momentum, energy, and chemical composition. In the following we briefly present these equations; a more detailed derivation is found in Appendix A and, for example, \cite{mckenzie84}, \cite{bercovici01a}, or \cite{rudge11}.

Consider a Cartesian coordinate system with the origin at the surface and the $z$-direction pointing vertically downwards in the direction of the gravity vector $\gvec = g \khat$. The following physical model describes the dynamics of a two-phase (solid--liquid) medium consisting of a contiguous skeleton or matrix of silicate minerals as the solid phase and an \rev{interconnected network of} silicate melt as the liquid phase. Both material phases are characterised by a multi-component chemical composition.

\subsubsection{Thermo-chemical model}
The thermochemical model enforces conservation of energy and mass for magmatic processes in the mantle. The model is expressed in terms of the temperature $T$, the mass fraction of melt $f$, and mass fractions of $n$ thermochemical components in the solid and liquid phase $c_{s,\ell}^i$, or in the two-phase mixture $\cbar$. \rev{Phase and component mass fractions are defined as the ratio of the mass of a specific phase or component to the total mass of all phases or components present in a control volume}.

Conservation of melt and component mass in terms of $f$ and $c^i_{s,\ell}$ is given by
\begin{linenomath*}%
  \begin{subequations}
    \label{eq:dimensional_governing_thermochem}
    \begin{align}
      \label{eq:MeltMassFinal}
      \matsdiff{f \rhobar}{t} &= \hspace{9pt} \Gamma + \fs \rhobar \Div \vs, \\
      \label{eq:LiquidCompMassFinal}
      f \rhobar \matldiff{\cli}{t} &= \hspace{9pt} \Gamma^i - \cli \Gamma + d_0 \Div f \Grad \cli, \\
      \label{eq:SolidCompMassFinal}
      \fs \rhobar \matsdiff{\csi}{t} &= -\Gamma^i + \csi \Gamma .
    \end{align}
  \end{subequations}%
\end{linenomath*}
Thus, $f$ evolves by advection, compaction and reaction, and components $c^i_{s,\ell}$ evolve by advection and reaction, along with diffusion/dispersion in the liquid phase. Note that component mass is advected by the corresponding phase velocity. Therefore the equations are written in terms of Lagrangian derivatives for solid and liquid materials, ${D_j()}/{Dt} = {\partial()}/{\partial t} + \mathbf{v}_j \cdot \Grad ()$. Pores between mineral grains are advected by the solid material, and thus the Lagrangian derivative associated with the solid velocity is applicable for melt fraction $f$. $d_0$ is the diffusive flux coefficient, taken to be constant for all components. \rev{As previously noted, natural component diffusivities may vary over orders of magnitude, leading to potentially interesting reaction and mass transport phenomena, as for example the rapid diffusion of sodium from melt channels into surrounding peridotite reported by \cite{lundstrom00}. However, such phenomena are most relevant on spatial scales not resolved in upper mantle scale continuum models.}

The fraction of the liquid material phase (melt) in a control volume can also be expressed as volume fraction $\phi$ or porosity. The two quantities are related to each other by $\rhobar f = \rhol \phi$ and $\rhobar \fs = \rhos \phis$. In this manuscript, we consistently express melt content as mass fraction $f$, except in melt-dependent constitutive laws of material parameters, which are expressed as functions of $\phi$ (see Table~\ref{tab:parameters}).

\begin{table}[htb]
	{\centering
	\caption{\textbf{Material parameters, definitions, and reference values}}
	\label{tab:parameters}
  	\begin{tabular}{llll}
	\hline
	Mixture density & $\rhobar$ & kg m$^{-3}$ & $\rho_0 - \phi \Deltarho_0$ \T \\
	Rock density constant & $\rho_0$ & kg m$^{-3}$ & 3200 \\
	Thermal expansivity & $\alpha$ & K$^{-1}$ & 2e-5 \\
	Specific heat capacity & $c_p$ & J kg$^{-1}$K$^{-1}$ & 1000 \\
	Latent heat of melting & $L^i$ & kJ kg$^{-1}$ & [600,450,350,350] \\
	Thermal conductivity & $k_0$ & J K$^{-1}$m$^{-1}$s$^{-1}$ & 2 \\
	Chemical diff./disp. flux constant & $d_0$ & kg m$^{-1}$s$^{-1}$ & 1e-4 \\
	Mantle upwelling rate & $u_m$ & cm yr$^{-1}$ & 5 \\
	Effective shear viscosity of rock & $\eta$ & Pa s & $(1-\phi) \eta_0 \exp(-\alpha_\phi \phi)$ \\
	Rock viscosity constant & $\eta_0$ & Pa s & 1e19 \\
	Rock viscosity melt-weakening factor & $\alpha_\phi$ & Pa s & 27 \\
	Effective compaction viscosity of rock & $\zeta$ & Pa s & $(1-\phi) \eta_0 \phi^{-p}$ \\
	Compaction viscosity exponent & $p$ & - & 1 \\
	Shear viscosity of melt & $\mu$ & Pa s & $\mu_0 \prod\limits \lambda_i^{\cli}$ \\
	Melt viscosity constant & $\mu_0$ & Pa s & 1 \\
	Melt viscosity compositional factors & $\lambda_i$ & - & [1,100,0.01,0.001] \\
	Permeability of rock & $K$ & m$^2$ & $K_\mathrm{ref} \phi^n$ \\
	Permeability constant & $K_\mathrm{ref}$ & m$^2$ & 1e-6 \\
	Reaction time & $\tau_\Gamma$ & yr & 500 (1D), 100 (2D) \\
	Assimilation ratio & $\chi$ & - & 1 \B \\\hline
	\end{tabular}}
\end{table}

Conservation of energy is formulated in terms of sensible heat $\rhobar c_p T$, which evolves in time due to advection, adiabatic heat exchange, thermal diffusion, latent heat exchange of reactions, and heating by viscous dissipation:
\begin{linenomath*}\begin{align}
	\label{eq:EnConsFinal}
	\rhobar c_p \matbdiff{T}{t} &= \alpha T g \rhobar \bar{w} + k_0 \Grad^2 T - \sum\limits_{i=1}^n L^i\Gamma^i + \Psi \: .
\end{align}\end{linenomath*}
Thermal material parameters are taken as constant and equal in all materials of all compositions. ${\bar{D}()}/{Dt} = \pdiff{()}{t} + \fs \vs \cdot \Grad () + f \vl \cdot \Grad ()$ is the material derivative of the two-phase mixture. The adiabatic exchange between sensible heat and compressive work depends on the vertical component of the mass-averaged velocity $\bar{w} = \fs w_s + f w_{\ell}$. The form of the viscous dissipation rate $\Psi$ is given in Appendix A. 

Substituting the potential temperature $\Tp = T \exp(\alpha g z / c_p)$ and cancelling all terms with $\alpha g / c_p \ll 1$, the above equation is rewritten in terms of $\Tp$ as
\begin{linenomath*}\begin{align}
	\label{eq:EnConsTpotA}
	  \rhobar c_p \matbdiff{\Tp}{t} &= k_0 \Grad^2 \Tp + \left( \Psi - \sum\limits_{i=1}^n L^i\Gamma^i \right) e^{-\alpha g z / c_p} \: .
\end{align}\end{linenomath*}
See Table~\ref{tab:parameters} for notation, definitions and typical parameter values.

\subsubsection{Fluid dynamics model}
Conservation of momentum and mass for a viscous two-phase flow model in the geodynamic limit ($\mu \ll \eta$) are given by the standard equations \citep{mckenzie84} written in the three-field form for solid velocity $\vs$, compaction pressure $p$ and dynamic Stokes pressure $P$ \citep{keller13, rhebergen14}:
\begin{linenomath*}%
  \begin{subequations}
    \label{eq:dimensional_governing_mechanics}
    \begin{align}
      \label{eq:BulkMomentumFinal}
      \Div 2\eta \mathbf{D}(\vs) + \Grad p - \Grad P &= 0, \\
      \label{eq:BulkMassFinal}	
      \Div \vs - \Div (K/\mu) \left(\Grad P + \Deltarho \gvec \right) - \Gamma \Delta(1/\rho)&= 0, \\
      \label{eq:CompactionFinal}
      \Div \vs + p/\zeta &= 0.
    \end{align}
  \end{subequations}%
\end{linenomath*}
The deviatoric symmetric gradient tensor operator is defined as
$\mathbf{D}() = \frac{1}{2}(\Grad() + \Grad()^T) - \frac{1}{3} \Div () \I$. Volumetric changes due to phase change reactions $\G$ depend on the inverse density difference $\Delta(1/\rho)=1/\rhos-1/\rhol$. Buoyancy is only considered as a driving force for melt segregation. The liquid velocity is not an independent variable in this problem and is obtained in a post-processing step from Darcy's law as $\vl = \vs - K \left(\Grad P + \Deltarho \gvec \right)/(\phi\mu) $. 

To close the system \eqref{eq:dimensional_governing_thermochem}, \eqref{eq:EnConsTpotA}, and \eqref{eq:dimensional_governing_mechanics} of 5 plus $2n$ equations, constitutive laws are required for various material properties. These are listed in Table~\ref{tab:parameters}, along with characteristic values for the mantle. The shear viscosity $\eta$ is taken to be independent of temperature and pressure, but is weakened by the presence of melt \cite{mei02}. The compaction viscosity $\zeta$ is defined relative to the shear viscosity and depends on melt fraction. The liquid viscosity varies with composition to capture leading order effects of the polymerisation state of a magma with variable silica and volatile content \citep{bottinga72,kushiro76,giordano03,hui07}. For simplicity, we choose a log-linear law to represent such compositional variations in melt viscosity. The melt-dependence of permeability is given by the low-porosity Kozeny-Carman relationship.

\subsubsection{Non-dimensionalisation \label{sect:nondim}}
With the scaling quantities and dimensionless numbers listed in Table~\ref{tab:scaling}, we write the governing equations \eqref{eq:dimensional_governing_thermochem}--\eqref{eq:dimensional_governing_mechanics} in non-dimensional form as
\begin{linenomath*}%
  \begin{subequations}
    \label{eq:nondim_governing_all}
    \begin{align}
      \label{eq:EnConsFinalNonDim}
      \rhobar' \matbdiff{\Tp}{t} - \Pet^{-1} \Grad^2 \Tp - \left( \Diss \Psi - \sum_{i=1}^n{\St^i \Gi} \right) e^{-\alpha g z / c_p} &= 0 , \\
      \label{eq:LiquidCompMassFinalNonDim}
      f \rhobar' \matldiff{\cli}{t} - \Pec^{-1} \Div f \Grad \cli + \cli \G - \Gi &= 0 , \\
      \label{eq:SolidCompMassFinalNonDim}
      (1-f) \rhobar' \matsdiff{\csi}{t} - \csi \G + \Gi &= 0 , \\
      \label{eq:MeltMassFinalNonDim}
      \matsdiff{f \rhobar'}{t} - (1-f) \: \rhobar' \: \Div \vs - \G &= 0 , \\
      \label{eq:BulkMomentumFinalNonDim}
      \Div 2\eta' \mathbf{D}(\vs) + \Grad p - \Grad P &= 0 , \\
      \label{eq:BulkMassFinalNonDim}	
      \Div \vs - \Div (K'/\mu') \left(\Grad P + \khat \right) - \Gamma \Delta(1/\rho') &= 0 , \\
      \label{eq:CompactionFinalNonDim}
      \Div \vs + p/\zeta' &= 0.
    \end{align}
  \end{subequations}%
\end{linenomath*}
Note that primes on non-dimensional quantities have been dropped, except for the non-dimensional material parameters, which are defined according to the constitutive laws given in Table \ref{tab:parameters} as:
\begin{subequations}
\label{eq:nondim_constitutive}
\begin{linenomath*}%
    \begin{align}
	\rhobar' &= \rho_0/\Deltarho_0 - \phi ,\\
	\eta' &= \phis \exp(-\alpha_\phi \phi) ,\\
	\zeta' &=  \phis r_{\zeta} \phi^{-p} ,\\
	K' &= \left( {\phi}/{\phi_0} \right)^n , \\
	\mu' &= \prod\limits_{i=1}^n \lambda_i^{\cli}.
    \end{align}
\end{linenomath*}
\end{subequations}

\begin{table}[htb]
\centering
	\caption{\textbf{Scaling quantities, dimensionless numbers, and values}}
	\label{tab:scaling}
	\begin{tabular}{lrll}
	\hline
	Temperature & $T_0 \:\:\: =$ & $T_m$ & 1350 $^{\circ}$C \T \\
	Density & $\Deltarho_0 \:\:\: =$ & $\rhos - \rhol$  & 500 kg m$^{-3}$ \\
	Rock viscosity & $\eta_0 \:\:\: =$ &  & 1e19 Pa s \\ 
	Melt viscosity & $\mu_0 \:\:\: =$ &  & 1 Pa s \\
	Permeability & $K_0 \:\:\: =$ & $K_\mathrm{ref} \phi_0^n$ & 1e-12 m$^2$ \\
	Melt fraction & $\phi_0 \:\:\: =$ &  & 0.01 \\
	Gravity & $g \:\:\: =$ & $|\gvec|$ & 9.81 m s$^{-2}$ \\
	Length & $\l_0 \:\:\: =$ & $ \sqrt{\eta_0 K_0~\mu_0^{-1}}$ & 3.162~km \\
	Pressure & $p_0 \:\:\: =$ & $\Deltarho_0 g l_0$ & 15.51 MPa \\
	Velocity & $u_0 \:\:\: =$ & $K_0 \Deltarho_0 g~\mu_0^{-1}$ & 15.48~cm yr$^{-1}$ \\
	Time & $t_0 \:\:\: =$ & $l_0~u_0^{-1}$ & 20.43 kyr \\
	Reaction rate & $\Gamma_0 \:\:\: =$ & $\Deltarho_0 u_0~l_0^{-1}$ & 24.5 g m$^{-3}$yr$^{-1}$ \\
	Dissipation rate & $\Psi_0 \:\:\: =$ & $\eta_0 u_0^2~l_0^{-2}$ & 2.406e-05 W m$^{-3}$ \\
	Thermal Peclet number & $\Pet \:\:\: =$ & $\Deltarho_0 u_0 l_0 c_p~k^{-1}$ & 4.265 \\
	Compositional Peclet number & $\Pec \:\:\: =$ & $\Deltarho_0 u_0 l_0~d^{-1}$ & 775.5 \\
	Dissipation number & $\Diss \:\:\: =$ & $\eta_0 u_0(c_p T_0 \Deltarho_0 l_0)^{-1}$ & 0.0209 \\
	Stefan numbers & $\St^i \:\:\: =$ & $L^i(c_p T_0)^{-1}$ & [0.404,0.303,0.303,0.236] \\
	Dahmkohler number & $\Da \:\:\: =$ & $\rho_0 u_0(\Deltarho_0 l_0 \tau_{\Gamma})^{-1}$ & 261.5 \\
	Assimilation ratio & $\chi \:\:\: =$ & $\tau_{\Gamma}~\tau_{\Delta}^{-1}$ & 0.1-1 \B \\\hline
	\end{tabular}
\end{table}

Non-dimensional reaction rates are governed by the Dahmkohler number $\Da$ and the assimilation ratio $\chi$ as
\begin{subequations}
\label{eq:nondim_reaction}
\begin{linenomath*}\begin{align}
	\Gi &= \cgi \G + \Di \:\: , \\
	\G  &= \Da (\feq - f) \:\: , \\
	\Di &= \Da \chi \left( \feq (\cleq - \cgi) - f (\cli - \cgi) \right).
\end{align}\end{linenomath*}
\end{subequations}

\subsection{Scaling analysis for reactive melting}
\rev{Assuming steady state ($\partial\cdot / \partial t = 0$) and full equilibration ($\csi = \cseq$, $\cli = \cleq$), and neglecting diffusion/dispersion ($d_0 = 0$), the sum of conservation equations for component mass in the solid and liquid phase in a 1D vertical column can be arranged to yield an expression for the net melting rate as a function of flow-induced disequilibrium in both phases \citep[cf.][]{aharonov95, spiegelman01, hewitt10}:}
\begin{linenomath*}\begin{align}
\label{eq:decomp_reaction}
	\G &= \dfrac{f \rhobar w_\ell \dfrac{\partial \cleq}{\partial z} - \fs \rhobar w_s \dfrac{\partial \cseq}{\partial z}} {(2c_r^i - \cseq - \cleq)} \:\: .
\end{align}\end{linenomath*}
\rev{Here, we have introduced the effective concentration of components in the reactively transferred mass $c_r^i = \Gi / \G$. To better understand the physical meaning of this expression, we perform a scaling analysis. With equilibrium expressed by partition coefficients $K^i$, and letting liquid concentration scale as $\cleq \sim c^i_0$, the solid concentration scales as $\cseq \sim K^i c^i_0$. The scaling for $c^i_r$ is expressed relative to the liquid concentration as $c^i_r \sim c^i_0 + \Delta c_r$. We further introduce solubility gradient scales in the solid and liquid phase, ${\partial \cleq}/{\partial z} \sim \delta_\ell^{i,eq}$ and $-{\partial \cseq}/{\partial z} \sim \delta_s^{i,eq}$, defined as positive to induce melting. Assuming that melt fractions in the upwelling mantle are small, we approximate $(1-f_0) \approx 1$. Lastly, we express the liquid scaling velocity $w_0$ as a function of solid scaling velocity $u_0$ as $w_0 = R u_0$, where the velocity ratio $R$ increases with increasing melt mobility \citep[cf.][]{liang10, hesse11}. With these simplifying assumptions, we write a scaling relation for \eqref{eq:decomp_reaction} as}
\begin{linenomath*}\begin{align}
\label{eq:decomp_reaction_simple}
	\G_0 &\sim \dfrac{\rho_0 u_0}{\Delta c^i_r} (R f_0 \delta_\ell^{i,eq} + \delta_s^{i,eq}) \:\: .
\end{align}\end{linenomath*}
\rev{Note, we have further simplified the scaling by absorbing a factor 1/2, and assuming in the denominator $(1-\frac{1}{2}(1+K^i)) \approx 0$. According to \eqref{eq:decomp_reaction_simple}, melting in an upwelling mantle column is driven by two processes. First, the upwelling flux of the solid mantle transports rock along solubility gradients in the mantle, leading to what decompression melting. This process scales as $\rho_0 u_0 \delta_s^{eq}$. Second, the flux of melt along solubility gradients in the liquid phase leads to reactive melting. This second process scales as $R f_0 \rho_0 u_0 \delta_\ell^{eq}$. We also note that melt productivity is inversely proportional to $\Delta c^i_r$, highlighting that the stoichiometry of reactions (here implied as proportion of components in reactive mass exchange) modulates melt productivity driven by a given flux-induced disequilibration rate.}

\rev{Theory predicts the emergence of reactive channels by the reactive infiltration instability \citep{chadam86}. Reactive channelling arises when perturbations in melt flux lead to locally increased flux melting relative to the laterally more homogeneous decompression melting. This condition is here expressed by the non-dimensional ratio of disequilibration rates induced by liquid and solid flow, $M_0$:}
\begin{linenomath*}\begin{align}
\label{eq:reactive_melting_number}
	M_0 
        =  R f_0 \dfrac{ \delta_\ell^{eq}}{\delta_s^{eq}} \:\: .
\end{align}\end{linenomath*}
\rev{For scenarios of mantle melting with $M_0 \ll 1$, we expect laterally homogeneous decompression melting and distributed melt percolation to dominate. However, for $M_0 \gtrsim 1$, the positive feedback between melt transport and melt production promotes the formation of reactive channels with a characteristic spacing proportional to the compaction length, and a width inversely proportional to the ratio of melt flux rates inside to outside these channels.}

\rev{Following from this scaling analysis, we understand reactive channelling to be favoured if the melt to solid velocity ratio is large, which is the case for high rock permeability, low liquid viscosity, or high melt buoyancy contrast. Furthermore, reactive channelling requires a large solubility gradient for major melt producing components in the liquid compared to the solid phase \citep{hesse11}. Although this simplified scaling analysis neglects various components of the coupled flow problem, such as latent heat consumption, compaction, and depletion of the soluble component \citep{hewitt10,weatherley12}, it will facilitate interpretation of the results presented below.}

\subsection{Mantle composition}
Water and carbon dioxide are known to depress the solidus of mantle peridotite, lower the depth of first melting by tens of kilometers, and stabilise small fractions of volatile-rich melt at depth. To reproduce these leading order features of hydrated and carbonated peridotite melting, we choose a four-component compositional space defined by four end-member compositions: \rev{(1) dunite, (2) mid-ocean ridge-type basalt (MORB), (3) hydrated basalt (hMORB), and (4) carbonated basalt (cMORB). These  components are chosen to represent groups of mantle minerals selected for their distinct behaviour in decompression melting of a hydrated and carbonated peridotite. The first two components represent the most refractory and fertile portions of volatile-free peridotite, respectively. The first component, dunite, represents the residue of decompression melting, whereas the second one, MORB, represents its product. Any volatile-free mantle composition is thus modelled by mixtures of various proportions of these two components.}

\rev{Note that we refrain from specifying mineral assemblages or oxide compositions to further characterise our choice of components. Our approach is designed for testing the dynamic consequences of simple hypotheses regarding mantle petrogenesis, and is not suited for resolving the actual multi-component thermodynamics of natural silicate materials. Here, our main focus is on the effect of water and carbon dioxide on mantle melting and melt transport, and therefore we choose the simplest possible representation of volatile-free basaltic decompression melting, which is a two-component mixture of melt residue and melt product, to which we now add the volatile-bearing components.}

\rev{Representing the mantle's state of hydration or carbonation in terms of the pure volatile compounds, water and carbon dioxide, is challenging conceptually as well as computationally. For our purposes, however, there is no need to consider these volatile compounds in their pure chemical form, which could theoretically be found in the mantle either as liquid/vapour phase or, in the case of carbon, as graphite/diamond. Instead, we include volatiles in the form of volatile-enriched basaltic components with a fixed volatile content. The chosen contents of 5~wt\% H$_2$O, and 20~wt\% CO$_2$ for hMORB and cMORB, respectively, are based on experimental constraints on the volatile content of low-degree, volatile-enriched near-solidus melts produced at depth \citep{hirschmann09, dasgupta13}. The component \rev{partition coefficients} $K^i$ of hMORB and cMORB are calibrated to values estimated for the solid-melt partitioning of H$_2$O and CO$_2$. We do not consider the stability of carbonatite melts with upwards of 30~wt\% CO$_2$ at high pressures. This choice is motivated by an assumption that the oxidation state of the mantle is sufficiently reducing at depths below 150--200~km that any carbon is bound in graphite or diamond, rather than forming carbonatite liquids. We therefore calibrate our model to a depth of first carbonated melting consistent with redox-melting reactions producing carbonated silicate melts near the solidus \citep[e.g.,][]{dasgupta06, hirschmann10, stagno10}. Table~\ref{tab:comp} summarises the definition of all components used to represent volatile-bearing mantle in this study.}

\begin{table}[htb]
\centering
	\caption{\textbf{Mantle composition}}
	\label{tab:comp}
	\begin{tabular}{llll}
	\multicolumn{4}{c}{\textbf{Components of mantle composition}\T\B} \\\hline
	  & \textbf{name} & \textbf{type} & \textbf{role in decompression melting} \T\B\\
	(1) & dunite & refractory & residue of basaltic melt production \\
	(2) & MORB & fertile & product of basaltic melt production \\
	(3) & hMORB & water-bearing & product of deep, hydrated melting \\
	(4)~~ & cMORB~~~ & carbon-bearing~~~ & product of deep, carbonated melting \B\\\hline\B
	\end{tabular}
	\begin{tabular}{llllll}
	\multicolumn{6}{c}{\textbf{Reference mantle compositions}\T\B} \\\hline
 	\textbf{type} & \textbf{dunite} & \textbf{MORB} & \textbf{hMORB} & \textbf{cMORB} & \textbf{volatile content} \T\B\\
 	& wt \% & wt \% & wt \% & wt \% & wt ppm \B\\
	volatile-free & 70 & 30 & 0 & 0 & 0 \\
	hydrated & 70 & 29.8 & 0.2 & 0 & 100 \\
	carbonated & 70 & 29.95 & 0 & 0.05 & 100 \B\\\hline
	\end{tabular}
\end{table}

\rev{Given this four-component model of mantle composition, we simulate volatile-free mantle melting as a mixture of dunite$+$MORB, whereas hydrated and carbonated mantle melting scenarios are simulated as mixtures of dunite$+$MORB$+$hMORB and dunite$+$MORB$+$cMORB, respectively. For reasons of computational expense, we do not currently consider any simulation with both water and carbon dioxide. For each of the three scenarios of volatile-free, hydrated, and carbonated mantle melting we define a reference mantle composition, which will be used in all simulations below, except where otherwise stated. Table~\ref{tab:comp} lists the component concentrations for these reference mantle compositions. Note that concentrations of hMORB and cMORB are chosen such that the resulting mantle composition contains the equivalent of 100~wt~ppm, or 0.01~wt~\% water or carbon dioxide. The necessary calculations are straightforward: 0.2~wt~\% (hMORB in mantle) $\times$ 5~wt~\% (H$_2$O in hMORB) = 0.01~wt~\% (H$_2$O in mantle); 0.05~wt~\% (cMORB in mantle) $\times$ 20~wt~\% (CO$_2$ in cMORB) = 0.01~wt~\% (CO$_2$ in mantle). We will continue to use these relations to convert between concentrations of volatile-bearing components hMORB and cMORB and the equivalent concentration of water and carbon dioxide, and most results will be presented in terms of the equivalent volatile content.}

\subsection{Model calibration and implementation \label{sect:ModelImplementation}}

\subsubsection{Melt model calibration}
\begin{figure}[htb]
	\centering
	\includegraphics[width=0.9\textwidth]{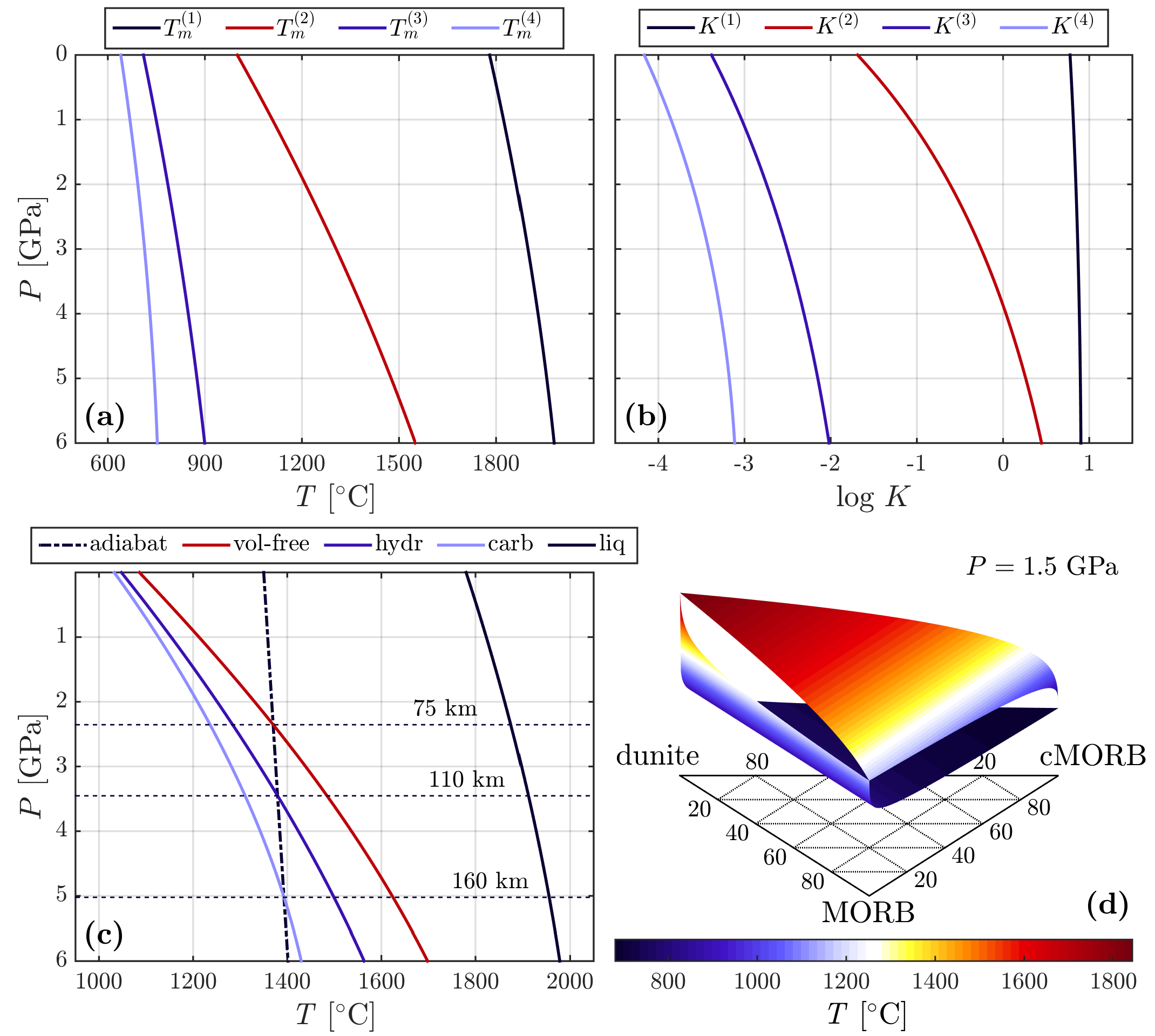}
	\caption{Calibration of the R\_DMC method to hydrated and carbonated mantle melting. \textbf{(a)} and \textbf{(b)} give component melting points $T_m^i$ with pressure and component \rev{partition coefficients} $K^i$ with pressure at reference mantle potential temperature $T_m=1350^\circ$C, respectively. \rev{Line colours denote components \textbf{(1)} dunite (black), \textbf{(2)} MORB (red), \textbf{(3)} hMORB (blue), and \textbf{(4)} cMORB (light blue). Mantle solidus curves for volatile-free (red), hydrated (blue), and carbonated (light blue) reference mantle compositions in \textbf{(c)}, alongside the mantle liquidus (black)}, and the reference adiabat at $T_m=1350^\circ$C (dash-dotted). Depths of first melting for volatile-free (75~km), hydrated (110~km) and carbonated (160~km) mantle indicated by dashed lines. \textbf{(d)} Solidus and liquidus $T$ [$^\circ$C] for the ternary compositional space of \rev{(dunite/MORB/cMORB)} calculated at $P$ = 1.5~GPa.}
	\label{fig:MeltModelCal}
\end{figure}

Figure \ref{fig:MeltModelCal} summarises some key characteristics of the R\_DMC method calibrated for mantle melting with volatiles (details of the calibration are given in Appendix~B). The calibrated component melting curves $\Tmi(P)$ are given in Fig.~\ref{fig:MeltModelCal}(a), and the component \rev{partition coefficients} $K^i(P,T)$ in Fig.~\ref{fig:MeltModelCal}(b). The dunite component has the highest melting point and \rev{preferentially partitions into the solid phase} over the whole pressure range. The MORB component has an intermediate melting point and transitions from being \rev{more stable in the solid phase at large pressures to being increasingly soluble at smaller pressures}, where most MORB-type melting is expected to occur. Both hMORB and cMORB components have low melting points and \rev{strongly partition into the melt phase at all pressures. In this, the volatile-bearing components behave similarly to incompatible trace elements, for which reason we will in the following refer to hMORB and cMORB as incompatible components}. The resulting solidus curves for volatile-free, hydrated, and carbonated reference mantle compositions are shown in Fig.~\ref{fig:MeltModelCal}(c), alongside the mantle liquidus, as well as a solid reference adiabat for the reference mantle potential temperature of $T_m=1350^\circ$C. The intersections between the three reference solidus curves and the adiabat mark the calibrated depths of first melting, which are 75~km for volatile-free, 110~km for wet, and 160~km for carbonated mantle melting.

The phase diagram for the \rev{carbonated} model at a pressure $P=1.5$~GPa is shown in Fig.~\ref{fig:MeltModelCal}(d). The solidus and liquidus planes are plotted over the ternary compositional space of \rev{dunite+MORB+cMORB}. The shape of the phase diagram demonstrates that this melt model \rev{is based on expressions derived from thermodynamics of ideal solutions, as seen by the solidus and liquidus planes forming phase loops spanning all corners of the ternary diagram.} The phase diagram of the \rev{hydrated system (dunite+MORB+hMORB)} is of the same general shape, but slightly shifted in temperature. The calibration parameters for the R\_DMC melt model employed here and in all simulations below are listed in Table~\ref{tab:meltpar}. A detailed discussion of the calibration procedure is provided in Appendix~B.

\begin{table}[htb]
\centering
	\caption{\textbf{R\_DMC calibration parameters}}
	\label{tab:meltpar}
	\begin{tabular}{lrrr}
	\multicolumn{4}{c}{\textbf{Component melting curves $T_m^i$}\B} \\\hline
	 & $T_{m,0}^i$ [$^{\circ}$C] 
	 & $A^i$ [$^{\circ}$C GPa$^{-1}$] 
	 & $B^i$ [$^{\circ}$C GPa$^{-2}$] \T \\
	(1) & 1780 &  45.0 & -2.00 \\
	(2) & 1000 & 112.0 & -3.37 \\
	(3) &  710 &  40.8 & -1.54 \\
	(4) &  640 &  30.1 & -1.88\B\\\hline
	\multicolumn{4}{c}{\textbf{Component distribution coefficients $K^i$}\T\B} \\\hline
 	 & $L^i$ [kJ kg$^{-1}$] & $r^i$ [J kg$^{-1}$K$^{-1}$] \T & \\
	(1) &  600 &  60 & \T \\
	(2) &  450 &  30 &  \\
	(3) &  350 &  30 &  \\
	(4) &  350 &  30 & \B \\\hline
	\end{tabular}
\end{table}

\begin{figure}[htb]
	\centering
	\includegraphics[width=0.85\textwidth]{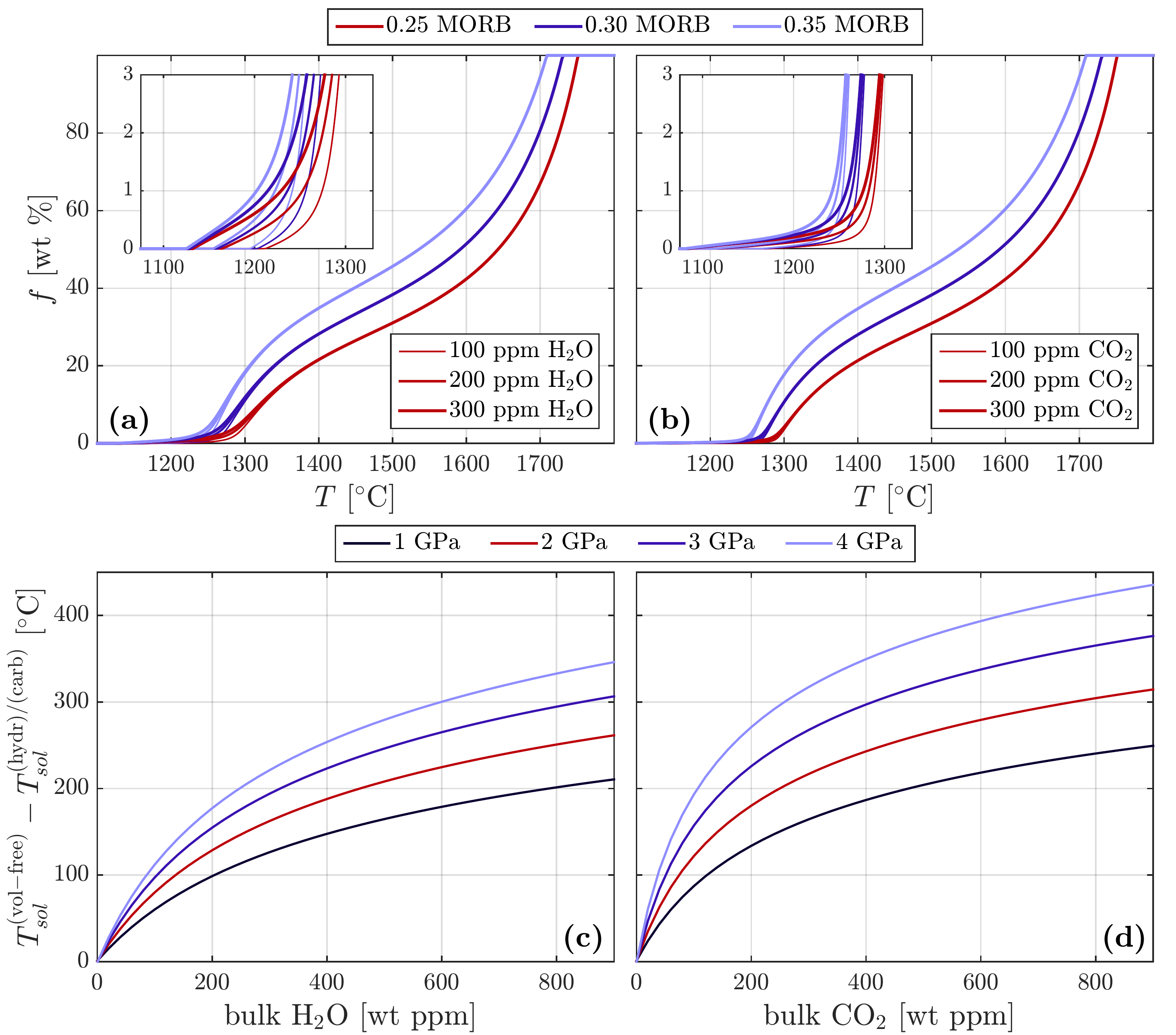}
	\caption{Isobaric melt production with volatiles. Melt fraction $f$ given with increasing $T$ at 1.5~GPa for bulk MORB fractions of 0.25, 0.3, and 0.35 \rev{(line colours)}, and 100, 200 and 300~ppm bulk volatile content (line weights). Hydrated melt production results given in \textbf{(a)}, carbonated melt production in \textbf{(b)}. Small panels show magnified view of onset of melting (axes same units). Solidus depression as a function of volatile content and pressure. Solidus depression measured as difference between volatile-free mantle solidus $T_\textrm{sol}^\textrm{(vol-free)}$ (0.3~MORB) and volatile-bearing solidus $T_\textrm{sol}^\textrm{(hydr)/(carb)}$ with bulk water contents of 0-1000~wt~ppm \textbf{(c)}; bulk carbon contents of 0-1000~wt~ppm \textbf{(d)}. \rev{Line colours} represents solidus depression at given pressures between 1--4~GPa.}
	\label{fig:MeltModelProp}
\end{figure}

Isobaric melt production curves calculated using the calibrated R\_DMC method, plotted in Fig.~\ref{fig:MeltModelProp}, are in broad agreement with previous parameterisations of mantle melting \citep[][and references therein]{katz03}. Fig.~\ref{fig:MeltModelProp}(a) shows melt production curves for hydrated mantle melting at a pressure of 1.5~GPa for bulk compositions of 0.25, 0.3, and 0.35~wt. fraction MORB, and 100, 200, and 300~ppm H$_2$O. Fig.~\ref{fig:MeltModelProp}(b) shows the same for carbonated mantle compositions with carbon dioxide contents of 100--300~ppm. The inset panels show a magnified view of the onset of melting, which occurs at lower $T$ with increasing volatile content. All melt production curves display the same general shape: near-solidus melting produces low fractions of melt dominated by volatile dissolution. Subsequently, the slope of the melt curves increases steeply as dissolution of the volatile-free MORB component sets in. As this component is exhausted, the melt production curves continue at shallower slope until temperatures are high enough to melt the mantle residual, at which point the curves steepen again to reach the liquidus.

The solidus depression caused by the presence of volatiles in the mantle is not uniform with pressure and composition, but rather increases with increasing pressure and volatile content (e.g., Fig.~3 of \cite{dasgupta13}). This property of mantle melting with volatiles is recovered by the calibrated R\_DMC method. Fig.~\ref{fig:MeltModelProp} shows the deviation of the volatile-bearing solidus from the volatile-free solidus at increasing pressures and with increasing bulk volatile content. Fig.~\ref{fig:MeltModelProp}(c) shows solidus depression curves for hydrated mantle compositions for the pressure range of 1--4~GPa, and water content of up to 1000~ppm bulk H$_2$O. Fig.~\ref{fig:MeltModelProp}(d) shows the same curves for a carbonated mantle with up to 1000~ppm bulk CO$_2$.

\subsubsection{Numerical implementation}
The non-dimensional system of equations is discretised by the finite difference method on a regular, rectangular, staggered grid \citep{katz07}. Nonlinearity of the discrete system is reduced by splitting the equations into two sub-sets, grouping together the hyperbolic thermo-chemical equations \eqref{eq:EnConsFinalNonDim}--\eqref{eq:MeltMassFinalNonDim}, and the elliptic fluid dynamics equations \eqref{eq:BulkMassFinalNonDim}--\eqref{eq:CompactionFinalNonDim}. Each subset is solved separately by an iterative Newton-Krylov method (inner iterations). Nonlinearities between the thermo-chemical and fluid dynamic sub-sets are resolved by a Picard fixed-point iterative scheme (outer iterations). The thermodynamic equilibrium state is calculated once in every outer iteration using a Newton method to solve the nonlinear equation \eqref{eq:EquilibriumMelt} for equilibrium melt fraction. The numerical implementation leverages tools of the PETSc toolkit \citep{petsc-web-page, petsc-user-ref, petsc-efficient, katz07}, providing full parallelism and efficient solution algorithms.

For the purpose of exposition, experimentation, and calibration of the R\_DMC method for mantle melting with volatiles, we have implemented a set of Matlab routines to plot phase diagrams, calculate equilibrium melt fraction and phase compositions at given conditions, and compute equilibration by reaction for closed-system scenarios. \rev{These routines are available online \citep{rdmc-repo} under an open source licence.}

\subsection{Column-model setup}
\begin{figure}[ppp]
	\centering
	\includegraphics[width=0.6\textwidth]{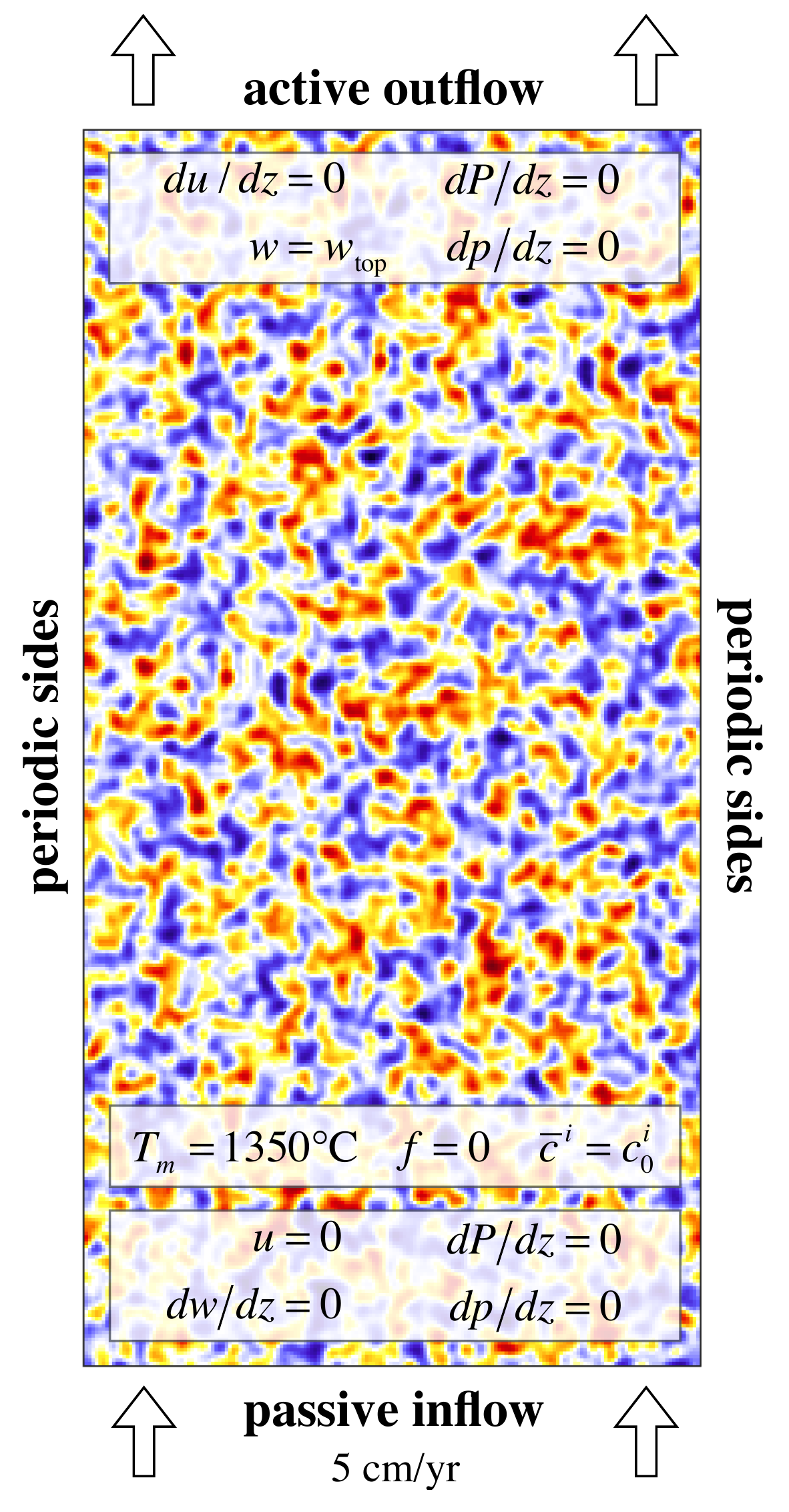}
	\caption{Model setup for two-dimensional upwelling column models. Boundary conditions for velocity and pressure components indicated along top and bottom boundary. $w_\mathrm{top}$ is chosen to fix passive inflow at bottom boundary at 5~cm/yr, compensating for active flow of two-phase mixture at top boundary. Side boundaries are periodic. The pressure $P$ is fixed in the lower left corner. Coloured pattern shows red-shifted random perturbations (arbitrary amplitude scale) used to initialise compositional heterogeneities.}
	\label{fig:ModelSetup}
\end{figure}

The most common cause for melting in the upper mantle is decompression due to upwelling mantle flow. Both one-dimensional (1D) and two-dimensional (2D) models of an upwelling column of mantle material are used here to investigate decompression melting and reactive flow in a volatile-bearing mantle. These mantle columns can be thought of as a simplified representation of the mantle beneath a MOR axis. The model setup is summarised in Fig.~\ref{fig:ModelSetup}.

\subsubsection{Boundary and initial conditions}
The domain has periodic sides. Boundary conditions enforce constant velocity at the top and zero normal stress at the bottom of the domain. The imposed velocity at the top $\vs = (0,w_\mathrm{top})$ is chosen such that a uniform mantle inflow of 5~cm/yr is induced at the lower boundary. Melt is allowed to flow out of the domain across the top boundary, driven by its buoyancy. Temperature, compositions, and melt fraction are passively advected across the outflow boundary at the top (zero normal gradient imposed). The inflowing mantle at the lower boundary has a constant temperature and specified composition $c_0^i(x,t)$, and zero melt fraction. The temperature is held fixed at the reference mantle potential temperature $T_m = 1350$~\degC.

The simulation is initialised at mantle potential temperature, but immediately reduced to the solidus $T_0 = \min(T_m,\Ts)$, such that the model domain is initially melt-free. Initial mantle composition for all reference runs is either volatile-free, hydrated, or carbonated reference mantle composition, as discussed above. To introduce some perturbation to the system, a red-shifted random field is generated, which is used both to initialise a randomly perturbed compositional state at the beginning of 2D simulations, as well as to continually feed in perturbations across the inflow boundary at the bottom of the domain. Fig.~\ref{fig:ModelSetup} shows the random perturbation pattern used for all simulations below (arbitrary amplitude). The approximate wave-length of perturbations is 5~km (10 grid cells).

Both 1D and 2D simulations are run with the same code. For 1D runs, the number of interior grid nodes in $x$-direction is set to unity, thus effectively rendering the solution one-dimensional (an additional 2 nodes are required on each boundary to handle boundary conditions). The model domain is specified as 240~km for 1D and 180~km for 2D runs; always deeper than the depth of onset of melting for all considered bulk compositions. The 2D model domain has a width of 90~km. Grid resolution is 0.5~km per grid cell in all runs. Numerical domain size therefore is 484$\times$5 for 1D, and 364$\times$184 for 2D simulations. These grid dimensions, with 10 degrees of freedom per grid node, result in a total number of degrees of freedom of 25,200 for 1D, and 669,760 for 2D runs. All simulations were performed on the Brutus cluster at ETH Zurich, Switzerland, using four nodes for 1D, and 32 nodes for 2D problems.

\subsubsection{Reference model parameters}
Reference model parameters are generally the same for 1D and 2D simulations and are chosen to represent typical asthenospheric conditions. All reference parameter values are reported in Table \ref{tab:parameters}. The most important parameter choices are discussed below.

The viscosity of silicate melt strongly depends on silica polymerisation, which is determined by melt composition. To a leading order, the variation of melt viscosity is strengthening with increasing silica content and weakening with increasing content of de-polymerising volatile compounds \citep{bottinga72,kushiro76,hui07}. Experimental evidence suggests that the weakening effect of carbon dioxide is at least as strong or stronger than the effect of water on a silica melt, in particular at depth where CO$_2$ is more soluble \citep{brearley89,giordano03,kono14}. To capture these trends, we introduce a log-linear variation of melt viscosity with melt composition. The compositional factors $\lambda_i$ take values of 1 (dunite), 100 (MORB), 0.01 (hMORB), 0.001 (cMORB) (see Table~\ref{tab:parameters}). Thus, an ultramafic melt will have a viscosity close to $\mu_0 = 1$~Pa-s. As more of the MORB-component is dissolved into the melt, $\mu$ will increase towards 10~Pa-s, whereas with dissolution of more hMORB or cMORB into the melt, $\mu$ is reduced to 0.1~Pa-s for typical hydrated melt, and 0.01~Pa-s for typical carbonated melt at the onset of volatile-enriched melting.

The reference permeability at 1\% porosity is set to $10^{-12}$~m$^2$. Both solid and liquid densities are held constant, with a density difference of $\Deltarho = 500$~kg-m$^{-3}$ driving melt segregation. The reference Dahmkohler number governing reaction rates is set to $\Da = 262$ for 1D, and $\Da = 1308$ for 2D, corresponding to a reaction time of $\tau_\Gamma = 500$ yrs, and 100 yrs, respectively. The reference assimilation ratio is set to $\chi=1$, such that melt is transported in (or very close to) thermochemical equilibrium. A list of all model parameters and reference values is given in Table~\ref{tab:parameters}.


\section{RESULTS \label{sect:Results}}

\subsection{One-dimensional column models \label{sec:one-D}}
One-dimensional simulations of decompression melting of upwelling mantle are employed to investigate the basic characteristics of the mantle melting model with volatiles. Parameter variations test the effects of changes in composition and volatile content, variable melt viscosity, and different styles of melting reactions on melt production and transport.

\subsubsection{Parameter variation - Mantle composition and volatile content}
\begin{figure}[ppp]
	\centering
	\includegraphics[width=\textwidth]{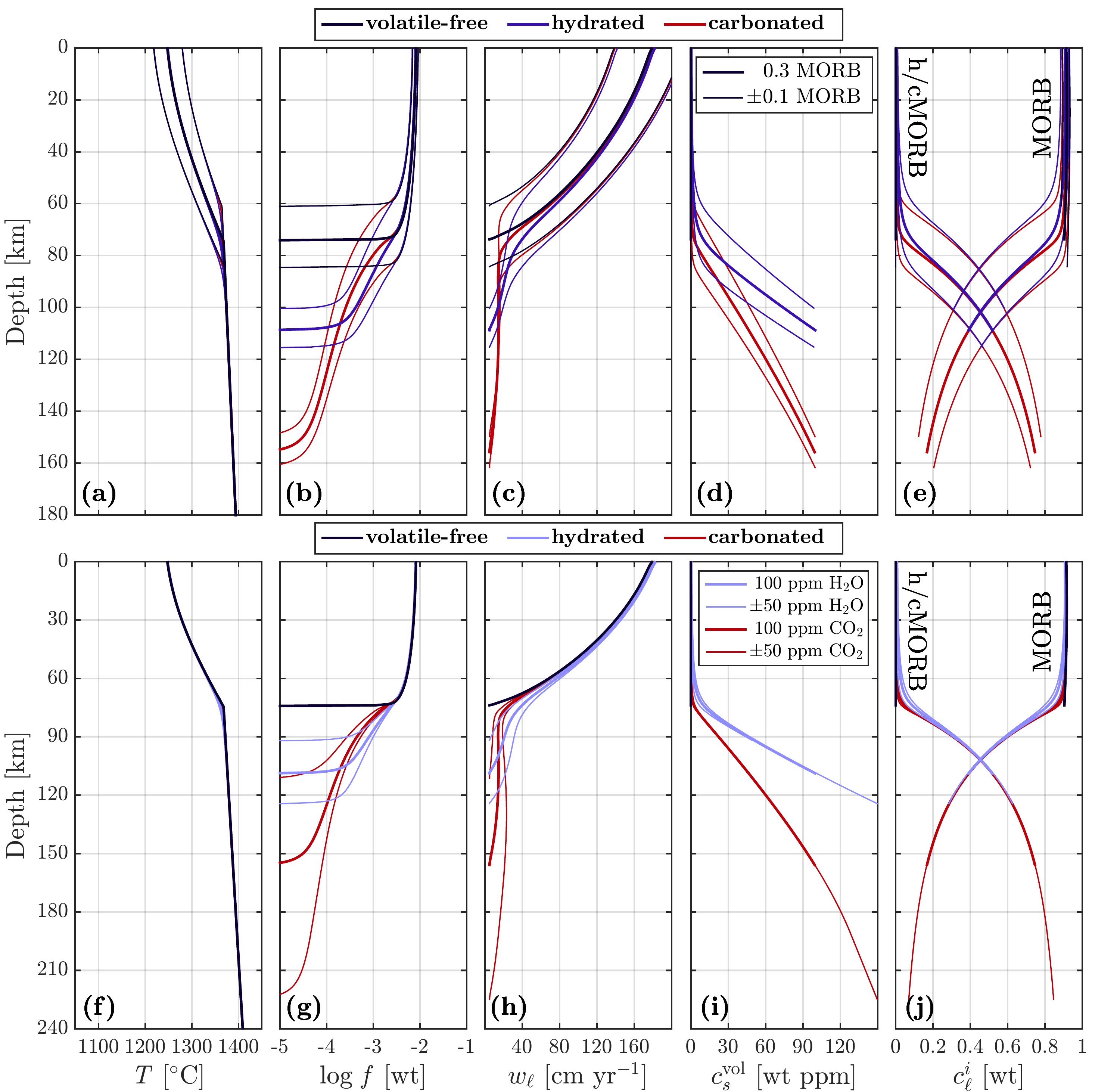}
	\caption{1D column model results for different bulk compositions. \rev{Line colours} represent volatile-free, hydrated, and carbonated model results. Results with 0.3$\pm$0.1 wt fraction bulk MORB content \textbf{(a)}--\textbf{(e)}; results with 100$\pm$50 wt~ppm bulk volatile content \textbf{(f)}--\textbf{(j)}. Profiles with depths of \textbf{(a)} \& \textbf{(f)} temperature; \textbf{(b)} \& \textbf{(g)} log of melt fraction; \textbf{(c)} \& \textbf{(h)} vertical component of liquid velocity; \textbf{(d)} \& \textbf{(i)} solid volatile content (H$_2$O for hydrated, CO$_2$ for carbonated mantle); \textbf{(e)} \& \textbf{(j)} melt composition (lines clustering to top left are MORB, to top right hMORB/cMORB). Fine lines bracketing bold lines denote variations of bulk composition from reference.}
	\label{fig:1DColBulkComp}
\end{figure}

Figure~\ref{fig:1DColBulkComp} compares the results of melt column models with different bulk compositions at steady state. The fertility (MORB) and volatile (hMORB, cMORB) content in these models are a key control on the extent and productivity of partial melting. The top row of panels shows depth profiles of clusters of three runs with different MORB content for volatile-free mantle melting, hydrated melting with 100~ppm H$_2$O, and carbonated melting with 100~ppm CO$_2$. Each cluster of lines consists of a bold line denoting a reference run with a MORB content of 30\%, bracketed by two fine lines showing model outcomes with a variation of $\pm$10\% MORB. In general, increasing the initial fertility of the mantle leads to a cooler mantle geotherm in Fig.~\ref{fig:1DColBulkComp}(a), and both a deeper onset and higher yield of melting in Fig.~\ref{fig:1DColBulkComp}(b). As a result of the higher melt productivity, vertical melt velocity in Fig.~\ref{fig:1DColBulkComp}(c) increases with increasing MORB content, in particular throughout the volatile-free melting region upwards of 75~km. 

The solid volatile content in Fig.~\ref{fig:1DColBulkComp}(d) shows a depleted profile. The volatile content in the mantle residue is depleted approximately linearly from the depth of first melting to near zero at the transition to volatile-free MORB melting. The composition of the melt shown in Fig.~\ref{fig:1DColBulkComp}(e) shows a general trend from volatile-enriched compositions at depth toward volatile-free MORB compositions above the volatile-free melting depth. Increasing the initial MORB content shifts the profiles down, while their general behaviour remains the same. A less fertile mantle results in a hotter geotherm, shallower onset and lower yield of melting, and consequently lower melt velocity. 

The bottom row of panels in Fig.~\ref{fig:1DColBulkComp} shows depth profiles of clusters of three runs with different volatile contents for hydrated and carbonated mantle models. The volatile-free reference case is shown for comparison. Bold lines again denote the results of hydrated and carbonated reference simulations, while finer lines show simulation data obtained with variations in H$_2$O and CO$_2$ content of $\pm$20~ppm. The effect of variable volatile content on the resulting temperature profile in Fig.~\ref{fig:1DColBulkComp}(f) is negligible. Fig.~\ref{fig:1DColBulkComp}(g) shows that the effect of variation in volatile content on the depth of first melting is significant for volatile-bearing mantle runs, where a variation in volatile content by 20~ppm leads to a downward shift of 16~km for hydrated, and more than 60~km for carbonated models. Significant variations in melt segregation rate with variable volatile content in Fig.~\ref{fig:1DColBulkComp}(h) are only observed below the volatile-free mantle solidus, associated with increased melt productivity and increased volatile concentrations there. The compositional evolution of the solid phase in Fig.~\ref{fig:1DColBulkComp}(i) and the melt in Fig.~\ref{fig:1DColBulkComp}(j) follow nearly identical pathways, apart from the variation in depth of first melting.

\subsubsection{Parameter variation - Melt viscosity}
In the absence of significant dynamic pressure gradients, melt segregation is driven by the density contrast between solid and liquid, $\Deltarho$. Phase separation flux scales linearly with the Darcy coefficient $K/\mu$. The effect of changing permeability $K$ on magma transport is well studied \citep[e.g.,][]{crowley15} and will not be further tested here. Melt viscosity $\mu$, however, has been treated as a constant in previous studies \citep[e.g.,][]{spiegelman01, katz08, hewitt10, liang10, weatherley12}. Here, we take melt viscosity to vary with composition to observe what effect volatile content has on the mobility of low-degree, volatile-enriched melt at depth.

\begin{figure}[htb]
	\centering
	\includegraphics[width=0.9\textwidth]{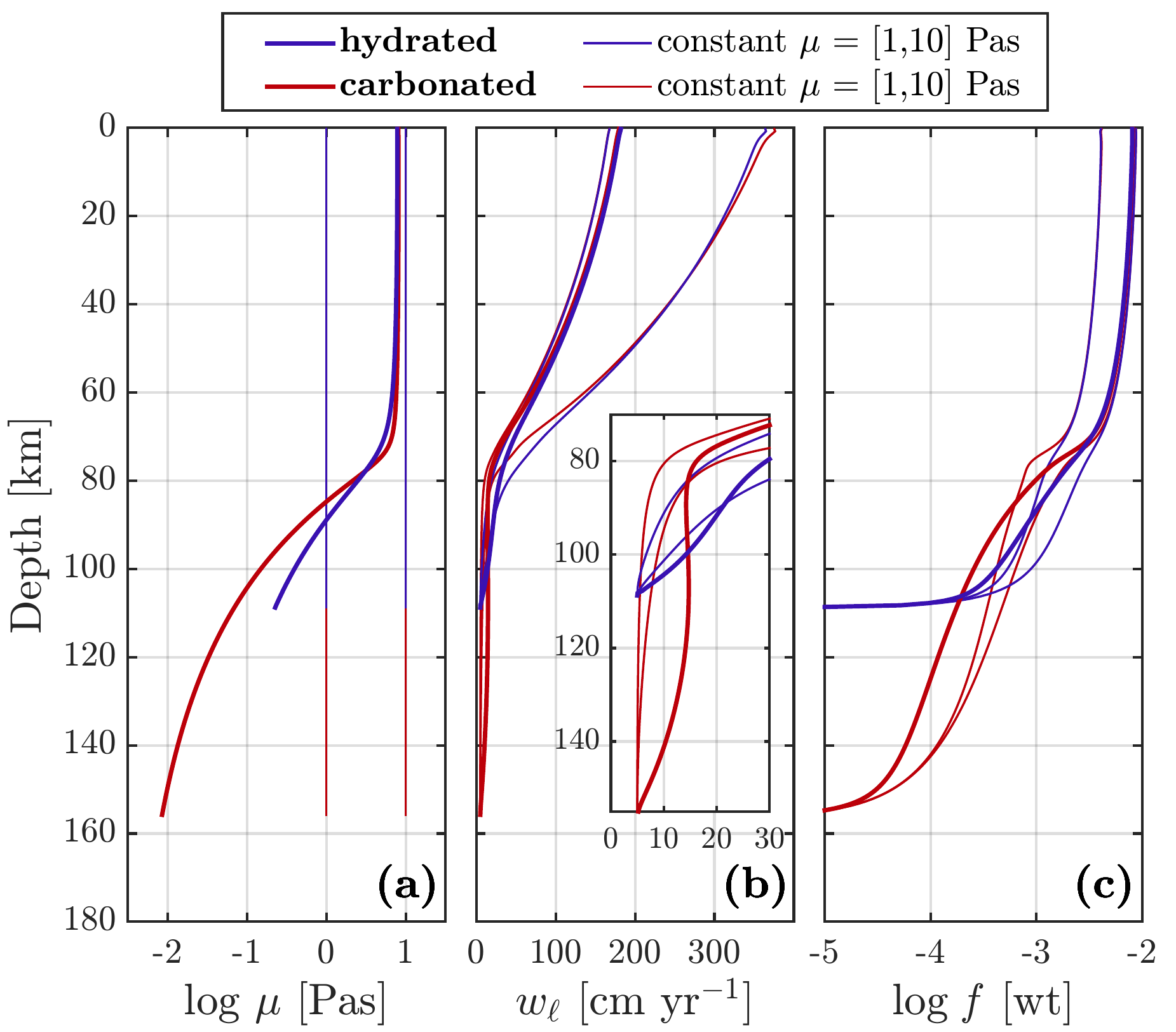}
	\caption{1D column model results with different melt viscosity models. Melt viscosity \textbf{(a)}, vertical melt velocity \textbf{(b)}, and melt fraction \textbf{(c)} for compositionally variable (bold) and constant melt viscosities of 1 and 10 Pa-s (fine). Small panel shows magnification of melt upwelling rate in the volatile-rich melting zone. \rev{Line colours denote results for hydrated (blue) and carbonated (red) models}.}
	\label{fig:1DColMu}
\end{figure}

Figure \ref{fig:1DColMu} shows results of the reference hydrated and carbonated models compared to two simulations with a constant melt viscosity $\mu = \mu_0 = [1,10]$ Pa-s. Fig.~\ref{fig:1DColMu}(c) shows the depth profiles of melt viscosity for the two volatile-bearing reference runs with variable $\mu$, and the viscosity magnitude of the two constant-viscosity runs, represented by vertical lines. These two values bracket the range of values for basaltic melt in the mantle. In our reference model, viscosity increases as MORB component is added, reflecting an increase in silica polymerisation due to increased silica content. Conversely, $\mu$ decreases as volatile-enriched silicate components hMORB and cMORB are added, reflecting an increase in de-polymerisation due to volatile content. The resulting viscosity variation in log-space thus reflects the shape of melt composition curves given in Fig.~\ref{fig:1DColBulkComp}(e) and (j), with hydrated and carbonated melt viscosities at depth taking values of just above 0.1 Pa-s, and 0.01 Pa-s, respectively.

Figure \ref{fig:1DColMu}(b) and (c) show the effect of variation in melt viscosity on melt segregation and on melt fraction. As expected from the volatile-induced viscosity variations, melt segregation is enhanced at depth, where melt composition is volatile-rich, and dampened towards the surface, where the melt consists mostly of volatile-free MORB. Melt flow rates rates are $\approx$15~cm/yr for carbonated melt and $\approx$25~cm/yr for hydrated melt below the volatile-free solidus depth of 75~km. 

The change of behaviour with compositionally dependent melt viscosity is most significant for carbonated mantle models, where melt fractions are well below 0.1~wt\%. Such melts would essentially be immobile with respect to mantle flow even at the relatively high reference permeability chosen here. However, even with a compositional change of viscosity of 4 orders of magnitude from volatile-free MORB to carbonated MORB with 20 \% CO$_2$ content \citep{kono14}, melt transport at depth is still an order of magnitude slower than in the volatile-free melting region.

\subsubsection{Parameter variation - Reactive model}
\begin{figure}[htb]
	\centering
	\includegraphics[width=\textwidth]{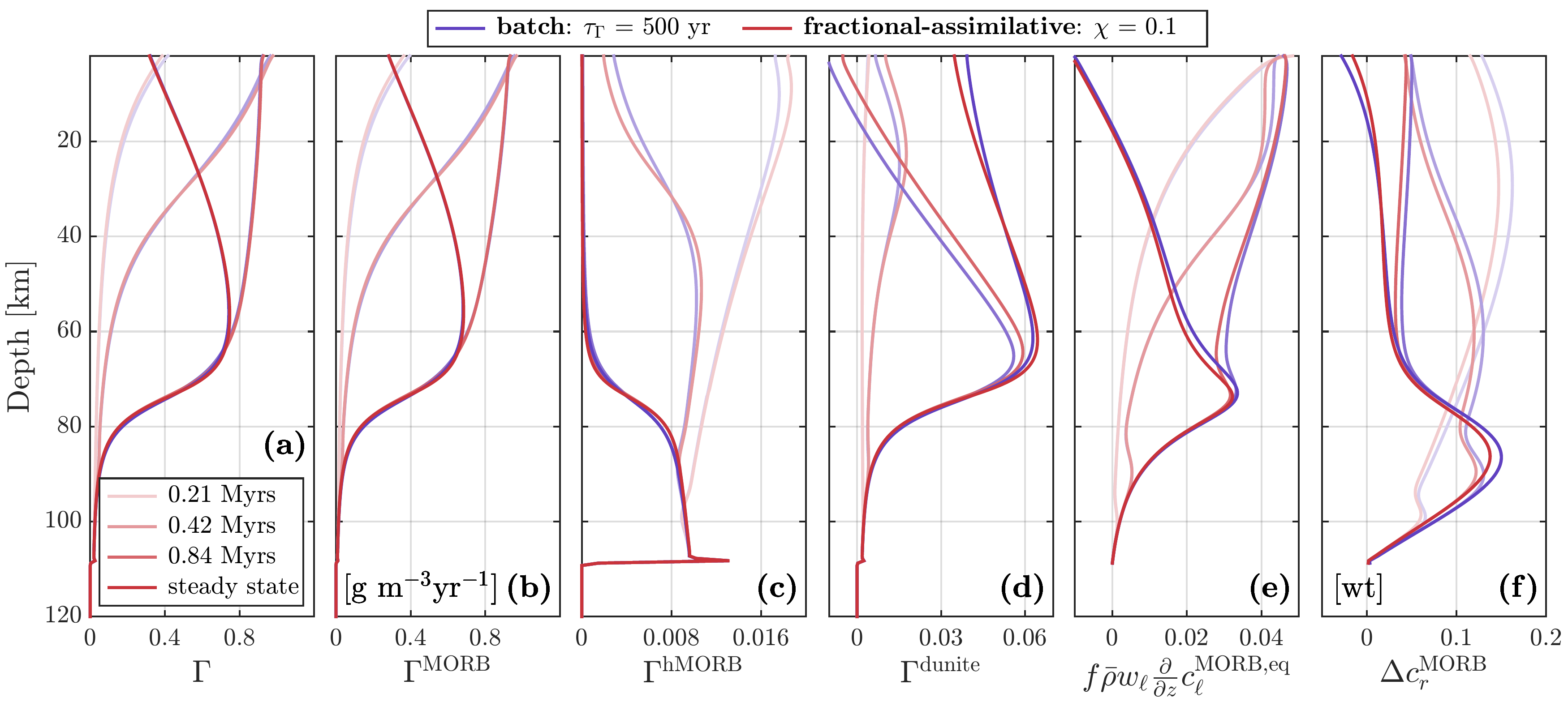}
	\caption{1D column model results for different reaction parameters. Batch reaction (blue), and fractional-assimilative reaction (red). Progress in model time shown as increasingly dark shading. Net reactive mass-transfer rate \textbf{(a)}; MORB \textbf{(b)}, hMORB \textbf{(c)}, and dunite \textbf{(d)} component mass-transfer rates; \rev{flux-induced liquid disequilibrium in MORB concentration \textbf{(e)}; difference between effective concentration reacting mass and liquid equilibrium for MORB \textbf{(f)}. Both calculations use reference hydrated mantle composition.}}
	\label{fig:1DColReactions}
\end{figure}

The main contribution of this work is the introduction of the R\_DMC method, which allows the testing of disequilibrium-driven reactive models. Fig.~\ref{fig:1DColReactions} shows results of the reference hydrated model in comparison to an additional simulation, in which the assimilation ratio was reduced by a factor 10 to $\chi=0.1$, while $\Da$ was kept at a value corresponding to the reference reaction time of $\tau_\Gamma=500$ yr. The reference case represents a batch-type reaction model, while the additional run represents a fractional-assimilative reaction model. In the latter case, assimilation reaction kinetics are set to be ten times slower than fractional melting kinetics. The figure panels display snapshots of the results at a sequence of model times to show the evolution towards steady state. The depth profiles at different times are drawn in increasingly dark shades with time.

Fig.~\ref{fig:1DColReactions}(a) shows the time evolution of the net melting rate $\Gamma$. Fig.~\ref{fig:1DColReactions}(b)--(d) show the component-wise reaction rates $\Gamma^i$ for MORB (b), dunite (c), and hMORB (d). The component reaction rates in (b)--(d) sum up to the net melting rate in (a). \rev{Lastly, Fig.~\ref{fig:1DColReactions}(e) shows the flux-induced disequilibrium along the vertical solubility gradient of the key melt-producing component, $f \bar{\rho} w_\ell \frac{\partial}{\partial z} c_\ell^\mathrm{MORB,eq}$.}

\rev{The transient behaviour shown in Fig.~\ref{fig:1DColReactions} is initially characterised by progressive dissolution and depletion of hMORB from the model domain, as observed in decreasing magnitudes of $\G^\mathrm{hMORB}$ in Fig.~\ref{fig:1DColReactions}(c). Following the volatile depletion, dissolution and depletion of MORB from the mantle column continues until steady state is reached. Net melt production is dominated by MORB dissolution throughout, as seen from the similar magnitudes of $\Gamma$ and $\Gamma^\mathrm{MORB}$ in Fig.~\ref{fig:1DColReactions}(a)--(b). The profiles of $\Gamma^\mathrm{MORB}$ and $\Gamma^\mathrm{hMORB}$ in steady state indicate that hydrated melting remains active at depths below the volatile-free solidus at 75~km, above which volatile-free MORB melting remains stable.}

\rev{The magnitude of $\Gamma^\mathrm{dunite}$ is about two orders of magnitude lower than MORB melting and follows a different transient evolution, where dunite reaction rates decrease to slightly negative values towards the top of the column during the model stage of dominant MORB melting and depletion. Although this effect is weak under the current model calibration, it serves as an example of an incongruent reaction where MORB dissolution coincides with dunite precipitation in a net melt-producing reactive system.}

\rev{The evolution of the flux-induced disequilibrium for the MORB component in Fig.~\ref{fig:1DColReactions}(e) shows the highest values around the transition between hydrated and volatile-free melting. This reactive transition at first occurs at the top of the model domain. From there, it spreads down through the column until it contracts to a depth-band around 60--80~km depth, where it remains stable at steady state. As the quantity displayed here is diagnostic of the systems propensity to reactive channelisation, areas experiencing a high flux-induced rate of disequilibration are the most likely locations for reactive channels to emerge in 2D simulations.}

\rev{The slight differences in reaction rates observed between the two types of melt reactions (batch/fractional-assimilative) are not straightforward to interpret. However, Fig.~\ref{fig:1DColReactions}(f) shows that the difference between the proportions of reacted mass and the equilibirum concentration of MORB in the melt is significantly lowered over much of the model evolution in the fractional-assimilative case. This is caused by the reduced assimilation ratio $\chi$ constraining the effective reactive composition $c^i_r$ to remain closer to the fractional composition $\cgi$ imposed on net-mass transferring reactions. Thus, melt producing reactions in the fractional-assimilative case are rendered less efficient, meaning that the same rate of flux disequilibration can drive an increased rate of melt production.}

\subsection{Two-dimensional column models}
Two-dimensional simulations of decompression melting and reactive flow in an upwelling mantle column are employed to investigate the coupling between melting reactions, thermochemical evolution, and melt-transport dynamics in models of mantle melting with volatiles. Parameter variations are employed to test the effects of the mantle upwelling rate, types and amplitudes of compositional heterogeneities, and different styles of melt reactions on melt production and transport. 

\subsubsection{Reference case}
The reference hydrated and carbonated 2D simulations are initialised with correlated compositional perturbations added to MORB and volatile content. The reference perturbation amplitudes are set to 1~wt\% and 10 wt~ppm for MORB and volatile content, respectively. Perfectly unperturbed 2D simulations result in the same solutions as reported for 1D geometry and are not considered further. One volatile-free simulation with 0.3 $\pm$ 0.01 MORB is employed for comparison, but the results are not shown here, as there is no significant difference to the reference, volatile-free, 1D simulation.

\begin{figure}[ppp]
	\centering
	\includegraphics[width=\textwidth]{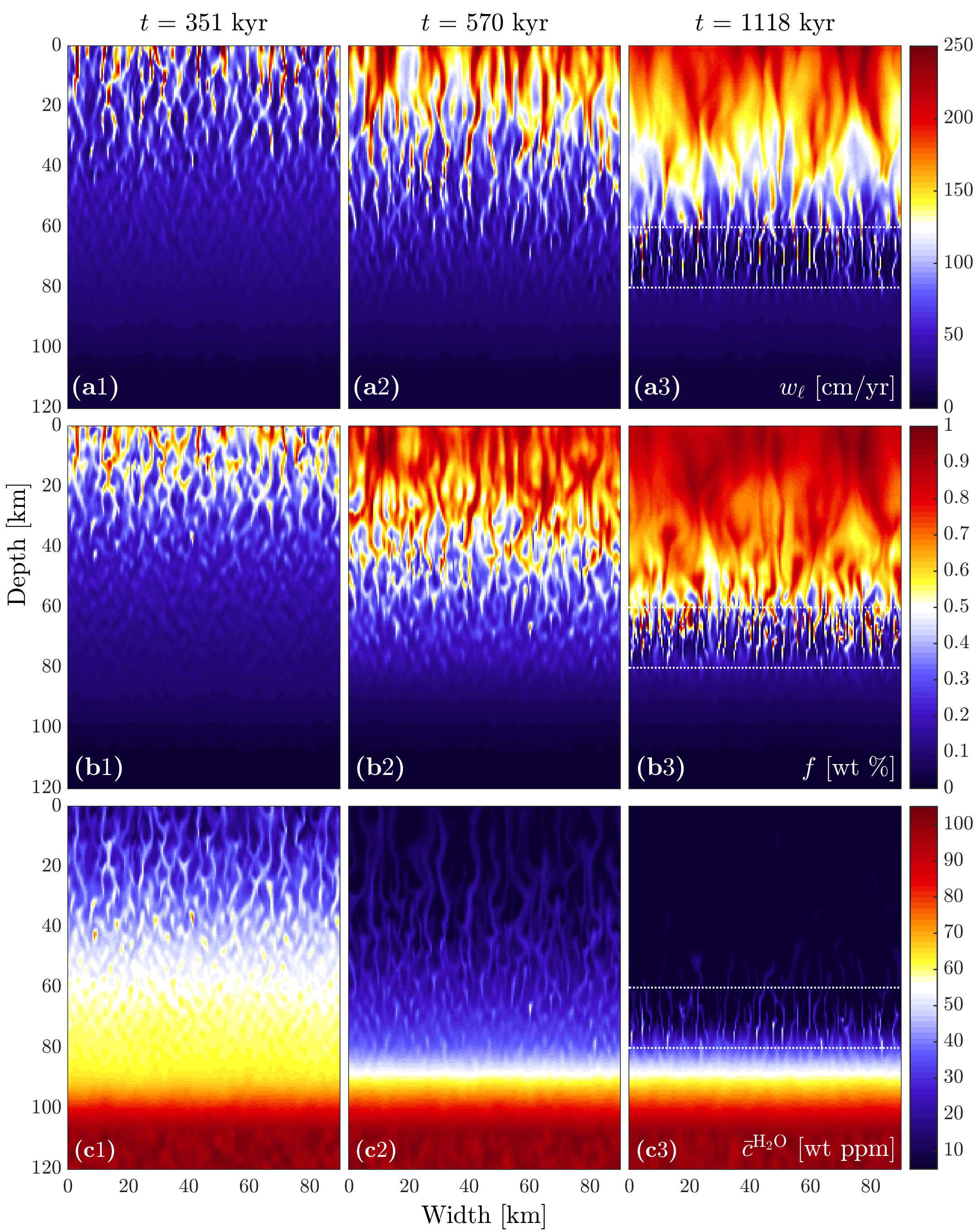}
	\caption{Time evolution of 2D column model for reference hydrated mantle. Vertical melt velocity \textbf{(a)}, melt fraction \textbf{(b)}, and bulk water content \textbf{(c)} at 351 kyr \textbf{(1)}, 570 kyr \textbf{(2)}, 1118 kyr \textbf{(3)}. Steady state channelling depth marked by white dotted lines. Model domain extends further to 180$\times$90~km.}
	\label{fig:2DColTimeEvoHydr}
\end{figure}

\begin{figure}[ppp]
	\centering
	\includegraphics[width=\textwidth]{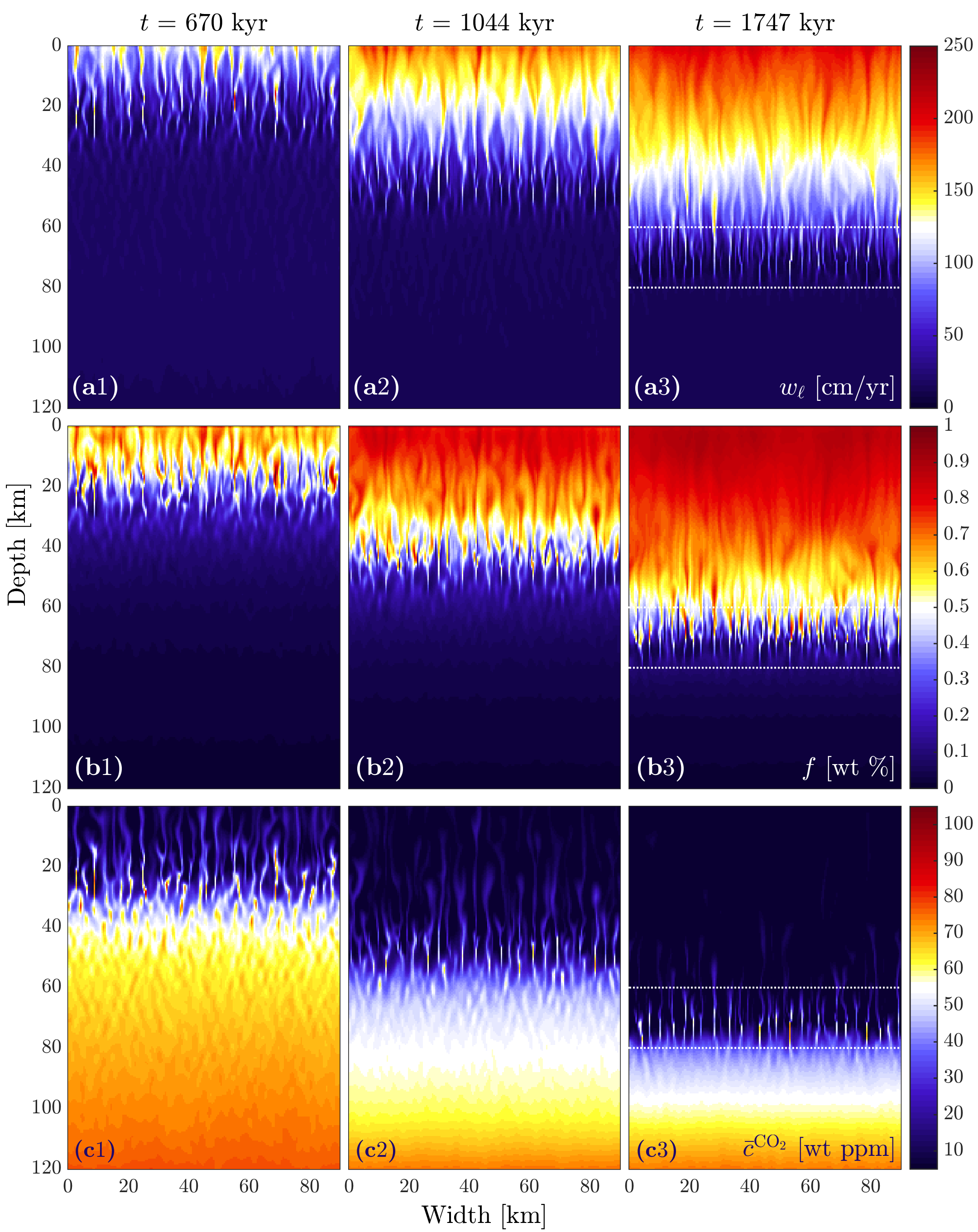}
	\caption{Time evolution of 2D column model for reference carbonated mantle. Vertical melt velocity \textbf{(a)}, melt fraction \textbf{(b)}, and bulk water content \textbf{(c)} at 670 kyr \textbf{(1)}, 1044 kyr \textbf{(2)}, 1747 kyr \textbf{(3)}. Steady state channelling depth marked by white dotted lines. Model domain extends further to 180$\times$90~km.}
	\label{fig:2DColTimeEvoCarb}
\end{figure}

Snapshots of the reference hydrated and carbonated mantle simulations are shown in Figs.~\ref{fig:2DColTimeEvoHydr} and \ref{fig:2DColTimeEvoCarb}. Vertical component of melt velocity $w_\ell$ (a), melt fraction $f$ (b), and bulk volatile content $\bar{c}^\mathrm{H_2O}$ or $\bar{c}^\mathrm{CO_2}$ (c) are given at model times corresponding to a distance of 75, 150, and 300 km covered by the fastest flowing melt. By the third snapshot, the models have reached a statistical steady-state, where further model evolution continues to exhibit stochastic fluctuations around the same steady solution found in the corresponding 1D column models.

In both the hydrated and carbonated mantle reference simulations, a transition zone of increasing melt velocity and melt fraction and decreasing \rev{volatile content} is observed to move down through the model domain with time. Recalling that these simulations are started from an initial condition of fertile mantle with a temperature at the pressure-dependent solidus, this transition corresponds to the transition of volatile-bearing to volatile-free melting \rev{identified in the 1D simulations above in Fig. \ref{fig:1DColReactions}}. Below this transition, melt velocity generally remains below 50~cm/yr, and melt fractions are of the order of 0.1~wt\%. Across the transition, melt velocity and melt fraction increase by up to an order of magnitude, while the bulk water content is depleted to below 10~ppm. Most importantly, the heterogeneity of melt distribution is increased across the transition, with localisation of melt transport leading to channel formation. The emerging channels exhibit increased melt velocity, melt fraction, and water content. As the models approach a statistical steady-state, channelling remains limited to an approximately 20~km long transition from 60 to 80~km depth, with melt distribution growing increasingly diffuse again above this transition and towards the top of the domain.

Comparing the hydrated to the carbonated mantle model, the transition to volatile-free melt production is sharper in the carbonated case, as seen by comparing the gradient of the volatile-depletion front in panels (c1) and (c2) of Figs~\ref{fig:2DColTimeEvoHydr}~\&~\ref{fig:2DColTimeEvoCarb}. \rev{On the one hand}, the volatile content in carbonated channels seems to be generally higher than in hydrated ones. On the other hand, heterogeneity in both melt velocity and melt fraction is more distinct in the hydrated model, in particular upwards of the main reactive transition. These differences are best understood in the context of the order-of-magnitude difference in \rev{partition coefficient} between the hMORB and cMORB components (Fig.~\ref{fig:MeltModelCal}). The more incompatible carbonated silicate component is more concentrated in the melt and segregates from its residue at a higher effective transport rate, thus creating a sharper reaction front compared to the less incompatible hydrated component. The lower volatile concentration in hMORB relative to cMORB means that more near-solidus melt is produced by hydrated relative to carbonated mantle melting, leading to stronger heterogeneity in melt flux caused by perturbations of water content than carbon content.

\begin{figure}[htb]
	\centering
	\includegraphics[width=\textwidth]{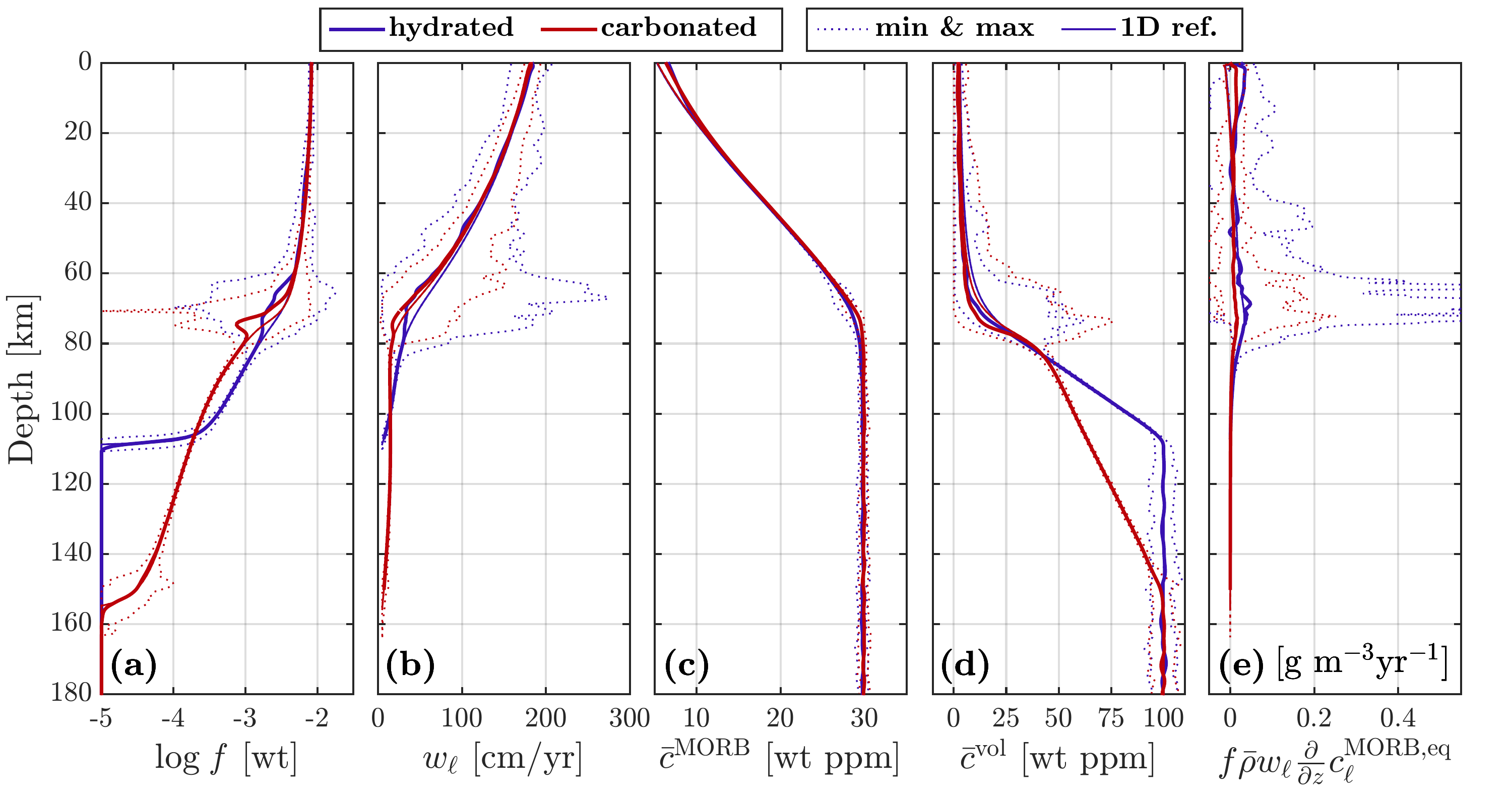}
	\caption{Comparison of 1D and 2D reference model results at steady state \rev{for hydrated (blue), and carbonated (red) mantle}. Melt fraction \textbf{(a)}, vertical melt velocity \textbf{(b)}, bulk MORB \textbf{(c)}, and bulk volatile content \textbf{(d)}. Depth profiles of horizontal mean (bold solid), minimum and maximum (dotted) of 2D, alongside 1D (fine solid) results.}
	\label{fig:2DColSteadyState}
\end{figure}

The steady-state solution of both 2D reference models are shown in comparison to that of their corresponding 1D runs in Fig.~\ref{fig:2DColSteadyState}. The 2D results are given as depth profiles of horizontal average, minimum and maximum of (a) melt fraction, (b) melt velocity, (c) bulk MORB content, (d) bulk volatile content, \rev{and (e) flux-induced MORB disequilibrium in the melt phase.} In general, the horizontal average recovers the 1D solution very well. However, in the depth-band where channelling remains active in steady state, horizontal average quantities are significantly altered by channelisation. The horizontal mean of both melt fraction and melt velocity is depressed by channel formation, while their maximum is found at up to an order of magnitude above the mean.
As mass flux along the column is conserved, this implies that the magma flux in channels must occur at transport rates up to an order of magnitude above values predicted by a laterally homogeneous flow model.

Flow localisation has little effect on the bulk MORB concentration in Fig.~\ref{fig:2DColSteadyState}(c). Horizontal variations of bulk MORB content due to channelisation are no more than 2--3 percent, with minimum values corresponding to channels. For the volatile bulk concentration in Fig.~\ref{fig:2DColSteadyState}(d), however, channelling introduces a significant horizontal variability. The concentration of a component in the melt is governed by its \rev{partition coefficient}. The incompatible volatile-bearing components have partition coefficients lower than 0.01, and are therefore enriched in the melt by more than two orders of magnitude. As melt flow is localised into channels, the incompatible components are enriched inside melt channels and depleted between channels. This effect is increased with a decreasing \rev{partition coefficient}, and with increasing melt localisation inside channels. It is interesting to note that distributed melt transport both below and above the channelling depth band leads to homogenisation of bulk composition, as observed by the horizontal mean, minimum, and maximum collapsing to the unperturbed 1D model state.

\rev{The disequilibration rate due to melt flux along the MORB solubility gradient in Fig.~\ref{fig:2DColSteadyState}(d) illustrates that channellisation is driven by reactive flux melting. The disequilibration rate for both hydrated and carbonated simulations peaks in the depth band from 60--80~km. This is the same depth band that was previously identified in 1D simulations as the most likely part of the domain for reactive channelling to emerge; it is now confirmed to be the area of sustained flow localisation in 2D simulations. Channelling ceases above that depth because volatiles are depleted in the residue and diluted in the melt; there is then no significant solubility gradient driving any further increase in dissolution of the MORB component.}

To combine compositional and dynamic aspects of these simulations and illustrate both the spatial distribution and rate of volatile transport through the upwelling mantle column, we calculate the bulk volatile mass flux as
\begin{subequations}
\begin{linenomath*}\begin{align}
	\label{eq:BulkVolatileFlux}
	Q^\mathrm{H_2O} &= 0.05 \times \rhobar \left( f c^\mathrm{hMORB}_\ell \vert \vl \vert + \fs c^\mathrm{hMORB}_s \vert \vs \vert \right) \: , \\
	Q^\mathrm{CO_2} &= 0.20 \times \rhobar \left( f c^\mathrm{cMORB}_\ell \vert \vl \vert + \fs c^\mathrm{cMORB}_s \vert \vs \vert \right) \: ,
\end{align}\end{linenomath*}
\end{subequations}
in units of [kg m$^{-2}$yr$^{-1}$].

\begin{figure}[htb]
	\centering
	\includegraphics[width=\textwidth]{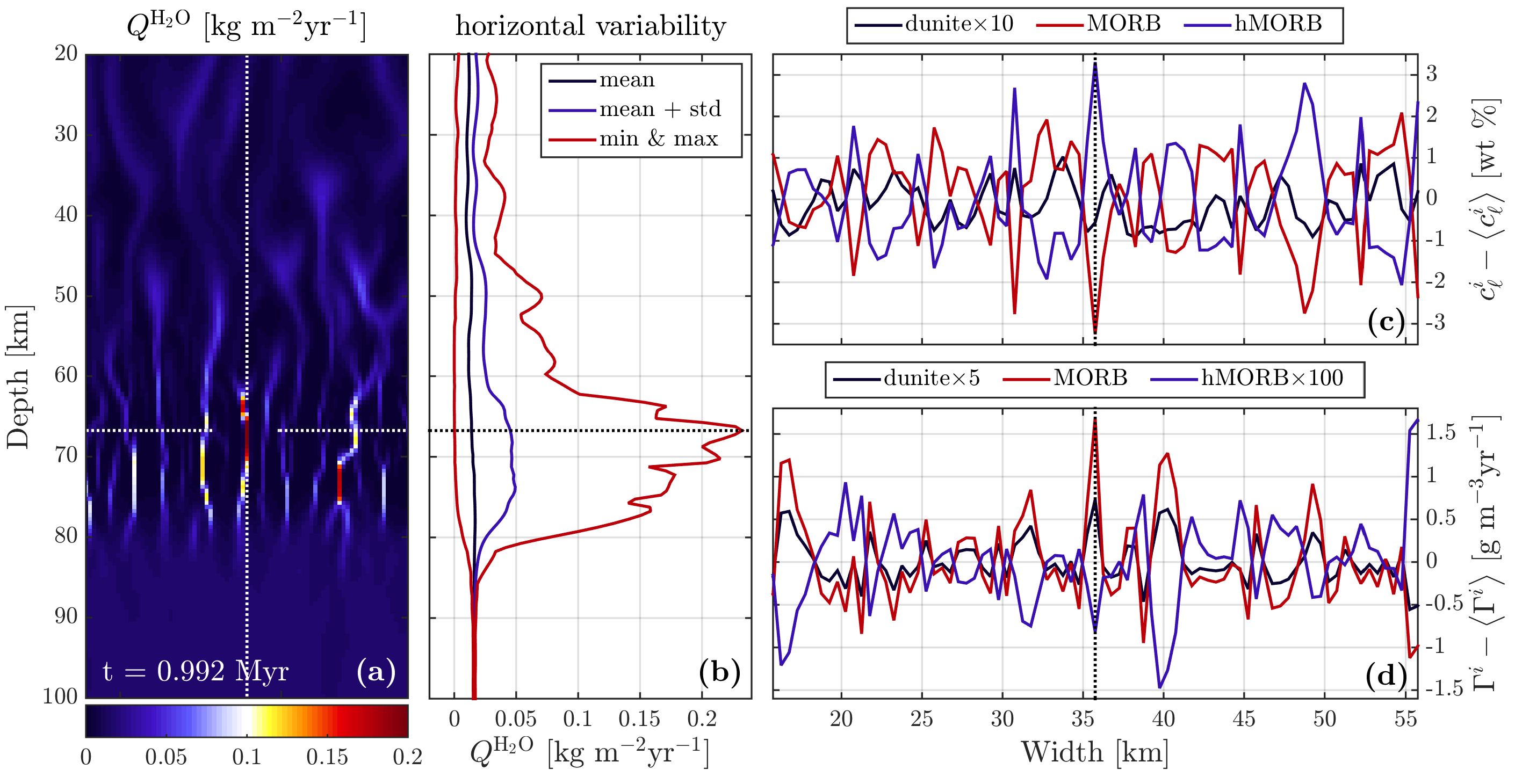}
	\caption{Bulk volatile flux in 2D column model of reference hydrated mantle. Color plot of bulk water flux $Q^\mathrm{H_2O}$ \textbf{(a)} focused in on highest flux channel (dotted lines in all panels). Horizontal variability of $Q^\mathrm{H_2O}$ in \textbf{(b)} with horizontal mean (black), mean plus standard deviation (blue), and minimum/maximum (red). Deviation from horizontal mean across max flux channel of melt composition \textbf{(c)}, and component reaction rates \textbf{(d)}. Line colours denote components of composition (scaling applied for visibility).}
	\label{fig:2DColVolFluxHydr}
\end{figure}

Figure \ref{fig:2DColVolFluxHydr} shows the bulk volatile mass flux at a selected model time of the hydrated mantle reference case. Fig.~\ref{fig:2DColVolFluxHydr}(a) is centred on the highest-flux channel in the model domain at that time, indicated by dotted lines. Depth profiles of the horizontal mean, standard deviation, minimum and maximum volatile flux are shown in Fig.~\ref{fig:2DColVolFluxHydr}(b). We note the following observations: First, the mean flux remains close to constant throughout the model box, which is required by mass conservation as this simulation is close to steady-state. Second, the minimum flux is less than a standard deviation below the mean flux, as the flux between channels approaches zero. In general, the standard deviation is approximately two times the mean flux throughout the channeling band (60--80~km depth). The maximum flux is recorded at more than a factor of ten above the mean flux. Upwards of the channeling band, the localised flux gradually disperses, as indicated by the decay of both maximum and standard deviation values towards the surface.

\begin{figure}[htb]
	\centering
	\includegraphics[width=\textwidth]{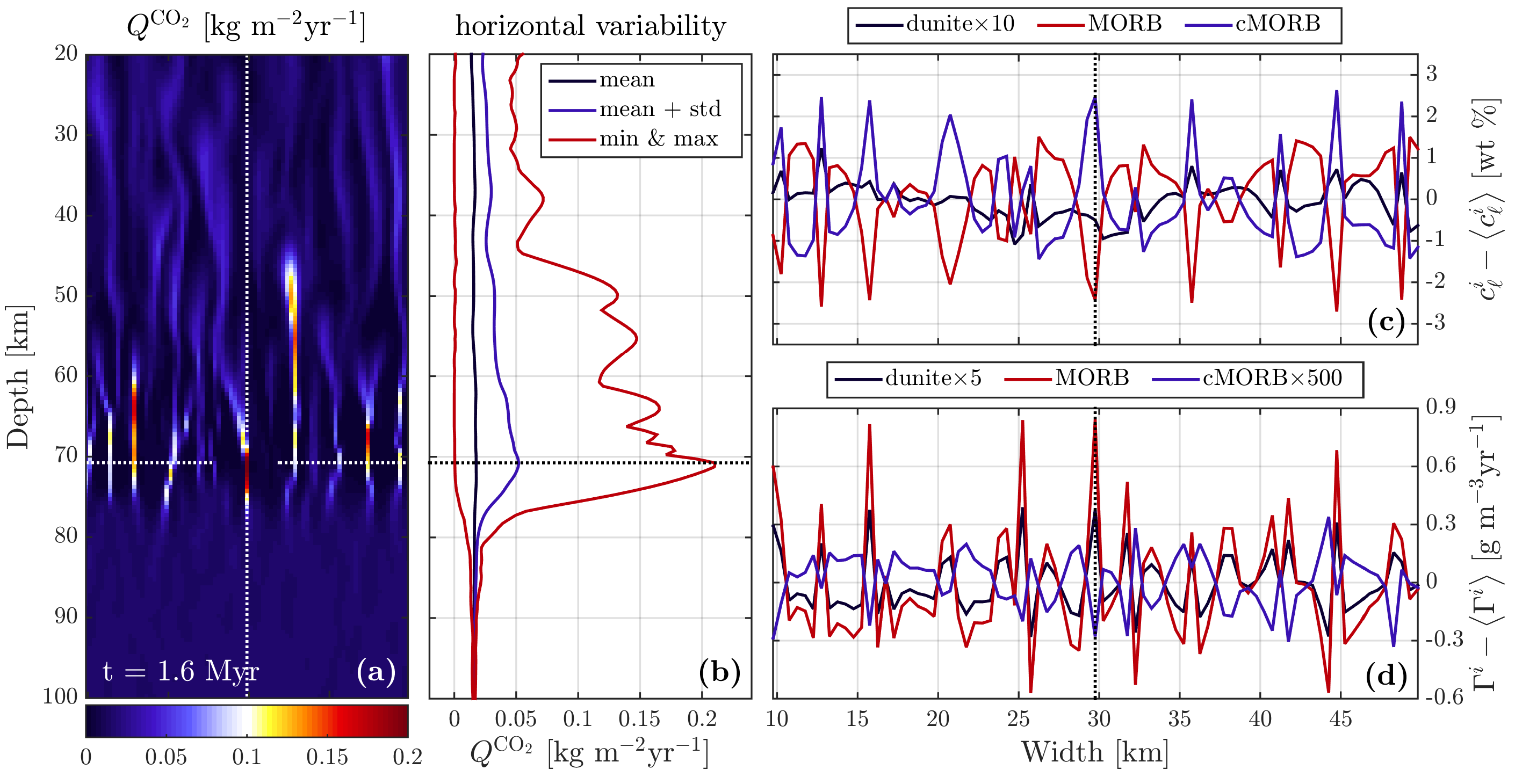}
	\caption{Bulk volatile flux in 2D column model of reference carbonated mantle. Panels and line colours the same as in Fig.~\ref{fig:2DColVolFluxHydr}.}
	\label{fig:2DColVolFluxCarb}
\end{figure}

The deviation from the horizontal mean across the maximum flux channel for component concentration in the liquid is shown in Fig.~\ref{fig:2DColVolFluxHydr}(c), and component-wise reaction rates in Fig.~\ref{fig:2DColVolFluxHydr}(d). The variation of these quantities around the mean sheds more light on the process of channel formation. Inside channels (max flux channels marked by dotted line), melt composition is enriched in the volatile-bearing hMORB component and depleted in the MORB component. The refractory dunite component is not correlated with channels. Component reaction rates show a relative increase in MORB dissolution inside channels. hMORB reaction rates are negatively perturbed across channels, while the dunite reaction rate correlates with the MORB component.

Fig.~\ref{fig:2DColVolFluxCarb} has the same form as Fig.~\ref{fig:2DColVolFluxHydr}, but applies to the carbonated mantle reference case. Volatile flux in the carbonated model is qualitatively very similar to the hydrated case. Channeling across the depth band of 60--80~km as seen in Fig.~\ref{fig:2DColVolFluxCarb}(a) leads to a variability of volatile flux with a standard deviation of approximately two times the mean and a maximum flux of up to an order of magnitude above the mean (Fig.~\ref{fig:2DColVolFluxCarb}(b)). The deviation from horizontal mean of melt composition (Fig.~\ref{fig:2DColVolFluxCarb}(c)) and component reaction rates (Fig.~\ref{fig:2DColVolFluxCarb}(d)) across the channels is again very similar to the hydrated case. The channels are arguably more well defined than in the hydrated case, which is likely a result of cMORB being the more incompatible component than hMORB.

The 2D reference models for a hydrated and carbonated  mantle column with compositional heterogeneities show that the flux of low-degree, volatile-enriched melt across the onset depth of major volatile-free silicate melting leads to volatile flux-enhanced melt production. As permeability is increased by these melting reactions, a positive reaction-transport feedback is created that leads to flow localisation \citep[e.g.,][]{chadam86}. Due to the qualitatively similar behaviour of hydrated and carbonated mantle models, we perform the following sensitivity analysis with parameter variations on hydrated simulations only.

\subsubsection{Parameter variation - Upwelling rate}
We test the sensitivity of our model to the imposed mantle upwelling rate by employing two additional simulations where the upwelling rate $u_m$ is set to 2 and 8~cm/yr compared to 5~cm/yr in the reference case. All other parameters are the same as in the hydrated reference simulation. \rev{This parameter range corresponds to upwelling rates characteristic for passive upwelling beneath medium-to-fast spreading mid-ocean ridges.}

\begin{figure}[htb]
	\centering
	\includegraphics[width=\textwidth]{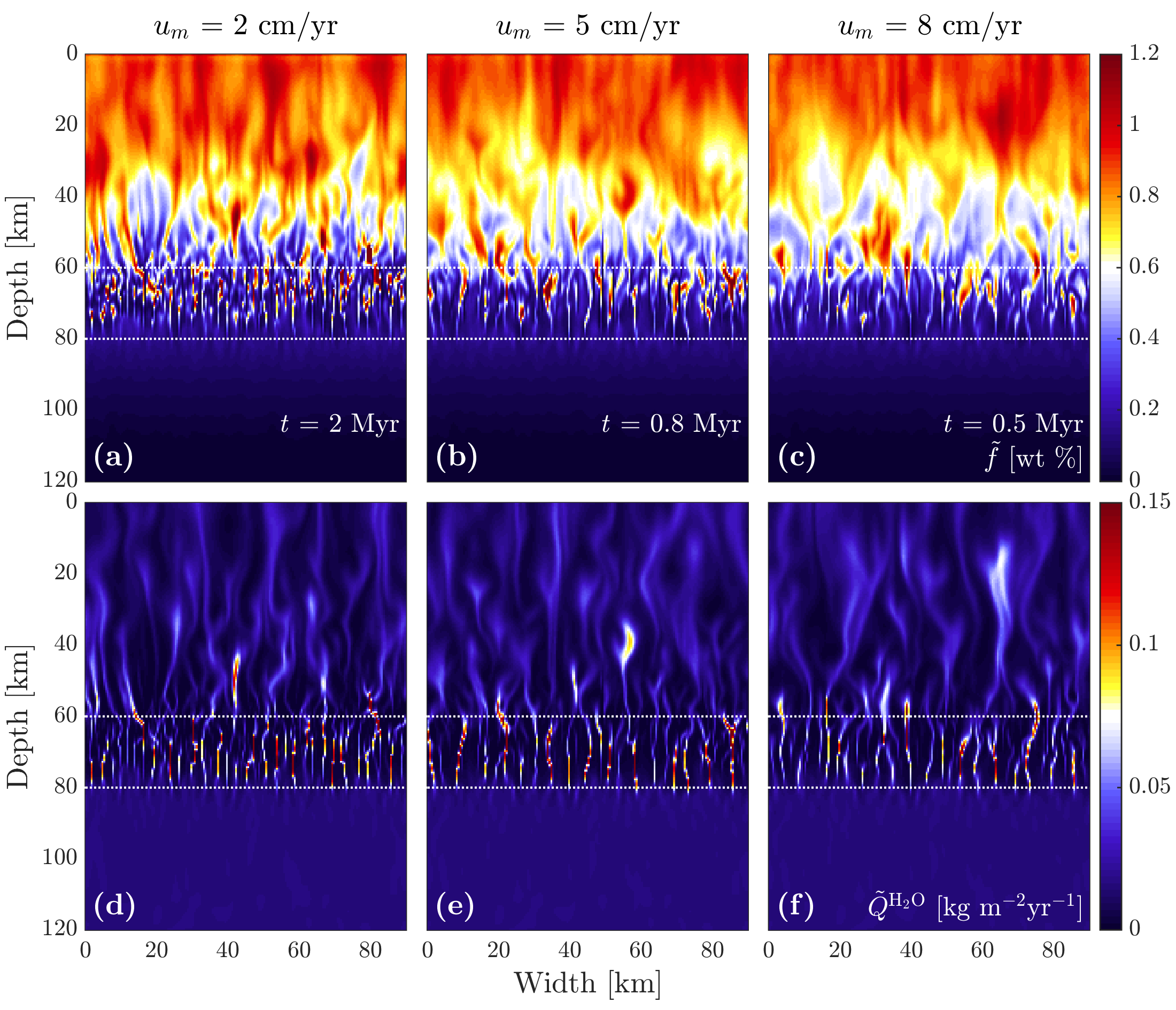}
	\caption{Effect of upwelling rate on reactive channelling in 2D column models of hydrated mantle. Scaled melt fraction $\tilde{f} = f \times (u_m^\mathrm{(ref)}/u_m)^{1/3}$ \textbf{(a)}--\textbf{(c)} and scaled bulk water flux $\tilde{Q}^\mathrm{H_2O} = Q^\mathrm{H_2O} \times u_m^\mathrm{(ref)}/u_m$ \textbf{(d)}--\textbf{(f)}. Scaling to highlight relative flow localisation. Results given at model times of 2 Myr \textbf{(a)} \& \textbf{(d)}, 0.8 Myr \textbf{(b)} \& \textbf{(e)}, and 0.5 Myr \textbf{(c)} \& \textbf{(f)}, corresponding to same solid upwelling distance of 40~km. Dotted lines indicate reference channelling depth band.}
	\label{fig:2DColUpwRate}
\end{figure}

The results of this parameter variation are shown in terms of scaled melt fraction and volatile flux in Fig.~\ref{fig:2DColUpwRate}, at times corresponding to a common solid upwelling distance of 40~km. Fig.~\ref{fig:2DColUpwRate}(a)--(c) show how relative perturbations in melt fraction are decreased as the upwelling rate is increased. A re-scaling of $\tilde{f} = f \times (u_m^\mathrm{(ref)}/u_m)^{1/3}$ applied in Fig.~\ref{fig:2DColUpwRate} demonstrates that the retained melt fraction scales with $u_m^{1/3}$, \rev{a consequence of the cubic power law chosen for permeability as a function of melt fraction.} Fig.~\ref{fig:2DColUpwRate}(d)--(f) show that localisation of volatile flux is strongest at the lowest upwelling rate. Here, the applied scaling is $\tilde{Q}^\mathrm{H_2O} = Q^\mathrm{H_2O} \times (u_m^\mathrm{(ref)}/u_m)$, as volatile mass flux scales linearly with upwelling rate. \rev{Furthermore, the slight increase in channel spacing with increasing upwelling rates illustrates that channel spacing is controlled by the compaction length, which depends linearly on the retained melt fraction and thus scales with $u_m^{1/3}$ as well.}

These results show that melt flow localisation is more pronounced at lower upwelling rates. This may be somewhat counter-intuitive, as increased melt production at higher $u_m$ enhances the melt segregation rate, which in turn allows for increased reactive melting. \rev{However, it is the ratio of melt to solid velocity that favours the stability of reactive channels, and not the absolute magnitude of melt segregation rates itself (cf.~\eqref{eq:reactive_melting_number}).} Thus, for increased upwelling rate, larger perturbations to melt segregation are required to trigger the instability. For a given amplitude of compositional heterogeneities that provide initial perturbations in melt segregation rate, growth of the reactive channelling instability is enhanced at lower mantle upwelling rates.

\subsubsection{Parameter variation - Perturbation type and amplitude}
Mantle heterogeneity is not sufficiently understood to constrain the spatial distribution and amplitude of compositional variations that are imposed on our model mantle. For example, it is not obvious if perturbations in major element composition are correlated, anti-correlated, or uncorrelated with perturbations in volatile content. Therefore, it is of interest to separately consider the effects of different types of compositional heterogeneities on melt production and transport. To this end, we have run three further simulations of hydrated mantle melting with perturbations of the same amplitude and distribution as in the reference case. The parameter variation runs have perturbations added: to the bulk MORB content only; to the bulk water content only; and to both combined. However for the combined heterogeneity, the MORB and water perturbations are anti-correlated, unlike in the reference case.

\begin{figure}[htb]
	\centering
	\includegraphics[width=0.75\textwidth]{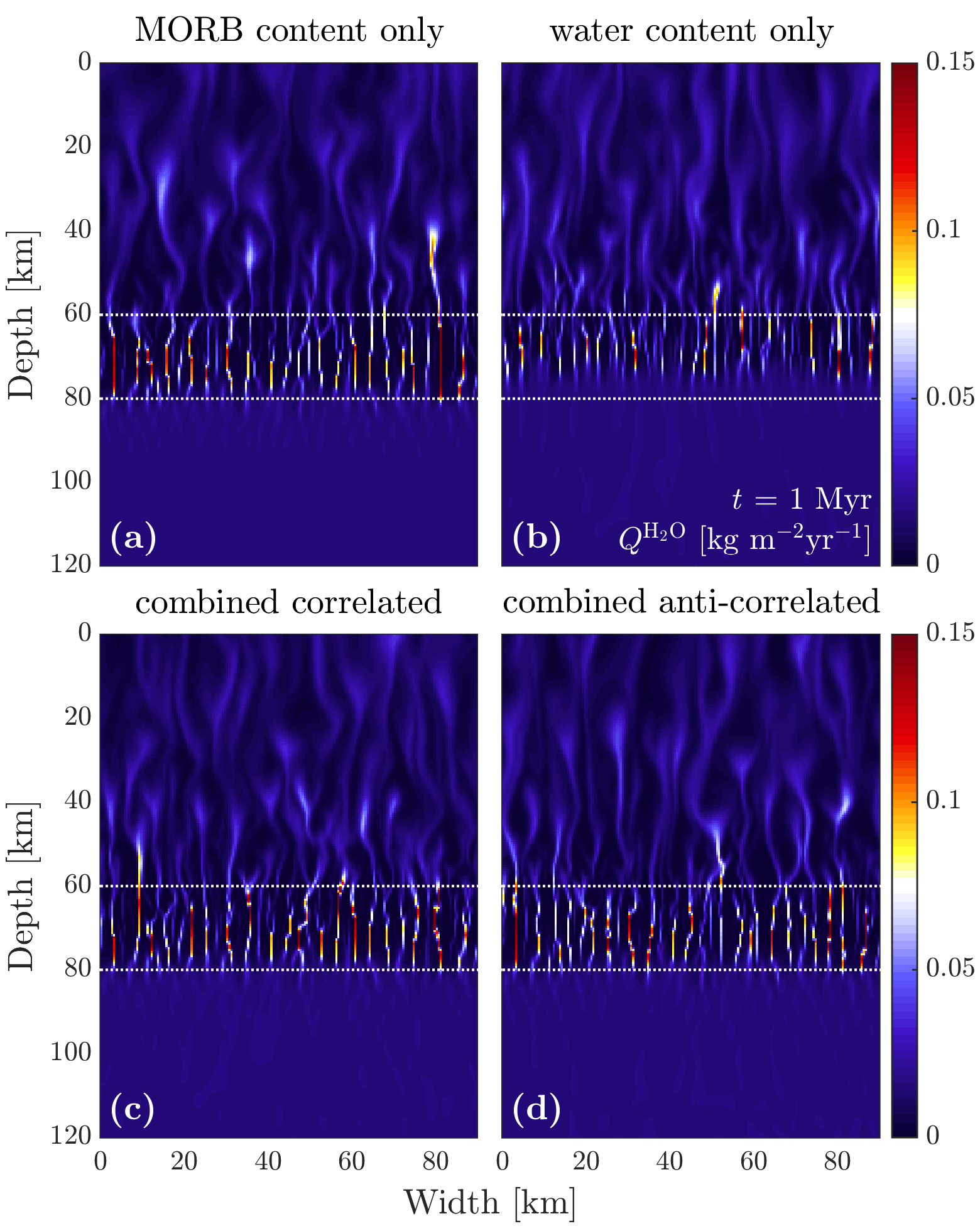}
	\caption{Effect of compositional perturbations on reactive channelling in 2D column models of hydrated mantle. Bulk water flux at model time of 1 Myr for models with smoothed random perturbations of $\pm$1~wt\% on bulk MORB content only \textbf{(a)}; perturbations of $\pm$10 wt~ppm on bulk water content only \textbf{(b)}; same amplitude of correlated \textbf{(c)} and anti-correlated \textbf{(d)} perturbations on bulk MORB and water content combined. Dotted lines indicate reference channelling depth band.}
	\label{fig:2DColPertType}
\end{figure}

The results of this parameter variation are shown in Fig.~\ref{fig:2DColPertType}. The figure shows model snapshots at 1 Myr, when models have reached their statistical steady-state behaviour, with the reference channelling band between 60 and 80~km depth indicated by dotted lines. Clearly, all types of perturbations to mantle composition lead to flow localisation expressed as channelised volatile flux. It is apparent, however, that simulations with perturbations to the MORB content (Fig.~\ref{fig:2DColPertType}(a), (c) \& (d)) exhibit more pronounced channelisation than the run in Fig.~\ref{fig:2DColPertType}(b), where bulk water content only is perturbed. In the latter case, fewer and shorter channels initiate at slightly shallower depth. 

There is no significant difference in behaviour between models where perturbations in volatile content are correlated (Fig.~\ref{fig:2DColPertType}(c)) and anti-correlated (Fig.~\ref{fig:2DColPertType}(d)) with major element heterogeneities. We therefore conclude that the major element distribution exerts a stronger control on the flow localisation than the distribution of volatiles. However, it is surprising to find that perturbations of as little as $\pm10$ wt~ppm water in the mantle give rise to only slightly less pronounced localisation of melt transport. In fact, variations in liquid composition arising from reactive channelling in the simulations in Fig.~\ref{fig:2DColPertType}(a) and (b) are of the same order of magnitude for both MORB and hMORB components, not depending on which of the components was initially perturbed. These findings suggest that perturbations in major elements lead to stronger initial perturbations in melt productivity, and consequently in melt segregation rate, leading to faster growth of reactive channels. However, variability of major element and volatile concentrations in the melt does not primarily depend on initial mantle heterogeneity, but rather on the presence of channelised reactive flow.

\begin{figure}[htb]
	\centering
	\includegraphics[width=\textwidth]{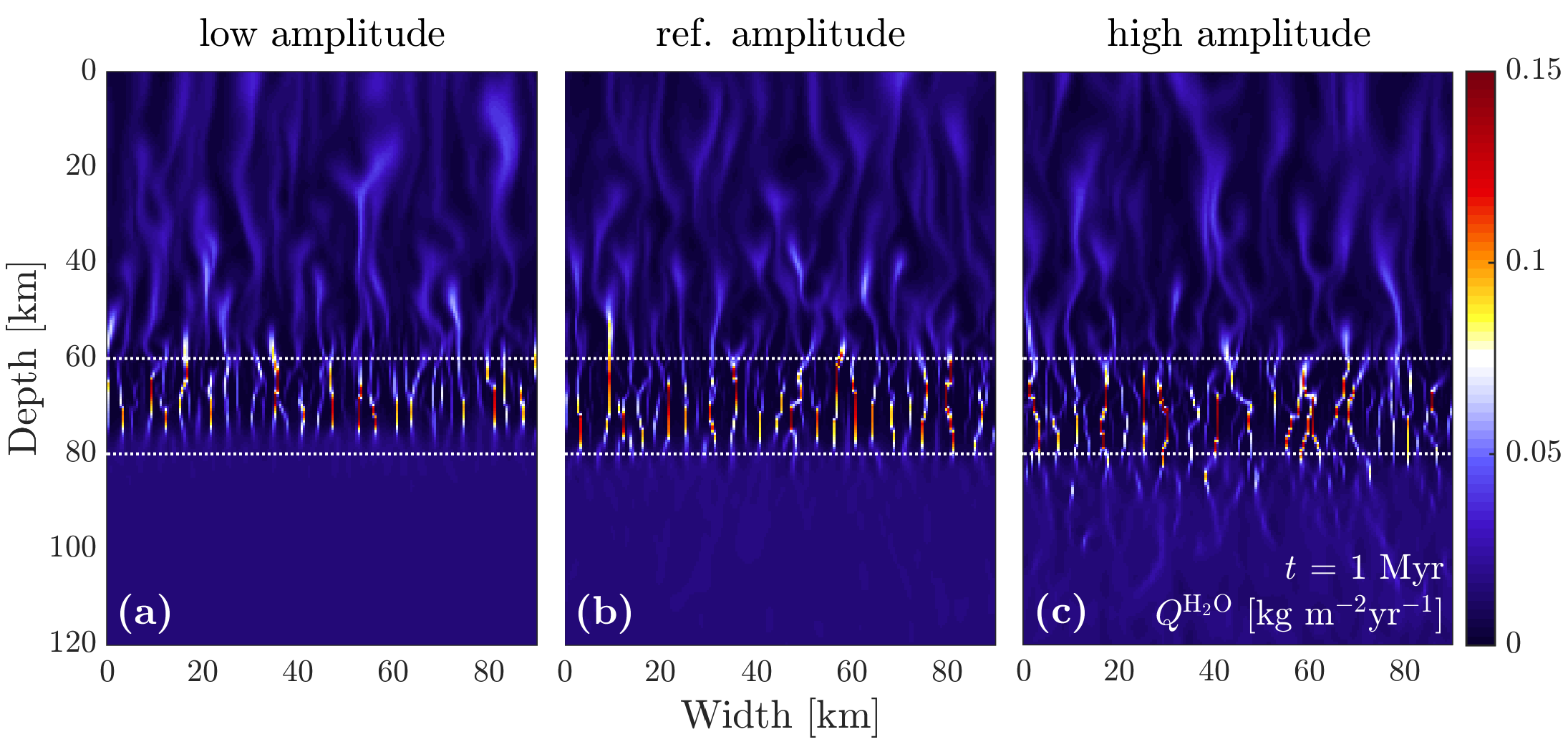}
	\caption{Effect of perturbation amplitude on reactive channelling in 2D column models of hydrated mantle. Bulk water flux at model time of 1 Myr for models with combined, correlated perturbations with amplitudes of $\pm0.2$ wt\% and $\pm2$ wt~ppm \textbf{(a)}, and $\pm5$ wt\% and $\pm50$ wt~ppm \textbf{(b)} on bulk MORB and bulk water content, respectively. Dotted lines indicate reference channelling depth band.}
	\label{fig:2DColPertAmpl}
\end{figure}

Having observed that any type of mantle heterogeneity may lead to significant perturbations in melt composition and localised melt transport in reactive channels, the question remains how robust these findings are for different amplitudes of perturbation. Two additional simulations are employed, where the perturbation amplitudes of combined, correlated heterogeneities are set to a factor of five smaller or larger than in the reference case. Perturbation amplitudes for the two parameter variation runs bracketing the reference value in the middle then become $\pm$[0.2,1,5]~wt\% on bulk MORB, and $\pm$[2,10,50] wt~ppm on bulk water content. 

The results at 1 Myr model time are reported in Fig.~\ref{fig:2DColPertAmpl} in terms of water mass flux $Q^\mathrm{H_2O}$. The reference channeling band between 60 and 80~km depth is again indicated by dotted lines. Clearly, flow localisation is observed in all three simulations. However, the amplitude of mantle heterogeneity has a significant control on the depth of onset, intensity, and pathways of flux channelling in the mantle column. Higher amplitude of perturbation leads to a deeper onset of flow localisation and results in larger amplitude channelised flow. The pathways of melt channels become more curved with increased perturbation amplitude, as channels interconnect the initially more fertile patches of mantle. These results reinforce the findings on perturbation types above: even the smallest perturbations lead to significant flow localisation in mantle models with volatiles.

\subsubsection{Parameter variation - Reactive model}
All results of 2D simulations so far have employed a batch reaction model, with an assimilation ratio of $\chi=1$, and a reaction time of $\tau_\Gamma=100$ yr. \rev{The faster reaction time in 2D simulations with respect to 1D simulations is chosen to ensure near complete equilibration even for channelised melt flow, which can be significantly faster than in 1D columns.} An additional simulation is employed here to test the effect of a different reactive model settings on reactive flow in the upwelling mantle. The assimilation ratio $\chi$ is reduced by a factor of ten to produce a fractional-assimilative model for comparison to the hydrated reference model.

\begin{figure}[htb]
	\centering
	\includegraphics[width=0.75\textwidth]{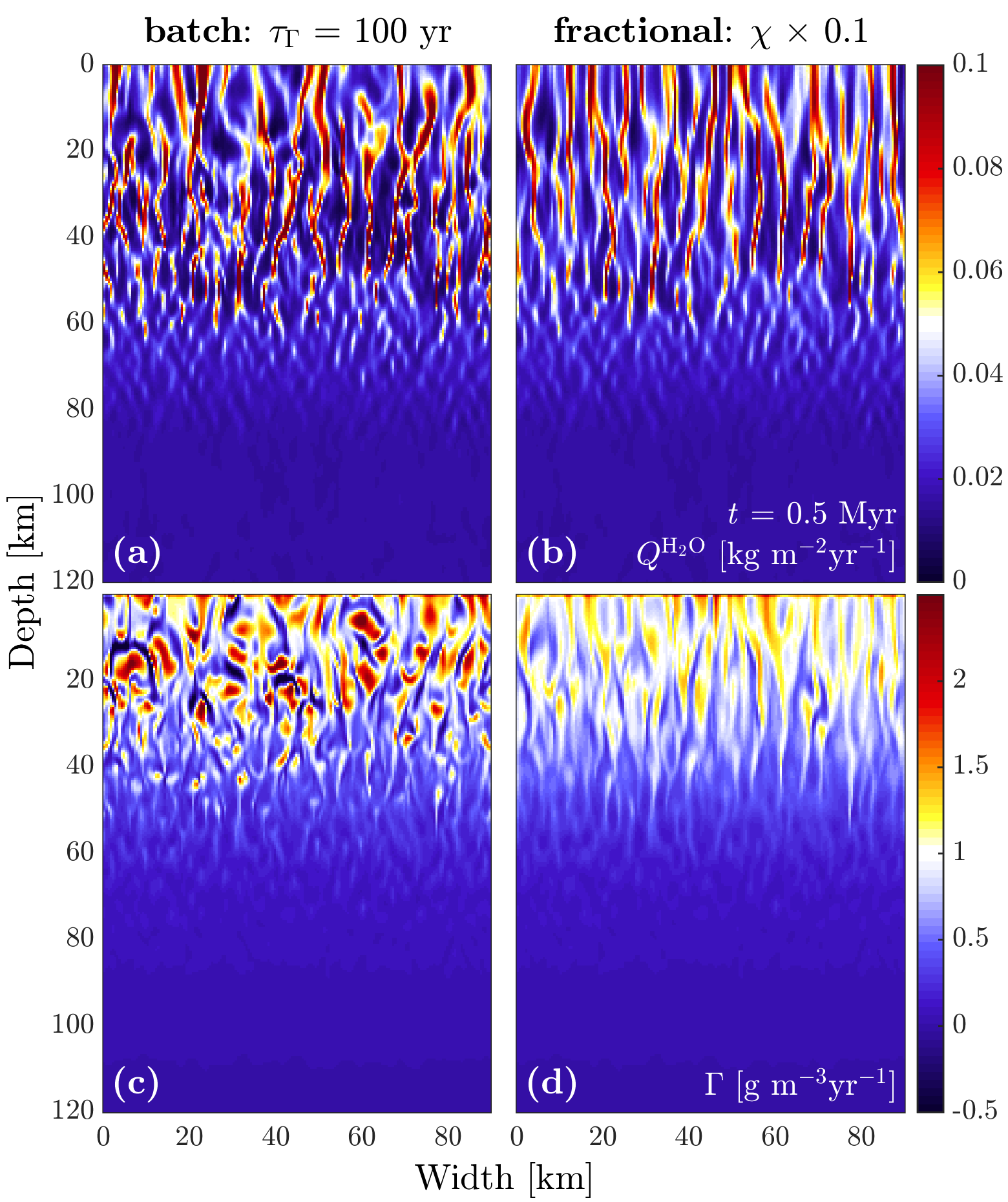}
	\caption{Effect of different reaction parameters on reactive channelling in 2D column models of hydrated mantle. Bulk water flux \textbf{(a)}--\textbf{(b)}, and net melting rate \textbf{(c)}--\textbf{(d)} at model time of 0.5 Myr for models with batch melting (reference $\tau_\Gamma$ = 100 yr) \textbf{(a)} \& \textbf{(c)}, and fractional-assimilative melting ($\chi$ = 0.1) \textbf{(b)} \& \textbf{(d)}. Colour scale is truncated to highlight flow and reaction patterns.}
	\label{fig:2DColReact}
\end{figure}

Figure \ref{fig:2DColReact} shows bulk water flux $Q^\mathrm{H_2O}$ and net melting rate $\Gamma$ after a model time of 0.5~Myr for the batch and fractional-assimilative models. The main difference arising from the variation in $\chi$ is the shape of localisation patterns in water flux and melting rate. In the reduced-$\chi$, fractional scenario, straighter and more elongated channels emerge. In the reference case, channels start dispersing upward of 20~km depth (panels (a) and (c)), whereas they persist to the top of the domain in the parameter variation scenario (panels (b) and (d)).

\rev{The flow regime in which vertical channels disperse into a more diffuse flow has characteristics of compaction--dissolution waves \citep{aharonov95, hesse11, liang11}. This type of reaction--transport feedback is predicted by stability analysis to be a stable regime for a range of conditions relevant to decompression melting in the upwelling mantle.} The fractional--assimilative reaction model favours stationary channelling over the travelling-wave instability. Fractional melting leads to a different proportion of components transferred by reactive mass exchange (see Fig.~\ref{fig:1DColReactions}), such that more mass transfer is required than in the batch model to achieve the same amount of compositional equilibration. Thus, more melt production can be driven by the same solubility gradient in the fractional model, strengthening reactive channellisation with respect to the compaction--dissolution instability \citep{hesse11}. \rev{Note, however, that this difference in behaviour is transient, and that, in steady state, melt transport above 60~km depth is found to be in the compaction--dissolution wave regime for all parameters tested here.}


\section{DISCUSSION  \label{sect:Discussion}}

\subsection{Volatile flux-induced reactive channelling}
Previous studies have found that reactive channelling is caused by the reactive infiltration instability \citep{chadam86}, where reactive melt production is increased in proportion to vertical melt flux \citep{aharonov95, spiegelman01, liang10}. However, background decompression melting, depletion of reacting species, the energetic cost of reactions, as well as compaction of the host rock may suppress flow instability within an upwelling mantle column \citep{hewitt10, weatherley12}. Unlike some previous studies, no boundary condition is applied here to introduce enhanced melt flow sufficient to cause instability. Instead, we explore the spontaneous emergence of reactive channelling from a volatile-bearing mantle column undergoing decompression and reactive melting. Energy conservation in our models includes consumption of latent heat, which acts to suppress channelisation by imposing an energetic cost to reactions. Additionally, we allow composition to evolve and be depleted by melting and reaction, which ultimately limits the availability of reactants. As known from analysis of trace element transport by percolating melt \citep{navon87}, more incompatible components of composition are transported away at an increased rate compared to more compatible ones. Here, this principle applies to both volatile and major element components, causing hydrated and carbonated components to be depleted from the mantle column \rev{at much faster rates} than the basaltic MORB component. As volatile flux fuels reactive channelling, exhaustion of available volatiles in the host rock and dilution of the melt volatile content by dissolution of volatile-free MORB both act to limit channelling \rev{to a depth band of 60--80 km in steady state}. Given these limiting effects, it is a significant finding that our models of mantle melting with volatiles exhibit robust reactive channelling over a range of realistic mantle conditions. Taking the ratio of vertical melt flux $q_z$ in the 2D reference simulations to that in the 1D reference models at steady state gives a melt flux in channels that is enhanced by about a factor of ten. 

The cause for enhanced flow focusing in our models under conditions where previous models predict distributed flow is the addition of volatiles. Water and carbon dioxide deepen the onset of decompression melting in the mantle. On the solidus, small fractions of volatile-enriched melt are produced, containing approximately 3~wt\% H$_2$O, or 15~wt\% CO$_2$, values close to recent estimates for near-solidus hydrated or carbonated mantle melts \citep{hirschmann09,dasgupta13}. Due to the depolymerising effect of volatiles on the viscosity of silicate melts, our model predicts these low degree melts to be mobile with velocities of order 10~cm/yr (Fig.~\ref{fig:1DColMu}). \cite{chadam91} suggested that the reactive infiltration instability may interact with viscous fingering \citep{saffman58} if the dissolving species causes an increase in liquid viscosity. This is the case in our model, where reactive melting means dissolution of MORB and thus an increase in liquid viscosity. However, there is no evidence that viscous fingering has a significant effect on flow focussing in our models. 

As the near-solidus, volatile-enriched melts flow upward, they approach the stability field of volatile-free MORB melting. As melt flows across the volatile-free solidus, the stable melt composition coexisting with peridotite changes rapidly from volatile-enriched to increasingly silica-rich and volatile-poor. Furthermore, the equilibrium melt fraction increases significantly across this transition. As reactions equilibrate melt that is flowing across this transition, regions of increased melt flux attain higher rates of silicate dissolution and therefore higher melt production. The resulting increase in melt fraction enhances permeability, allowing for further increase in flux of volatile-enriched melt from below. For transient magmatic systems in previously undepleted rock, this transition may occur at much shallower depth than the volatile-free solidus (Figs.~\ref{fig:2DColTimeEvoHydr} \& \ref{fig:2DColTimeEvoCarb}). This feedback of enhanced volatile flux leading to enhanced basaltic melt production is the underlying cause for flow instability in our models.

The reduced melt viscosity with volatile content (see Fig.~\ref{fig:1DColMu}) does not appear to have a strong effect on these dynamics, as channelling only sets in as the volatile-free silicate melt production picks up. However, due to the dependence of the reactive infiltration instability on the melt flow rate, a decrease in liquid viscosity or an increase in permeability should generally enhance reactive channelling. This expected behaviour was confirmed in a series of additional simulations, which are not presented here for reasons of brevity.

We do not find a significant difference in the effects of hydrated versus carbonated melt production and transport (Figs.~\ref{fig:2DColVolFluxHydr} \& \ref{fig:2DColVolFluxCarb}). This may be due to the simplifications incorporated into the R\_DMC petrogenesis model with volatiles. Here, hydrated and carbonated components are distinguished only by a difference in melting point and \rev{partition coefficient}. Even though these differences modulate the depth of first melting, the concentration of the volatile compound in near-solidus melt, and the melt productivity at depths below the volatile-free solidus, they do not significantly modify volatile flux-induced reactive channelling. \rev{Therefore, we conclude that the flux of any low-degree melt enriched in incompatible components into the base of the major silicate melting region may promote a similar reactive channelling instability. The exact properties of the incompatible component under consideration are of secondary importance, as long as such a component significantly depresses the mantle solidus and strongly partitions into the melt. This implies that models including both water and carbon dioxide in the mantle column would likely see even more pronounced reactive channelling. Furthermore, we expect that the inclusion of an alkali-rich component would have a similar effect, possibly extending channelling to shallower depth than observed here.}

\subsection{Implications for magma/mantle dynamics and petrology}
Considering these results in various geodynamic contexts highlights some key implications. Beneath mid-ocean ridges, decompression melting takes place in a roughly triangular zone extending down from the ridge axis. The shape of the melt region is controlled by the extent of upwelling flow of the mantle in response to plate spreading, as well as the cooling of the plates away from the axis. As the melting region widens with depth, low-degree volatile-enriched melts may be produced over a broad lateral extent of asthenospheric mantle. As these melts cross the volatile-free solidus they will trigger reactive channelling. In a steady-state system, our results show that such channelling is limited to a depth range of 60--80~km. However, in the MOR system melt is focussed towards the ridge axis, thereby increasing melt flux, which may contribute to enhancement of melt channelisation. Our results therefore support the hypothesis of melt transport beneath MORs being controlled by a network of high-flux channels originating at depths around the volatile-free solidus.

In geodynamic contexts where basaltic melt is produced beneath an extending lithosphere, the depth band of reactive channelling observed here is likely to be directly beneath the lithosphere-asthenosphere boundary. Melt channelling across the volatile-free solidus depth is therefore expected to significantly control the distribution and composition of melt arriving at the base of the lithosphere, where rheological channelling or fracturing processes across the ductile-brittle transition may connect reactively channelised melt pathways in the mantle with crustal magma-processing systems \citep{havlin13}. In a subduction context, water- and carbon-bearing fluids released from dehydration of subducting oceanic lithosphere may lead to strong channelisation as they induce volatile-flux melting in the mantle wedge. The interaction of such channels with the subduction-related flow of the mantle wedge will likely control the extraction pathways of hydrous slab and mantle melts. 

\rev{As discussed above, reactive channelling observed in our models is driven by volatile flux-induced dissolution of MORB. Therefore these channels should not be directly identified with orthopyroxene-dissolution channels thought to create dunite bodies observed in ophiolites. A volatile-free, 2D control simulation (results not shown above) did not exhibit any significant reactive channelisation related to dissolution of MORB coinciding with precipitation of dunite at shallow pressures}. Therefore, the question remains what bearing our results have on the formation of dunite bodies as a possible geologic signature of channelised melt extraction in the near-surface asthenosphere beneath a MOR. \rev{The reasons that our models do not predict the formation of replacive dunite channels is understood from the scaling analysis of reactive melting in \eqref{eq:decomp_reaction_simple}, which states that reactive flow localisation depends on disequilibration by melt flux along a vertical solubility gradient of a fusible component. Localisation only occurs if melt can be sufficiently disequilibrated by melt segregation and if this disequilibrium drives a sufficient increase in melt fraction. The propensity of a reactive system to channelisation is affected by the stoichiometry of the reaction, which is here implicitly expressed by the concentration of components in reactively transferred mass.}

\rev{As seen in Fig.~\ref{fig:1DColReactions}, the rate of disequilibration by vertical melt flux along the solubility gradient of MORB is high within the zone of transition from volatile-enriched to volatile-free melting but drops to low levels again throughout the volatile-free melting region. This transition region corresponds to the depth band of 60--80~km within which sustained reactive channelling is observed in all 2D simulations. Even though melt flux is moderate across this depth range, it is combined with an increase in MORB solubility that is sufficient to drive reactive channelling. At depths above 60~km, channelling ceases even though melt flux continues to increase towards the surface. Instead, it is the lack of a significant solubility gradient of the MORB component towards the surface that is responsible for the transition from reactive channelling to a compaction-dissolution wave regime at shallower depths.}

\rev{Our model choice of representing volatile-free decompression melting as a two-component process of dunite+MORB and the resulting calibration of the R\_DMC method do not adequately approximate the complexities of basaltic melt production. A different multi-component representation of this process may yield solubility gradients of fertile components that extend to shallower pressures. In such a model, we would expect coupled magma/mantle flow to exhibit at least two stages of interconnected, reactive channelling. The first stage would be induced by flux of volatile- and incompatible-enriched and relatively Si-poor melt across the volatile-free solidus; The second stage would be driven by ongoing dissolution of basaltic components, possibly accompanied by precipitation of dunite component, extending further towards the surface.} The geologic signature of such channels would be predicted to take the shape of shallow dunite bodies transitioning to harzburgitic composition with depth, as proposed in previous studies \citep{kelemen95a, liang10}.

\subsection{Model limitations}
There are four key model limitations. First, to reduce computational expense, models presented here are run with only three chemical components. Capturing the complex mantle mineralogy with volatiles through only three effective components necessarily requires some restrictive model choices. Details of the mineralogy such as the solid-solution of the Mg-Fe-olivine system, the distinctive behaviour of clinopyroxene and orthopyroxene in the mantle, the role of high-Al phases, or the difference in volatile partitioning with different mineral assemblages at various pressures, are not currently represented by our thermochemical model. Also, the saturation of volatiles in the melt phase as it undergoes decompression is not treated here. As a consequence, the solidus curves of volatile-bearing mantle do not exhibit the characteristic sloping back to higher temperatures at low pressures found for hydrated and carbonated peridotite studies \citep{hirschmann99, katz03, dasgupta13}. This latter simplification is not considered of major importance to this study, however, as the main behaviour discussed here is channelised flow at pressures well above the level for fluid exsolution.

Second, our choice of linear kinetic reactive model is not rigorously constrained by theory or experiments. Rather, it is a model choice aimed at providing a simple closure condition, while retaining consistency with key petrological assumptions. In particular, our model choices based on \cite{rudge11} allow distinction between batch and fractional-assimilative types of reactions by adjusting a single model parameter, $\chi$. It should be noted, however, that the present approach does not allow models to evolve far from thermodynamic equilibrium. Furthermore, the possible dependence of reaction rates on temperature, pressure, component-specific activation energy, available reactive surface area, or reacting component properties (e.g.,~diffusivity) are not considered here, to avoid overcomplicating the model description.

Third, the rheology applied to these models is linear viscous, neglecting known \rev{effects} such as the dependence of mantle viscosity on temperature, pressure, grain size, water content, and shear stress \citep{hirth03}, stress-induced anisotropy of viscous shear and compaction flow in partially molten rock \citep{takei10}, and the limited yield strength of mantle rock under tensile stress conditions (i.e., decompaction under liquid over-pressure) \citep{murrell64}. Some of these additional complexities in mantle rheology are known or hypothesised to cause localisation of melt flow in the partially molten asthenosphere \citep{stevenson89, spiegelman03b, katz06, connolly07, keller13, takei13}. As the aim of this contribution is to highlight the interaction between petrogenesis and magma/mantle dynamics, we neglected all rheological feedbacks, allowing as to focus on reactive channelling. Enabling both rheological and reactive channelling in combination will likely lead to interesting, coupled behaviour. However, the investigation of such effects is beyond the scope of the present study, and will be left for future work.

Fourth, numerical resolution of these models is limited by computational cost. \rev{The characteristic spacing of channels is of order of the compaction length $l_0$, which varies over 1--5~km across the parameter space tested here (see Table~\ref{tab:scaling}). Considering mass flux conservation in channelised flow, the width of channels is predicted to scale as the channel spacing divided by the flux enhancement by channelling, hence predicting an equilibrium width of channels of order 100--500~m in our models. It follows therefore that these features seek a length scale just below what the present numerical discretisation allows (grid spacing is 500~m) and are therefore not sufficiently resolved.}  Varying grid resolution within the currently accessible range shows that the results do not change qualitatively with grid refinement. However, as flow enhancement is a function of localisation intensity, our model likely underestimates the effect of channelling on magma transport in natural systems.

\section{CONCLUSIONS  \label{sect:Conclusions}}
We have presented simulations that combine conservation of energy, mass and momentum with a simplified thermodynamic framework for disequilibrium-driven multi-component reactions. Calibrating the R\_DMC method to reproduce leading-order features of mantle melting with volatiles allows our simulations to make predictions about the formation and segregation of deep, low-degree, volatile-enriched magma. The flux of volatiles \rev{transported by} these magmas towards the stability field of volatile-free basaltic melting is found to trigger reactive channelisation. Magma flux is localised due to a positive reaction-transport feedback, whereby an increase in volatile-enriched melt flux enhances basaltic melt production, which in turn increases permeability and thus further enhances melt flux. These results couple the thermodynamics of multi-component petrogenesis to the dynamics of magmatic two-phase flow. We find that a bulk mantle volatile content of the order of 100~ppm allows for enough deep, volatile-enriched melt to be formed to fuel this flow instability over a depth band of 60--80~km. As a consequence, \rev{volatile-enriched} magma is transported, at least through parts of the mantle, in high-flux channels. Previous physical scaling arguments and simple two-phase models, which do not take into account the effective permeability enhancement provided by reactive channelling, may have underestimated melt transport rates in the upper mantle by up to an order of magnitude. This study adds to the evidence that magma transport arising from decompression melting of the upwelling mantle occurs as localised flow in a network of magmatic channels initiated around the depth of the volatile-free solidus. Also, the role of volatiles on mantle melting is demonstrated to be significant not only by depressing the mantle solidus and stabilising melt at higher pressures, but also by its effect on the dynamics of magma transport. While the results presented here are aligned with leading-order observations of mantle melting with volatiles, more research is needed to develop, calibrate, and validate this simplified reactive flow model in comparison to field and laboratory observations.

\subsection*{Acknowledgements}
The research leading to these results has received funding from the European Research Council under the European Union's Seventh Framework Programme (FP7/2007--2013)/ERC grant agreement number 279925. The authors thank the Geophysical Fluid Dynamics group for access to the BRUTUS cluster at ETH Zurich, Switzerland, where all simulations in this study were computed. The authors also thank members of the FoaLab for helpful discussions, and P.~Asimow, O.~Shorttle, J.~Rudge, M.~Hirschmann, D.~Bercovici, and an anonymous reviewer for their insightful comments. Katz is grateful for the support of the Leverhulme Trust.


\singlespacing
\bibliographystyle{jpet}
\bibliography{manuscript}
\onehalfspacing
\pagebreak


\cleardoublepage
\appendix

\section{APPENDIX A: CONSERVATION EQUATIONS \label{sect:AppendixA}}
\renewcommand{\theequation}{A\arabic{equation}}
\setcounter{equation}{0}
\renewcommand{\thefigure}{A\arabic{figure}}
\setcounter{figure}{0}

\subsection{Fluid dynamics model}

\subsubsection{Conservation of mass and momentum}
Conservation of mass in the solid and liquid phase is given in terms of mass fraction of melt $f$:
\begin{subequations}
  \label{eq:MassConsA}
  \begin{linenomath*}\begin{align}
    \label{eq:SolidMassA}
    \pdiff{\fs \rhobar}{t} + \Div (\fs \rhobar \vs) &= -\Gamma \: , \\
    \label{eq:LiquidMassA}
    \pdiff{f \rhobar}{t} + \Div (f \rhobar \vl) &= \Gamma \: ,
  \end{align}\end{linenomath*}
\end{subequations}
where $\vs$ and $\vl$ are the solid and liquid velocities and $\Gamma$ [kg~m$^{-3}$] is the mass-transfer rate from  solid to liquid \citep{mckenzie84, bercovici01a}. Given these statements of mass conservation, the following expressions hold for any scalar quantity $a$ transported by the solid and liquid phases, respectively
\begin{subequations}
  \label{eq:MassConsSubsA}
  \begin{linenomath*}\begin{align}
    \label{eq:SolidMassSubsA}
    \pdiff{\fs \rhobar a_s}{t} + \Div (\fs \rhobar a_s \vs) &= \fs \rhobar \matsdiff{a_s}{t} - a_s \Gamma \: ; \\
    \label{eq:LiquidMassSubsA}
    \pdiff{f \rhobar a_{\ell}}{t} + \Div (f \rhobar a_{\ell} \vl) &= f \rhobar \matldiff{a_{\ell}}{t}+ a_{\ell} \Gamma \: .
  \end{align}\end{linenomath*}
\end{subequations}

Summing equations \eqref{eq:MassConsA} yields an expression for mass conservation of the bulk mixture
\begin{linenomath*}\begin{align}
	\label{eq:BulkMassA}
	\Div \vbar = \Gamma \Delta(1/\rho) \: ,
\end{align}\end{linenomath*}
where $\vbar = \phi \vl + (1-\phi) \vs$ is the mixture velocity.  Here we have applied the Boussinesq approximation to each phase independently, assuming that $\rhos = \rho_0$, $\rhol = \rho_0-\Deltarho_0$. As a result, the mixture density only varies with volume fraction of melt as $\rhobar = \rho_0 - \phi \Deltarho$. The term on the right hand side of \eqref{eq:BulkMassA} represents the volume changes associated with phase change reactions. It has a negligible effect on large-scale flow, but is required for accurate mass conservation --- in particular in a reactive flow model with strongly incompatible components like volatiles.

The mixture velocity can be decomposed as $\vbar = \vs + \q$ using the phase separation flux,
\begin{linenomath*}\begin{align}
	\label{eq:DarcysLawA}
	\q = -\phi \Delta \mathbf{v} = - K \mu^{-1} \left(\Grad P_{\ell} - \rho_{\ell} \gvec \right) \:.
\end{align}\end{linenomath*}
This is the standard form of Darcy's law for a two-phase medium \citep{mckenzie84}, stating that the separation of melt and solid is driven by pressure gradients and modulated by the permeability $K$ and the liquid viscosity $\mu$. Substituting this flow decomposition into \eqref{eq:BulkMassA} gives
\begin{linenomath*}\begin{align}
	\label{eq:BulkMass2A}
	\Div \vs - \Div (K/\mu) \left(\Grad P_{\ell} - \rhol \gvec \right) - \Gamma \Delta(1/\rho) = 0 \:.
\end{align}\end{linenomath*}
This equation states that divergence in the solid flow is balanced by convergence in the phase separation flux, and that in the absence of phase change reactions, the mixture flow field is divergence free.

Momentum conservation in the two-phase mixture takes the form of the Stokes equation
\begin{linenomath*}\begin{align}
	\label{eq:BulkMomentumA}
	\Div \taubar - \Grad \Pbar + \rhobar \gvec = 0 \: ,
\end{align}\end{linenomath*}
where $\taubar$ is the phase-averaged deviatoric stress tensor and $\Pbar$ the phase-averaged pressure. In the following, buoyancy forces will be dropped from the bulk momentum conservation. The focus of this study is to investigate reaction transport feedbacks, therefore active flow of the matrix will be suppressed here. It has proved convenient to introduce the pressure decomposition 
\begin{linenomath*}\begin{align}
	\label{eq:PresDecompositionA}
	\Pbar = P + p + P_\textrm{lith} \: , 
\end{align}\end{linenomath*}
with $P = P_{\ell} - P_\textrm{lith}$ being the dynamic Stokes pressure, $p = (1-\phi) \Delta P$ the compaction pressure \citep{katz07, keller13}, and $P_\textrm{lith} = \rho_0 g z$ a lithostatic reference pressure corresponding to the solid density. The compaction pressure $p$ is the volumetric component of stress related to compaction deformation. 

\subsubsection{Shear and compaction flow laws}
We write viscous constitutive laws for shear stress $\taubar$ (neglecting shear stresses in the liquid phase \citep{sleep84, mckenzie84}) and compaction pressure $p$ of standard form
\begin{linenomath*}\begin{align}
	\label{eq:ShearFlowLawA}
	\taubar &= \hspace{9pt} 2\eta \mathbf{D}(\vs) \: , \\
	\label{eq:CmpctFlowLawA}	
	p &= - \zeta \Div \vs \: .
\end{align}\end{linenomath*}
Here, $\eta = (1-\phi) \eta_0 \exp{(-\alpha\phi)}$ the effective shear viscosity and $\zeta = (1-\phi) \eta_0 \phi^{-1}$ the effective compaction viscosity, where $\eta_0$ is the intrinsic shear viscosity of the solid phase, and $$\mathbf{D}() = \frac{1}{2}(\Grad() + \Grad()^T) - \frac{1}{3} \Div () \I$$ is the deviatoric symmetric gradient tensor operator.

With the pressure decomposition and the viscous flow laws defined as above, we finally write the conservation of bulk momentum and mass in the two-phase mixture in the three-field form \citep{keller13, rhebergen14} in terms of the variables $\vs$, $P$ and $p$ as:
\begin{subequations}
  \begin{linenomath*}\begin{align}
    \label{eq:BulkMomentumFinalA}
    \Div 2\eta \mathbf{D}(\vs) - \Grad p - \Grad P &= 0 \: , \\
    \label{eq:BulkMassFinalA}	
    \Div \vs - \Div (K/\mu) \left(\Grad P + \Deltarho g z \right) - \Gamma \Delta(1/\rho) &= 0 \: , \\
    \label{eq:CompactionFinalA}
    \Div \vs + p/\zeta &= 0 \: .
  \end{align}\end{linenomath*}
\end{subequations}

\subsection{Thermo-chemical model}

\subsubsection{Conservation of melt and component mass}
To state a conservation law for the mass of melt per unit volume, we rewrite the mass conservation equation for the solid phase \eqref{eq:MassConsA} as
\begin{linenomath*}\begin{align}
	\label{eq:MeltMassA}
	\matsdiff{f \rhobar}{t} - \fs \rhobar \Div \vs = \Gamma \: .
\end{align}\end{linenomath*}
Thus $f$ evolves in time due to advection, compaction, and inter-phase mass transfer.

Conservation laws for component mass $c_{s,\ell}^i$ in the solid and liquid phase, respectively, are given by
\begin{subequations}
  \begin{linenomath*}\begin{align}
    \label{eq:LiquidCompMassA}
    \pdiff{f \rhobar \cli}{t} + \Div (f \rhobar \cli \vl) &= \hspace{9pt} \Gi + d_0 \Div f \Grad \cli \: , \\
    \label{eq:SolidCompMassA}
    \pdiff{\fs \rhobar \csi}{t} + \Div (\fs \rhobar \csi \vs) &= -\Gi \: .
  \end{align}\end{linenomath*}
\end{subequations}
$\Gi$ is the mass transfer rate of the $i$th component from solid to liquid phase due to reaction. Note that we only consider chemical diffusion/dispersion in the liquid phase, where it is controlled by the diffusivity/dispersivity $d_0$, taken as constant for all components.

Substituting eqns.~\eqref{eq:MassConsSubsA} gives a more convenient form for the conservation of component mass,
\begin{subequations}
  \begin{linenomath*}\begin{align}
    \label{eq:LiquidCompMassFinalA}
    f \rhobar \matldiff{\cli}{t} &= \hspace{9pt} \Gamma^i - \cli \Gamma + d_0 \Div f \Grad \cli \: , \\
    \label{eq:SolidCompMassFinalA}
    \fs \rhobar \matsdiff{ \csi}{t} &= -\Gamma^i + \csi \Gamma \: .
  \end{align}\end{linenomath*}
\end{subequations}
Thus, the chemical composition of the solid and liquid phase evolves in time due to advection, reaction, and diffusion/dispersion in the liquid.

\subsubsection{Conservation of energy}
Conservation of internal energy is formulated in terms of the specific internal energy [J/kg] of each phase, which is taken as the sum of the specific phase enthalpies $h_{s,\ell}$ and the potential energy due to gravity (kinetic and chemical potential energy are neglected)
\begin{linenomath*}\begin{align}
	\label{eq:IntEnergyA}
	e_{s,\ell} = h_{s,\ell} - g z \: .
\end{align}\end{linenomath*}
Infinitesimal changes in specific phase enthalpies are due to changes in sensible and latent heat (with $L^i$ the latent heat of fusion of the $i$th component), and the volumetric work of thermal expansivity $\alpha$ (taken as constant for all phases and components)
\begin{subequations}
  \label{eq:InfEnthA}
  \begin{linenomath*}\begin{align}
    \label{eq:InfEnthSolidA}
    \infd h_s &= c_p \infd T + (1-\alpha T) g \infd z \: , \\
    \label{eq:InfEnthLiquidA}
    \infd h_\ell &= c_p \infd T + (1-\alpha T) g \infd z + \sum\limits_{i=1}^n L^i \infd \cli \:,
  \end{align}\end{linenomath*}
\end{subequations}
where we have assumed that specific heat $c_p$ and thermal expansivity $\alpha$ are constant and equal between phases.

Conservation laws for internal energy in the solid and liquid phase are given by
\begin{linenomath*}\begin{align}
	\label{eq:EnConsLiquidA}
	  \pdiff{f \rhobar h_{\ell}}{t} + \Div (f \rhobar h_{\ell} \vl) 
	+ \pdiff{f \rhobar g z}{t} + \Div (f \rhobar g z \vl) &= \Div \q^e_{\ell} + Q^e_{\ell} \: , \\
	\label{eq:EnConsSolidA}
	  \pdiff{\fs \rhobar h_s}{t} + \Div (\fs \rhobar h_s \vs) 
	+ \pdiff{\fs \rhobar g z}{t} + \Div (\fs \rhobar g z \vs)&= \Div \q^e_s + Q^e_s  \: .
\end{align}\end{linenomath*}
The terms introduced on the right hand side include the divergence of diffusive fluxes $\q^e_{s,\ell}$, and source terms $Q^e_{s,\ell}$.

After substituting eqns.~\eqref{eq:MassConsSubsA}, collecting and simplifying of terms, energy conservation is expressed in terms of changes of enthalpy over time as
\begin{linenomath*}\begin{align}
	\label{eq:EnConsLiquidA2}
	  f \rhobar \matldiff{h_{\ell}}{t} - f \rhobar g w_{\ell} + (h_{\ell} - g z) \G &= \Div \q^e_{\ell} + Q^e_{\ell} \:, \\
	\label{eq:EnConsSolidA2}
	  \fs \rhobar \matsdiff{h_s}{t} - \fs \rhobar g w_s - (h_s - g z) \G &= \Div \q^e_s + Q^e_s  \: .
\end{align}\end{linenomath*}
Here, $w_{s,\ell}$ are the vertical components of solid and liquid velocity, respectively.

The phase contributions to the total internal energy of the two-phase mixture are now added to yield a statement for the conservation of bulk energy
\begin{linenomath*}\begin{align}
	\label{eq:EnConsSumA}
	  f \rhobar \matldiff{h_{\ell}}{t} + \fs \rhobar \matsdiff{h_s}{t} &= g \rhobar \bar{w} + k_0 \Grad^2 T - \G \sum\limits_{i=1}^n L^i \cli  + \Psi \: .
\end{align}\end{linenomath*}
To arrive at this relation, we have substituted the enthalpy phase difference $h_\ell - h_s = \sum\limits_{i=1}^n L^i \cli$, and simplified the sum of diffusive flux terms $\q^e_{s,\ell}$ to be the total heat diffusion depending on a constant mixture conductivity $k_0$. The sum of source terms $Q^e_{s,\ell}$ we define to be the total viscous dissipation in the mixture
$$\Psi = 2\eta\Vert\mathbf{D}(\vs)\Vert^2 + \zeta(\Div \vs)^2 + (\mu/K) \vert\q\vert^2,$$
which is the sum of work rates due to shear and compaction deformation, and phase separation flux \citep{bercovici03, sramek07, rudge11}. Here, $\Vert\cdot\Vert$ and $\vert\cdot\vert$ denote tensor and vector magnitudes, respectively.

To write the conservation of energy in terms of mixture temperature $T$ (assuming thermal equilibrium between the phases), eqns.~\eqref{eq:InfEnthA} are substituted in \eqref{eq:EnConsSumA}, resulting in
\begin{linenomath*}\begin{align}
	\label{eq:EnConsFinalA}
	   \rhobar c_p \matbdiff{T}{t} &= \alpha T g \rhobar \bar{w} + k \Grad^2 T - \sum\limits_{i=1}^n L^i\Gamma^i + \Psi \: ,
\end{align}\end{linenomath*}
stating that potential temperature evolves in time due to advection, adiabatic heat exchange, thermal diffusion, heating by viscous dissipation, and exchange of sensible to latent heat by reaction. $$\matbdiff{()}{t} = \pdiff{()}{t} + \fs \vs \cdot \Grad () + f \vl \cdot \Grad ()$$ is the material derivative in the two-phase mixture.

\begin{figure}[htb]
	\centering
	\includegraphics[width=0.8\textwidth]{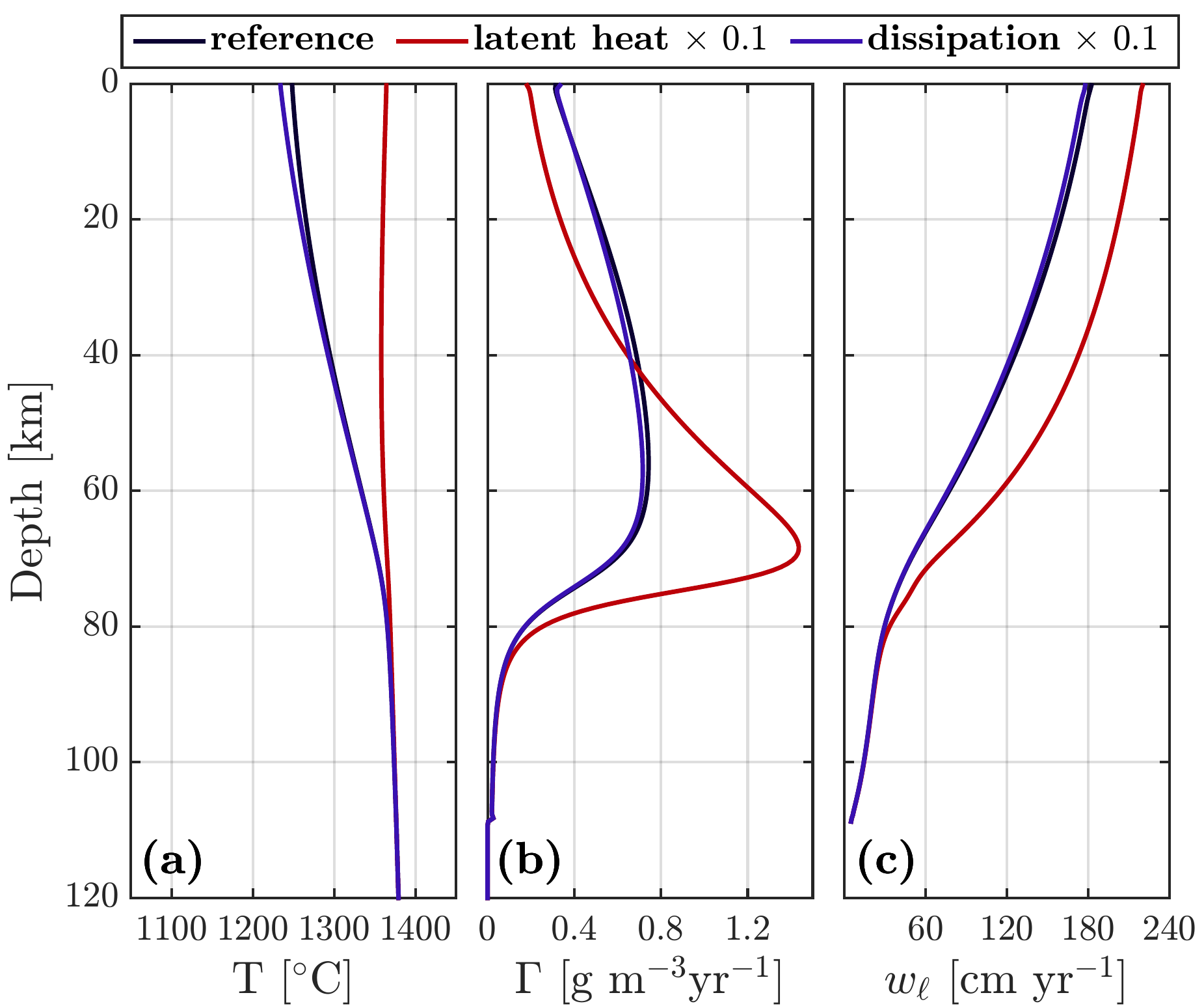}
	\caption{Effects of latent heat and viscous dissipation in 1D column model of hydrated mantle. Temperature \textbf{(a)}, net melting rate \textbf{(b)}, and vertical melt velocity \textbf{(c)} profiles with depth. \rev{Line colours represent results of reference model (black), model with reduced latent heat exchange ($\times$ 0.1, red), and reduced dissipative heating ($\times$ 0.1, blue).}}
	\label{fig:1DColEnergy}
\end{figure}

\subsubsection{Additional results}
Some consequences of energy conservation in a two-phase reactive flow model are demonstrated by some additional 1D and 2D column model results. Fig.~\ref{fig:1DColEnergy} shows the results of the reference model of hydrated mantle melting in comparison with an additional model where the latent heat exchange term is reduced by a factor of ten, and a second one where viscous dissipation is reduced by a factor of ten. Reducing the latent heat coupling leads to a mantle temperature profile similar to a solid reference adiabat, as little heat is consumed to facilitate melting. As a consequence, melting rates are increased by almost a factor of two at the base of the volatile-free melting region. The increased melt production causes an increase in melt velocity. Reducing the viscous dissipation has a less significant effect, but causes a 4\% reduction in melt production and a 10\% lower mantle temperature at the top of the melting column. The instantaneous melt fraction is barely affected by either variation in energy conservation. 

\begin{figure}[htb]
	\centering
	\includegraphics[width=0.75\textwidth]{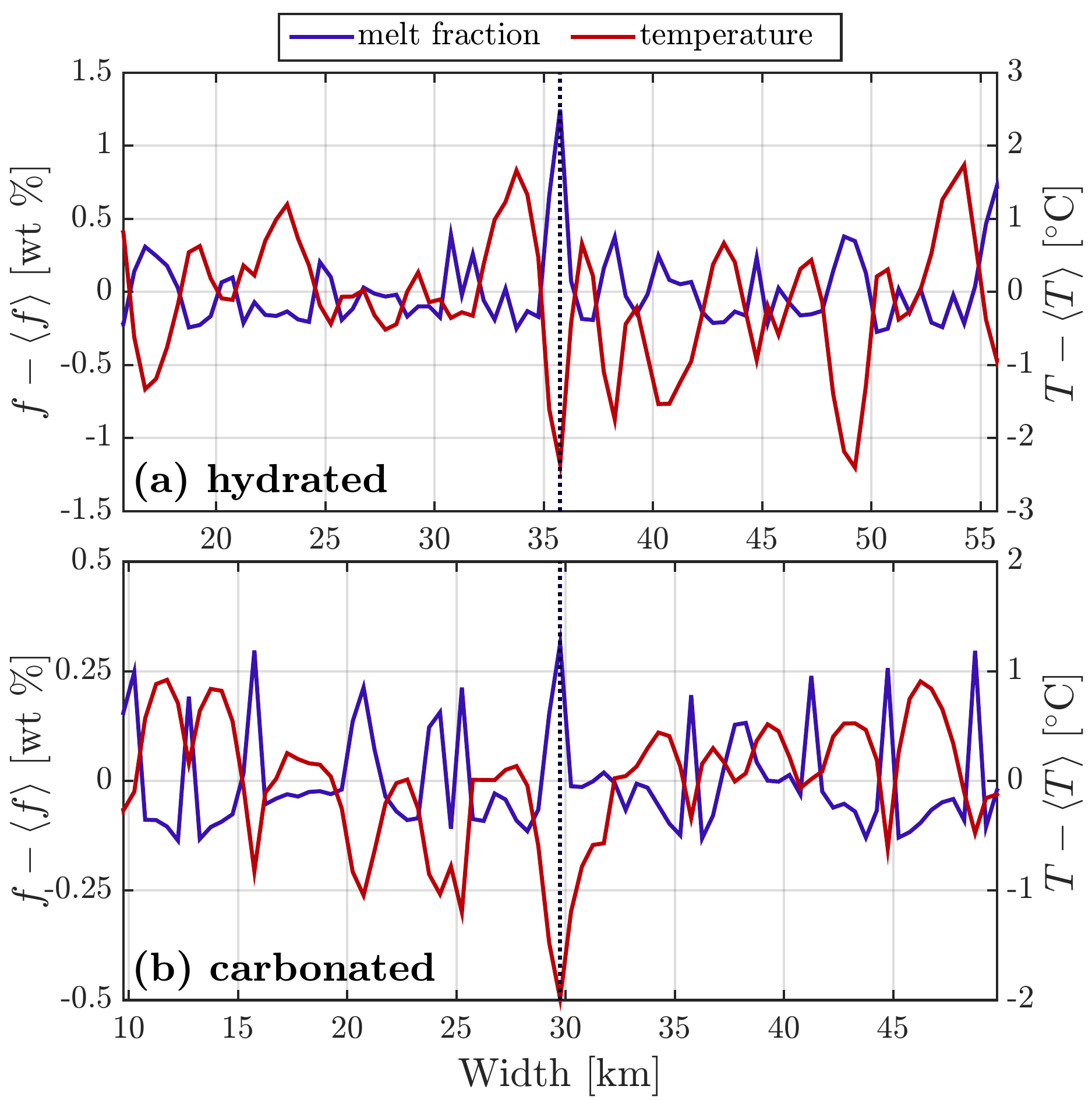}
	\caption{Thermo-chemical effects of channelized magma transport in reference 2D column models of hydrated \textbf{(a)} and carbonated \textbf{(b)} mantle. Deviation from horizontal mean of melt fraction \rev{(blue)} and temperature \rev{(red)} across max flux channels (dotted lines). Plots focused on same channels as in Figs.~\ref{fig:2DColVolFluxHydr} and \ref{fig:2DColVolFluxCarb}).}
	\label{fig:2DColThermoChem}
\end{figure}

Figure \ref{fig:2DColThermoChem} shows a section of a horizontal transect at 70~km depth for the hydrated and carbonated 2D simulations presented in section \ref{sect:Results}, giving deviations from horizontal mean for temperature and melt fraction. As a consequence of increased melting inside reactive channels, the temperature is anti-correlated to the melt fraction. Channels are marked by a relative increase in melt fraction of the order of 1~wt\% over the horizontal average, while temperatures are depressed by up to 2 degrees below the mean. Consumption of latent heat is therefore confirmed as a significant factor in limiting reactive channelisation, whereas viscous dissipation works in favour of channeling, but has a much smaller effect. This brief demonstration illustrates that an adequate statement of energy conservation has strong control over the outcome of reactive flow models.

\section{APPENDIX B: R\_DMC MODEL CALIBRATION \label{sect:AppendixB}}
\renewcommand{\theequation}{B\arabic{equation}}
\setcounter{equation}{0}
\renewcommand{\thefigure}{B\arabic{figure}}
\setcounter{figure}{0}

\subsection{Experimental constraints and calibration procedure}
To calibrate the R\_DMC method with the four effective components introduced above to the melting behaviour of volatile-enriched peridotite, we fit the $P$-dependence of component melting points $\Tmi$ and the $T$-dependence of component \rev{partition coefficients} $K^i$ to a number of experimental constraints by way of manual data assimilation. We choose to keep the number of data constraints relatively small in order to focus on coupled flow models with the R\_DMC method rather than on a sophisticated data fitting routine. 

The most important constraints on mantle melting are experimentally determined solidus curves for volatile-free, hydrated and carbonated peridotite, along with solidus data on a MORB-type pyroxenite and a mantle liquidus constraint. Complementary to these, we use estimates of the latent heat of melting of the four components and constraints on the depth of first melting beneath mid-ocean ridges, as well as on the \rev{partition coefficients} for water and carbon dioxide. The free parameters to be determined in calibrating the R\_DMC method are the ones introduced in eqns.~\eqref{eq:CompDistrCoeff} and \eqref{eq:CompMeltingPoint}: the latent heat capacities $L^i$, tuning parameters $r^i$, surface melting points $T_{m,0}^i$, and polynomial coefficients $A^i$, $B^i$.

The expressions for $K^i$ adapted from \cite{rudge11} lead to solidus and liquidus planes \rev{forming phase loops between all components}. Whereas the melting curves $T_m^i(P)$ determine the end-point temperatures of these phase loops at a given pressure, the ratio of latent heat to tuning parameter $L^i/r^i$ determine the amount of opening and curvature of the phase loops in $T,C$-space. As there are five degrees of freedom for each component, the same number of independent data constraints per component would be required to determine a unique data fit for all parameters. Often, data coverage of melting experiments will not be sufficient to fulfill this requirement. In the following, we demonstrate how an adequate fit is achieved from a small number of constraints. This procedure will, at the same time, highlight some of the important dependencies and trade-offs inherent to this parameterisation. 

First, we define the component-wise latent heat within a conservative range of values, assuming that more \rev{refractory} components with higher melting point exhibit a higher latent heat of melting. \rev{This is justified to a degree by the fact that the enthalpy of melting is thought to be characterised by the gain in disorder between crystal lattices and liquids, $\Delta S_\mathrm{fus}$, and thus for equilibrium melting $L^i = T^i_m \Delta S_\mathrm{fus}$.} The chosen values are $L^{(1)}=600$ kJ/mol for dunite (similar to forsterite), $L^{(2)}=450$ kJ/mol for MORB (within range of peridotite estimates), and $L^{(3)}=L^{(4)}=350$ kJ/mol for hydrated and carbonated MORB. For the rest of the calibration procedure, these values will be held constant. Second, the melting point of the dunite component $T_m^{(1)}$ is chosen to be equal to the quadratic polynomial given by \cite{katz03} for the mantle (harzburgite) liquidus.

\begin{figure}[htb]
	\centering
	\includegraphics[width=\textwidth]{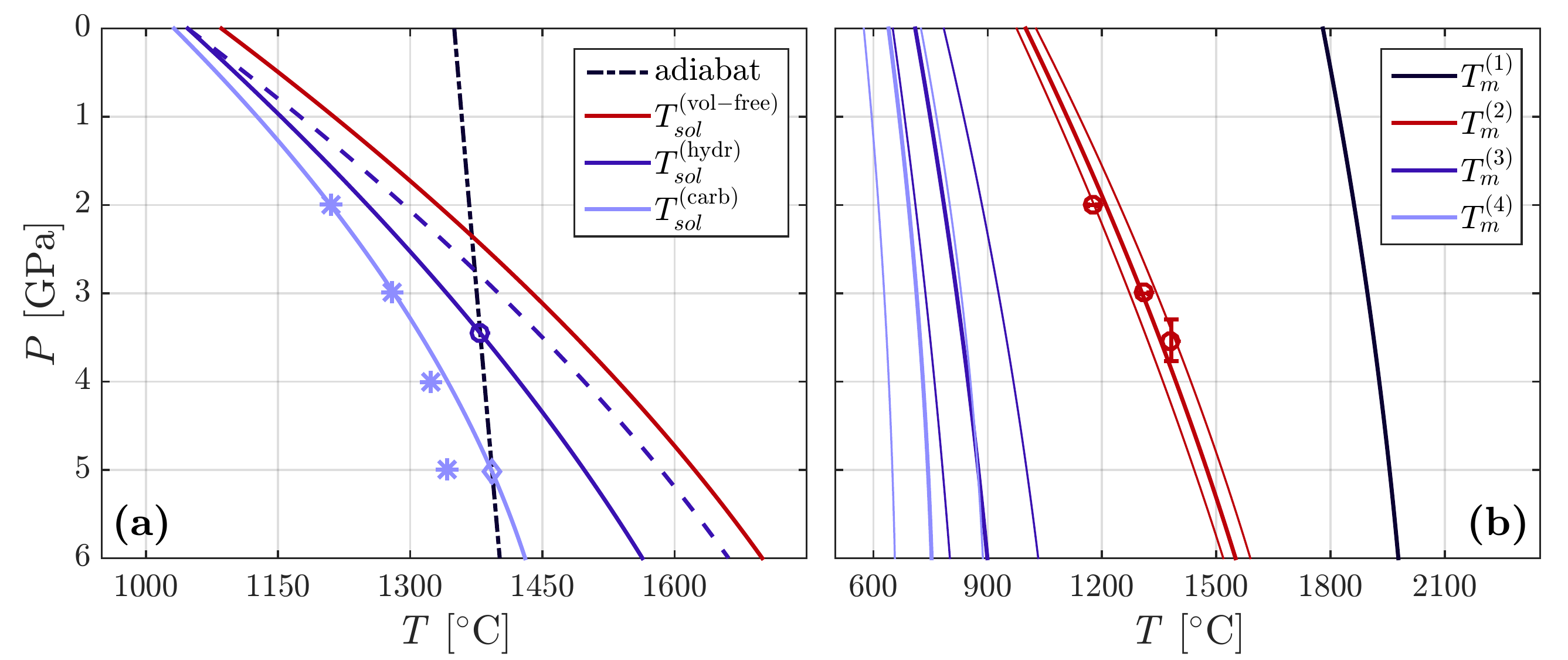}
	\caption{Constraints for calibration of component melting points $T_m^i(P)$ of mantle melting model. \textbf{(a)}: \rev{Reference polynomials for volatile-free, hydrated and carbonated fertile peridotite. Volatile-free mantle solidus from \cite{katz03} (red). Hydrated mantle solidus (blue), 100~wt~ppm bulk H$_2$O, coincides with wet solidus (dashed), 5~wt\% melt H$_2$O, by \cite{katz03} and crosses the solid adiabat (dash-dotted) at $P$ equivalent to 110~km, similar to \cite{hirth96} (circle). Carbonated mantle solidus (light blue)}, 100~wt~ppm bulk CO$_2$, calibrated to fit solidus depression data between 2 and 5~GPa from \cite{dasgupta13} (asterisks) and cross the solid adiabat at $P$ equivalent to 160~km (diamond). \textbf{(b)}: \rev{Component melting points $T_m^i$ for dunite \textbf{(1)} (black), MORB \textbf{(2)} (red), hMORB \textbf{(3)} (blue), cMORB \textbf{(4)} (light blue)}. $T_m^{(1)}$ equal to mantle liquidus by \cite{katz03}. Circles mark solidus constraints for MORB-like pyroxenite by \cite{pertermann03a}. Fine curves give high/low $r^i$ calibrations for comparison.}
	\label{fig:SolLiq}
\end{figure}

Third, we choose three reference polynomials representing the solidus curves corresponding to the three reference mantle compositions defined in section \ref{sect:Method}. Fig.~\ref{fig:SolLiq}(a) shows these quadratic polynomials for a fertile volatile-free peridotite with a bulk composition $\cbar=[0.7,0.3,0,0]$ (red); a hydrated peridotite with a water content of 100~ppm H$_2$O, at $\cbar=[0.7,0.298,0.2,0]$ (blue); a carbonated peridotite with a carbon content of 100~ppm CO$_2$, at $\cbar=[0.7,0.2995,0,0.05]$ (light blue). Shown for reference is the solid reference adiabat $T_a = T_m \exp[P \alpha/(\rho_0 c_p)]$ for a constant mantle potential temperature of $T_m=1350^{\circ}$C (dash-dotted). The reference volatile-free solidus in Fig.~\ref{fig:SolLiq} is that given by \cite{katz03}. The reference hydrated solidus is chosen to coincide at zero pressure with the wet solidus of a melt water content of 5~wt\% given by \cite{katz03} (dashed), and to intersect the solid adiabat at a pressure corresponding to a depth of 110~km (circle), similar to the 115~km for 125~ppm H$_2$0 proposed by \cite{hirth96}. The reference carbonated solidus is chosen to fit the solidus depression data given by \cite{dasgupta13} for 2--3~GPa (asterisks), and to intersect the mantle adiabat at 160~km (diamond). This depth is less well-constrained, but falls within the estimated range for carbonated peridotite \citep{dasgupta13}. Also, deeper onset of carbon dissolution melting may be inhibited by redox changes in the mantle column with depth.

\begin{figure}[htb]
	\centering
	\includegraphics[width=\textwidth]{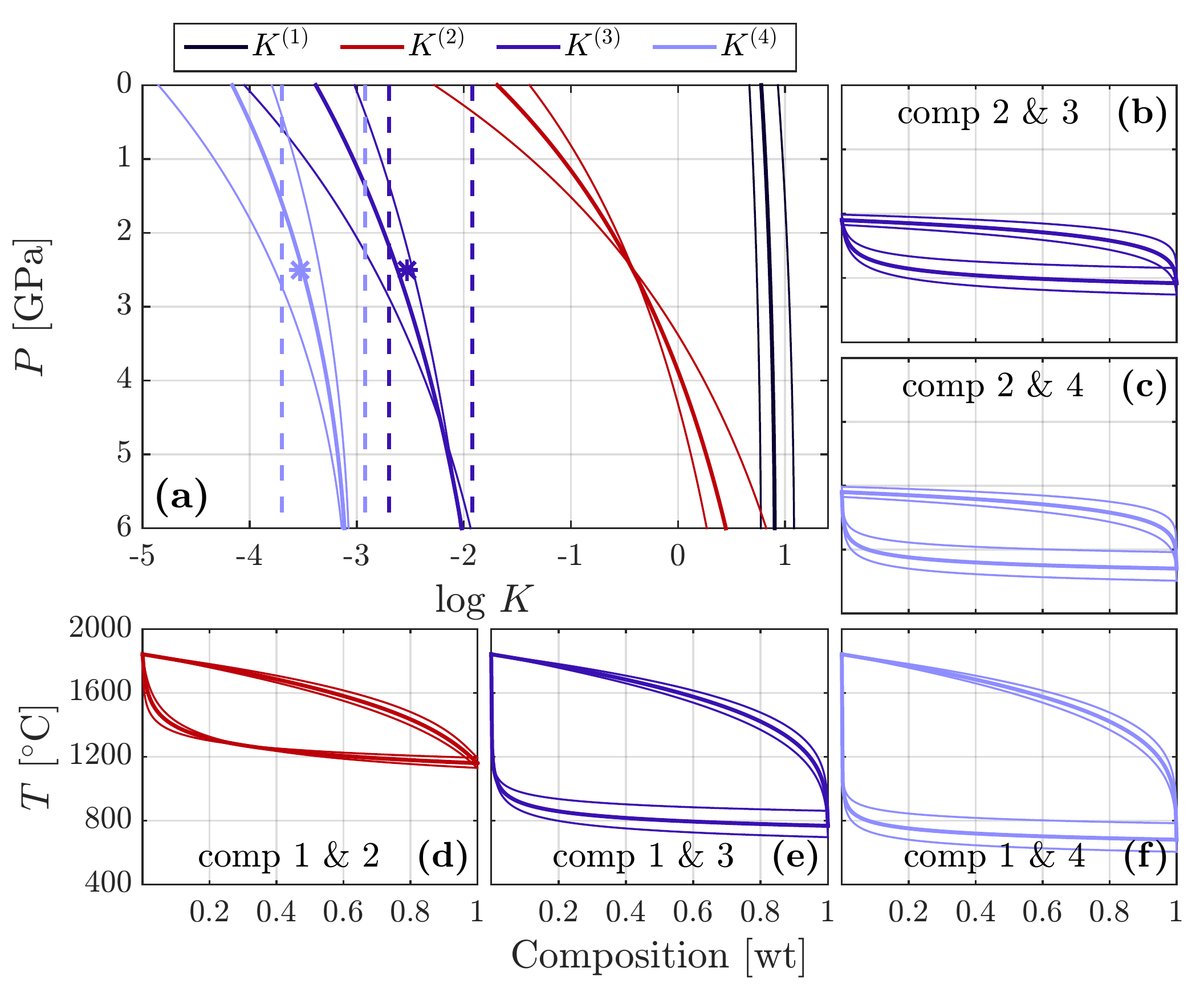}
	\caption{Constraints for calibration of component \rev{partition coefficients} $K^i(P,T)$, and resulting phase loops in $T,C$-space. \textbf{(a)}: Component \rev{partition coefficients} for \rev{dunite \textbf{(1)} (black), MORB \textbf{(2)} (red), hMORB \textbf{(3)} (blue), and cMORB \textbf{(4)} (light blue)}. Calculated $K^{\mathrm{H_2O}}$ at 2.5~GPa from \cite{hirth96} \rev{(blue asterisk)} and range of $K^{\mathrm{H_2O}}$ with depth from \cite{hirschmann09} \rev{(blue dashed)} for comparison. Same constraints shifted down by one order of magnitude \rev{(light blue asterisk, dashed)} as estimates for range of $K^{\mathrm{CO_2}}$. Binary phase loops between MORB \& hMORB \textbf{(b)}, MORB \& cMORB \textbf{(c)}, dunite \& MORB \textbf{(d)}, dunite \& hMORB \textbf{(e)}, and dunite \& cMORB \textbf{(f)} show compositional dependence of solidus and liquidus across range of bulk compositions. Fine curves show high/low $r^i$ calibrations for comparison.}
	\label{fig:KiLoops}
\end{figure}

Fourth, we seek to adjust the remaining melt model parameters $T_{m,0}^i$, $A^i$, $B^i$ and $r^i$ in such a way that the calculated solidus curves for volatile-free, hydrated and carbonated reference compositions fit the three reference polynomials discussed above. Doing so requires negotiation of a particular trade-off, where the curvature of phase loops in $T,C$-space governed by the \rev{partition coefficients} $K^i$, combined with the curvature of melting points $T_m^i$ in $P,T$-space, allows for non-unique fits of solidus temperature at a given pressure and mixed-component bulk composition. In particular, it is not immediately obvious how to choose values for fitting parameters $r^i$. Values of $r^i$ for which the properties of the calculated phase diagrams come within general expectations are $r^i$ = [60,30,30,30]. Given these values for $r^i$, the remaining free parameters $T_{m,0}^i$, $A^i$, $B^i$ are manually adjusted such that the $P$-dependent solidus curves at volatile-free, hydrated and carbonated reference compositions fit the three solidus reference polynomials. The results of this procedure are shown in Fig.~\ref{fig:SolLiq}(b) and Fig.~\ref{fig:KiLoops}. in terms of $P$-dependent profiles of $T_m^i$ and $K^i$, along with phase loops in $T,C$-space (given at $P$=1.5~GPa). To highlight the effect of a particular choice of $r^i$, these plots include results from two additional calibrations performed at $r^i$ of $\pm$10 compared to preferred values.

The following additional constraints helped to select this preferred calibration. The circles in Fig.~\ref{fig:SolLiq}(b) are solidus data points obtained by \cite{pertermann03a} from melting experiments on a MORB-type pyroxenite. Our preferred calibration of $T_m^{(2)}$ falls within the range of these constraints. The dark blue asterisk in the same plot gives the value of the \rev{partition coefficient} of water, $K^\mathrm{H_2O}$, calculated at $P$ = 2.5~GPa by \cite{hirth96}. \rev{Since that study, others have produced refined estimates. Most importantly, $K^\mathrm{H_2O}$ is found to strongly decrease with decompression, decrease with increasing activity of water \citep{asimow04}, and vary with mineral mode \citep{hirschmann09}.} The dark blue dashed lines in Fig.~\ref{fig:KiLoops}(a) outline the range of values for $K^\mathrm{H_2O}$ given by \cite{hirschmann09}. Our calibration of $K^{(3)}$ is such that it falls within the range of these constraints across the hydrated melting $P$-range of 2.2--3.6~GPa. Experimental constraints on the partitioning of carbon in mantle melting are not as readily available, but it is generally assumed that $K^\mathrm{CO_2}$ is about an order of magnitude smaller than $K^\mathrm{H_2O}$. Light blue dashed lines and asterisk in Fig.~\ref{fig:KiLoops}(a) are the same constraints as for $K^\mathrm{H_2O}$, but shifted down by one order of magnitude. Again, our preferred calibration is chosen such that $K^{(4)}$ fits these constraint within the carbonated melting pressure range of 3.6--5~GPa.

\subsection{Four-component models of hydrated and carbonated mantle melting}
As mentioned above, due to computational limitations we only employ simulations for two and three component mixtures, representing either volatile-free, hydrated or carbonated mantle melting. For each component added to the compositional space in a simulation, two more degrees of freedom need to be solved for (i.e., $\csi$ and $\cli$ for $i=1,...,n$). The R\_DMC method, however, is general for $n$ components. As computational methods become more efficient, it will be possible to extend simulations to a mantle that is both hydrated and carbonated (or any other melting system involving four or more components of composition).

\begin{figure}[htb]
	\centering
	\includegraphics[width=0.75\textwidth]{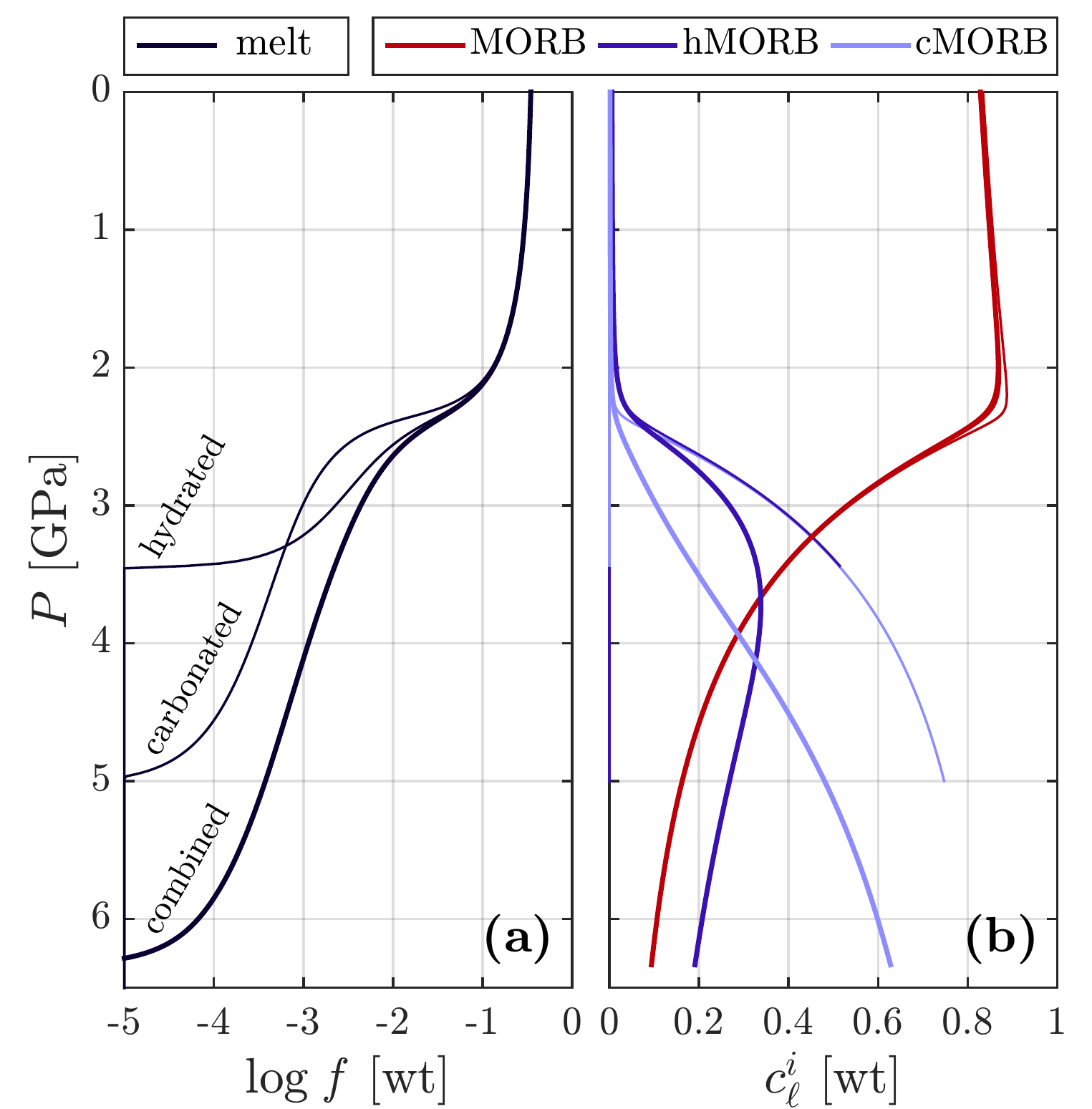}
	\caption{Four component compositional model including both water and carbon content. Equilibrium state computed along mantle adiabat at $T_m=1350^{\circ}$C over a $P$ range of 0-6.5~GPa. Equilibrium melt fraction \textbf{(a)}, and melt composition \textbf{(b)} for bulk volatile content of 100~ppm H$_2$O and CO$_2$ combined (bold), and same amount of either component for reference (fine).}
	\label{fig:FourComp}
\end{figure}

Figure \ref{fig:FourComp} provides a first look at the melting behaviour and compositional evolution that is predicted by phase relations computed for a four component mixture of hydrated and carbonated peridotite with a bulk composition of $\cbar=[0.7,0.2975,0.2,0.05]$, corresponding to 100 wt~ppm of H$_2$O and CO$_2$. These results are obtained by reading out equilibrium melt fraction and phase compositions over a range of $P,T$-conditions by means of the Matlab routines provided \rev{in the online distribution of the R\_DMC method \citep{rdmc-repo}}. As predicted \citep{dasgupta13}, the presence of water and carbon dioxide together lower the onset of melting further than either the water or carbon content alone would do. 

At depth, only small fractions of around 0.01\% melt are produced. The onset of melting is predicted to occur at 200~km depth. With decreasing pressure, gradually higher melt fractions of up to 1\% are stabilised towards the volatile-free solidus depth. The first melt produced at depth mostly consists of carbonated silicate ($\sim$0.6 cMORB). At pressures corresponding to the wet melting region (115-75~km), the equilibrium melt composition becomes increasingly enriched in the hydrated silicate component ($\sim$0.3 hMORB). Crossing into the pressure range of below 75~km, melt compositions rapidly transition to dominantly MORB-type basaltic melts ($\sim$0.8 MORB), with very small fractions of both volatile-bearing components stable in the melt.

These first predictions indicate that dynamic models including all four effective components defined here will be able to capture a number of leading-order characteristics of mantle melting in the presence of water and carbon dioxide. Similarly, the R\_DMC method may be applied to other multi-component systems in the future to study other aspects of magmatic systems in the mantle and crust.

\end{document}